\documentclass[11pt,a4paper]{article}

\pdfoutput=1

\RequirePackage[OT1]{fontenc}
\RequirePackage{amsthm,amsmath}
\RequirePackage[authoryear,round]{natbib}
\RequirePackage[dvipsnames]{xcolor}
\RequirePackage[colorlinks,citecolor=RoyalBlue,urlcolor=RoyalBlue,linkcolor=RoyalBlue]{hyperref}

\usepackage[margin=2.7cm]{geometry}
\usepackage[titletoc,title]{appendix}

\usepackage{amsfonts,amssymb}
\usepackage{mathtools}

\usepackage{multirow}
\usepackage{graphicx}
\usepackage{booktabs}

\usepackage{eucal}
\usepackage{color}
\usepackage{mathpazo}

\usepackage[caption=false]{subfig}
\usepackage{caption}

\allowdisplaybreaks

\usepackage{mathsMacros}

\graphicspath{{Figures/}}

\usepackage{pifont}
\newcommand{\cmark}{\ding{51}}
\newcommand{\xmark}{\ding{55}}

\title{Lateral transfer in Stochastic Dollo models\thanks{Corresponding author: Luke J. Kelly, \texttt{kelly@stats.ox.ac.uk}.}}
\author{Luke J. Kelly\thanks{Supported in part by the St John's College and Engineering and Physical Sciences Research Council partnership award EP/J500495/1.} \quad Geoff K. Nicholls \\ ~ \\ {\small Department of Statistics, University of Oxford}}
\date{}

\begin{document}

\maketitle

\begin{abstract}

Lateral transfer, a process whereby species exchange evolutionary traits through non-ancestral relationships, is a frequent source of model misspecification in phylogenetic inference. Lateral transfer obscures the phylogenetic signal in the data as the histories of affected traits are mosaics of the overall phylogeny. We control for the effect of lateral transfer in a Stochastic Dollo model and a Bayesian setting. Our likelihood is highly intractable as the parameters are the solution of a sequence of large systems of differential equations representing the expected evolution of traits along a tree. We illustrate our method on a data set of lexical traits in Eastern Polynesian languages and obtain an improved fit over the corresponding model without lateral transfer.

\end{abstract}

\section{Introduction}
\label{sec:intro}

Evolutionary traits used to infer the ancestry of a set of taxa take many forms beside DNA base values in sequence data. For example, \citet{cybis15} study the spread of antibiotic resistance in Salmonella strains and \citet{jofre17} estimate the shared ancestry of twenty-two stars from measurements on seventeen elements in their chemical composition. In this paper, we infer the shared ancestry of languages from lexical trait data.

When species evolve in isolation, we commonly assume that traits pass vertically from one generation to the next through ancestral relationships. A phylogenetic tree describes the shared ancestry of taxa which evolve in this manner: branches represent evolving species, internal nodes depict speciation events, and leaf nodes correspond to observed taxa. In this paper, we wish to infer the phylogeny of taxa which evolved through a combination of vertical and \emph{lateral} trait transfer. Lateral transfer, such as \emph{horizontal gene transfer} in biology or \emph{borrowing} in linguistics, is an evolutionary process whereby species acquire traits through non-vertical relationships.

Lateral transfer distorts the phylogenetic signal of the speciation events in the data as the histories of affected traits may conflict with the overall taxa phylogeny. Models based solely on vertical trait inheritance are misspecified in this setting and, in our experience, this model error can result in overly high levels of confidence in poorly fitting trees. In this article, we develop a fully model-based Bayesian method for trait presence/absence data which explicitly accounts for lateral transfer in reconstructing dated phylogenies.

To illustrate our method, we analyse a data set of lexical traits in Eastern Polynesian languages. There have been many previous phylogenetic studies of languages and language families, including Austronesian \citep{gray09}, Indo-European \citep{gray03, nicholls08, ryder11, bouckaert12, chang15}, Linear B \citep{skelton08} and Semitic \citep{kitchen09}. Lateral transfer is a frequent occurrence in language diversification \citep{greenhill09}, yet a common theme of the above studies is that the authors do not control for it in their fitted models. Typically, known-transferred traits are discarded and a model for vertical trait transfer is fit to the remainder (\citealp{gray03, bouckaert12}; and many others). This approach is problematic as recently transferred traits are more readily identified, so earlier transfers remain in the data set.

There exist various methods which test for evidence of lateral transfer in data but do not estimate a phylogeny. \citet{patterson12} review various tests for admixture in allele frequency data, while \citet{daubin02}, \citet{beiko06} and \citet{abby10} describe similar tests for sequence data which compare gene trees to a species tree constructed \emph{a priori}. Similarly, internal nodes in implicit phylogenetic networks accommodate incompatibilities in the data with the assumption of an underlying species tree but do not necessarily represent the evolutionary history of the taxa \citep{huson06, oldman16}. Under the assumption of random trait transfer between lineages, \citet{roch13} demonstrate that a number of nonparametric reconstruction methods recover the true phylogeny with high probability when the expected number of transfer events is bounded.

The problem of controlling for lateral transfer in inference for dated phylogenies has received little attention in the statistics literature. In particular, there are few fully likelihood-based inference schemes for dated phylogenies which control for lateral transfer for any data type. Parametric inference for the underlying phylogeny with an explicit model for lateral transfer is a difficult computational problem. This is due to the near intractability of the likelihood calculation as pruning \citep{felsenstein81} is no longer directly applicable in integrating over unobserved trait histories. Approximate Bayesian computation, although a useful tool for estimating demographic parameters in complex models \citep{tavare97} or selecting a particular tree from a restricted set of alternatives \citep{veeramah15}, does not help here as a summary statistic which informs a dated phylogeny has to be relatively high dimensional, thereby leading to low acceptance rates in simulation.

\citet{lathrop82} and \citet{pickrell12} describe methods to infer explicit phylogenetic networks of population splits and instantaneous hybridisation events from allele frequency data. For input gene trees inferred \emph{a priori}, \citet{kubatko09} investigates the support for the hybridisation events in a given hybrid phylogeny under the multispecies coalescent model \citep{rannala03}. \citet{wen16} describe a Bayesian method to infer an explicit phylogenetic network under the multispecies coalescent model for the input gene trees.

From a set of input gene trees, \citet{szollosi12} seek the species tree which maximises the likelihood of reconciling the gene trees under their model incorporating lateral transfer, trait gain and loss. The authors discretise time on the tree, limiting the number of transfer events which may occur, so that their computation is tractable. This allows them to consider many more taxa than \citet{wen16}, for example. In addition, their method returns a time-ordering of the internal nodes rather than a fully dated tree. We summarise these model-based methods in Table~\ref{tab:criteria}.

In this paper, we describe a novel method for inferring dated phylogenies from trait presence/absence data. This research is motivated by problems such as the example in Section~\ref{sec:applications} where we estimate a \emph{language tree} from lexical trait data. These data sets are gathered under a different experimental design to sequence data used to infer gene trees. In collecting trait presence/absence data, we choose a trait and record which taxa display it; the patterns of presence and absence of traits across taxa are informative of the tree. Gene content data is defined similarly \citep{huson04}. In contrast, in the design for gene tree data, we choose a gene and sequence homologs of that gene in each individual corresponding to a leaf; a gene is a complex trait and the displayed characters inform the gene tree. In the context of our application in Section~\ref{sec:applications}, it may be tempting to think of lexical traits as genes and languages as biological species. The analogy does not hold as the objects in trait presence/absence data and gene tree data have different meanings due to the different experimental designs. For these and other reasons summarised in Table~\ref{tab:criteria}, the model-based methods we cite above are not directly applicable to the problem at hand.

\begin{table}[t]
\centering
\captionsetup{singlelinecheck=off}
\caption[]{Model-based phylogenetic methods which incorporate lateral transfer. The criteria are:
\begin{itemize}
\item[(A)] The method infers dated phylogenies controlling for lateral transfer (does not require known species phylogeny or gene trees as input).
\item[(B)] The method quantifies uncertainty in parameters, tree structure and node times.
\item[(C)] The method uses exact model-based inference (up to Monte Carlo error) or an explicitly quantified approximation.
\item[(D)] The model is a generative description of the observation process for the data the authors analyse, with physically meaningful parameters.
\item[(E)] The approach is directly applicable to our binary Dollo trait data.
\end{itemize}}
\label{tab:criteria}
\begin{tabular*}{\textwidth}{ll @{\extracolsep{\fill}} ccccc} \toprule
\multirow{2}{*}{Method}	&	\multirow{2}{*}{Input}	&	\multicolumn{5}{c}{Criteria} \\ \cmidrule{3-7}
	&	&	(A)		&	(B)		&	(C)		&	(D)		&	(E)	 \\ \midrule
\citet{pickrell12}	&	Allele frequencies
		&	\cmark	&	\cmark	&	\xmark	&	\cmark	&	\xmark	\\
\citet{lathrop82}	&	Allele frequencies
		&	\cmark	&	\cmark	&	\cmark	&	\cmark	&	\xmark	\\
\citet{kubatko09}	&	Gene trees, species tree
		&	\xmark	&	\cmark	&	\cmark	&	\cmark	&	\xmark	\\
\citet{szollosi12}	&	Gene trees
		&	\xmark	&	\cmark	&	\xmark	&	\cmark	&	\xmark \\
\citet{wen16}		&	Gene trees
		&	\xmark	&	\cmark	&	\cmark	&	\cmark	&	\xmark \\ \bottomrule
\end{tabular*}
\end{table}

We take the \emph{Stochastic Dollo} model of \citet{nicholls08} (SD) for unordered sets of trait presence/absence data as the starting point for our lateral transfer model. The SD model posits a birth-death process of traits along each branch of the tree, with parent traits copied into offspring at a branching event. The basic process respects \emph{Dollo parsimony}: each trait is born exactly once, and once a trait is extinct, it remains so. \citet{alekseyenko08} extend the SD model for multiple character states and \citet{ryder11} introduce missing data and rate heterogeneity. \citet{bouchard13} describe a sequence-valued counterpart to the SD model. The SD model has been implemented in the popular phylogenetics software packages \texttt{BEAST} \citep{drummond12} and \texttt{BEAST 2} \citep{beast2}. In a recent study, \citet{mcpherson16} use the SD model to infer cancer clone phylogenies from tumour samples. Simulation studies of the SD model show that topology estimates are robust to moderate levels of random lateral transfer when the underlying topology is balanced \citep{greenhill09} but the root time is typically biased towards the present \citep{nicholls08, ryder11}.

\citet{nicholls08} describe how to simulate lateral transfer in the basic SD model whereby each species randomly acquires copies of traits from its contemporaries. No previous attempt has been made to fit this model incorporating lateral trait transfer. We perform exact likelihood-based inference under this model. Our lateral transfer process is ultimately defined by the description in Section~\ref{sec:model} and we do not attempt to model specific processes such as incomplete lineage sorting, hybridisation or gene introgression directly. While our model can generate the trait histories which arise in these processes, it also generates many others and we recommend further case-specific modelling. We do not infer trait trees in advance then reconcile them to form a species tree; rather, we use Markov chain Monte Carlo methods to sample species trees and parameters, and integrate over all possible trait histories under our model in computing the likelihood. In contrast to \citet{szollosi12}, our method operates in a continuous-time setting and we are able to infer the timing of speciation events.

The SD model with lateral transfer will be misspecified for lexical trait data in many ways. Trait birth, death and transfer events will be correlated in complex ways due to real-world processes that we do not model. We are particularly interested in misspecification-induced bias impacting branching time and tree topology estimates. We test for this bias by removing information constraints on known leaf ages and checking that they are correctly reconstructed. This is a test for evidence against the model akin to a pure test for significance in a frequentist setting. These tests demonstrate that \emph{whatever} the misspecification, there is no evidence that it is impacting our estimates. The SD model is a special case of our model and is a natural basis for assessing the effect of controlling for lateral transfer at the expense of an increase in computational cost. We show that the SD model fails these misspecification tests.

To summarise our approach, we build a detailed \emph{ab initio} model of trait and tree dynamics which fully describes the data-observation process. In doing so, we do not compromise the model to make it easier to fit. The price we pay is a massive integration over the unobserved trait histories. In looking for competing methods, we focus on methods which infer dated trees, can quantify the uncertainty in their estimates and perform exact inference or use explicitly quantified approximations. For the lexical trait data we consider, there are no obvious benchmarks among the competing model-based inference schemes discussed above and summarised in Table~\ref{tab:criteria}. Our method satisfies each of these criteria.

We describe our binary trait data in Section~\ref{sec:data} and introduce our lateral transfer model in Section~\ref{sec:model}. We describe the likelihood calculation in Section~\ref{sec:like} and extensions to the model in Section~\ref{sec:modExt}. We discuss our inference method in Section~\ref{sec:inference} and tests to validate our computer implementation in Section~\ref{sec:testing}. We illustrate our model on a data set of lexical traits in Eastern Polynesian languages in Section~\ref{sec:applications} and conclude in Section~\ref{sec:conclusions} with a discussion of the model and possible directions for future research.

\section{Homologous trait data}
\label{sec:data}

Homologous traits derive from a common ancestral trait through a combination of vertical inheritance and lateral transfer events. We assign each set of homologous traits a unique common label from the set of trait labels, $ \cZ $. A set of trait categories is chosen and, for each taxon in the study, we gather instances of traits in each category. We record the status of trait \emph{h} in taxon \emph{i} as
\[
d_i^h =
\left\{
\begin{array}{ll}
1, \quad & \text{trait \emph{h} is present in taxon \emph{i},} \vspace{1pt}\\
0, \quad & \text{trait \emph{h} is absent in taxon \emph{i},} \vspace{1pt}\\
?, \quad & \text{the status of trait \emph{h} in taxon \emph{i} is unknown.} \\
\end{array}
\right.
\]
We denote by \textbf{D} the array recording the status of each trait across the observed taxa. A column $ \bd^h $ of $ \bD $ is a \emph{site-pattern} recording the status of trait \emph{h} across the taxa. These patterns of trait presence and absence, which we assume are independent, exchangeable entities, shall form the basis of our model.

In the analysis in Section~\ref{sec:applications}, each trait is a word in one of 210 meaning categories and each taxon is an Eastern Polynesian language. For example, the Maori and Hawaiian words for \emph{woman} and \emph{wife}, both \emph{wahine}, derive from a common ancestor \emph{h}, say, so $ d_{\mathrm{Maori}}^h = d_{\mathrm{Hawaiian}}^h = 1 $. On the other hand, the Maori word for \emph{mother}, \emph{whaea}, is not related to its Hawaiian counterpart, \emph{makuahine}, so we record zeros in the respective entries of the data array.

\section{Generative model}
\label{sec:model}

\begin{figure}[t]
\centering
\includegraphics[width=\textwidth]{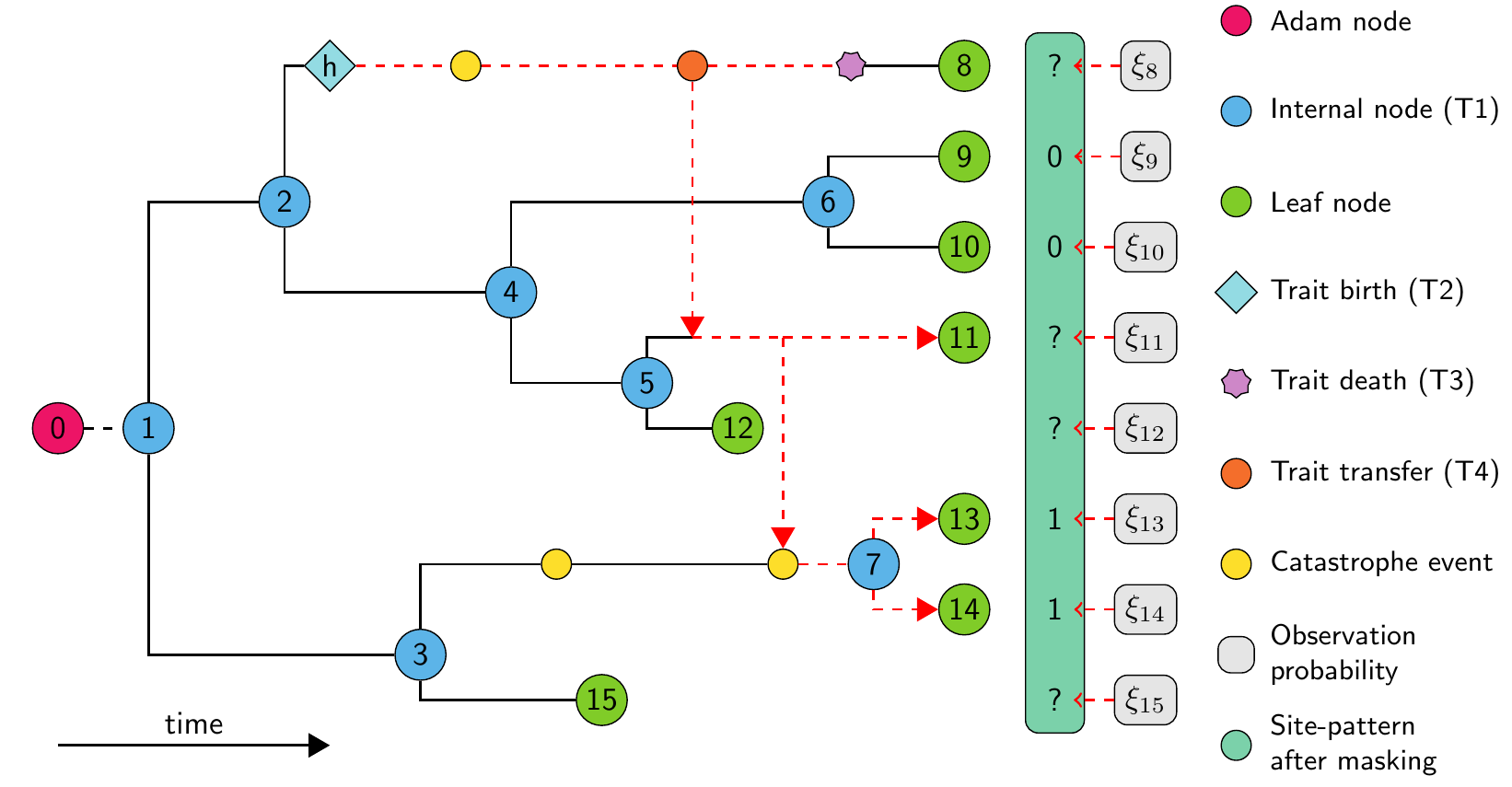}
\caption{Illustration of the Stochastic Dollo with Lateral Transfer model. Dashed lines represent the history of a trait \emph{h}. We describe catastrophes, missing data and offset leaves in Section~\ref{sec:modExt}.}
\label{fig:phyTree}
\end{figure}

A branching process on sets of traits determines the phylogeny of the observed taxa. The set contents diversify according to a process of trait birth, death and lateral transfer events. We describe these events in greater detail below. Figure~\ref{fig:phyTree} depicts a realisation of the model and the history of a single trait. The trait history bears little resemblance to the underlying phylogeny as a consequence of trait death and lateral transfer events.

We first define our model and inference method in terms of binary patterns of trait presence and absence in taxa which are recorded simultaneously. In Section~\ref{sec:modExt}, we extend the model to incorporate missing data and different leaf sampling times.

A rooted phylogenetic tree $ g = (V, E, T) $ on \emph{L} leaves is a connected, acyclic graph with node set $ V = \{0, 1, \dotsc, 2L - 1\} $, directed edge set \emph{E} and node times $ T \in \{-\infty\} \times \R^{2L - 1} $. The node set \emph{V} comprises one \emph{Adam} node labelled 0 of degree 1, the internal nodes $ V_A = \{1, 2, \dotsc, L - 1\} $ of degree 3, and the leaf nodes $ V_L = \{L, L + 1, \dotsc, 2L - 1\} $ of degree 1. Node $ i \in V $ arises at time $ t_i \in T $ denoting when the corresponding event occurred relative to the current time, 0. For convenience, we label the internal nodes $ V_A $ in such a way that $ t_1, \dotsc, t_{L - 1} $ is a strictly increasing sequence of node times. We observe the taxa at time 0 so $ t_i = 0 $ for $ i \in V_L $.

Edges represent evolving species and are directed forwards in time. We label each edge by its offspring node: if $ \pa(i) $ denotes the parent of node $ i \in V \setminus \{0\} $, edge $ i \in E $ runs from node $ \pa(i) $ at time $ t_{\pa(i)} $ to \emph{i} at time $ t_i $. We assume that the Adam node arose at time $ t_0 = -\infty $, so a branch of infinite length connects it to the \emph{root} node 1 at time $ t_1 $. If we slice the tree at time \emph{t}, there are $ \Lt $ species labelled $ \kt = (i \in E : t_{\pa(i)} \leq t < t_i) $. In Figure~\ref{fig:phyTree}, there are $ L^{(t_2)} = 3 $ species labelled $ \bk^{(t_2)} = (8, 4, 3) $ immediately after the speciation event at time $ t_2 $, for example.

Let $ H_i(t) \subset \cZ $ denote the set of traits possessed by species $ i \in \kt $ at time \emph{t}. We now define four properties of the set-valued evolutionary process $ H(t) = \{H_i(t) : i \in \kt\} $ for $ t \in (-\infty, 0] $.

\begin{myprop}[Set branching event]
\label{tp:branch}
Species $ i \in \bk^{(t_i^-)} $ branches at time $ t_i $ and is replaced by two identical offspring, \emph{j} and $ k \in \bk^{(t_i)} $,
\begin{align*}
H_j(t_i) &\leftarrow H_i(t_i^-), \\
H_k(t_i) &\leftarrow H_i(t_i^-),
\end{align*}
where $ t_i^- $ denotes the time just before the branching event.
\end{myprop}

\begin{myprop}[Trait birth]
\label{tp:birth}
New traits are born at rate $ \lambda $ over time in each extant species. If trait $ h \in \cZ $ is born in species $ i $ at time \emph{t}, then
\[
H_i(t) \leftarrow H_i(t^-) \cup \{h\}.
\]
\end{myprop}

\begin{myprop}[Trait death]
\label{tp:death}
A species kills off each trait it possesses independently at rate $ \mu $. If trait $ h \in H_i(t^-) $ in species $ i $ dies at time \emph{t}, then
\[
H_i(t) \leftarrow H_i(t^-) \setminus \{h\}.
\]
\end{myprop}

\begin{myprop}[Lateral trait transfer]
\label{tp:transfer}
Each instance of a trait attempts to transfer at rate $ \beta $. Equivalently, a species acquires a copy of a trait by lateral transfer at rate $ \beta $ scaled by the fraction of extant species which possess it. If species \emph{i} acquires a copy of trait $ h \in \cHtm = \bigcup_{i \in \ktm} H_i(t^-) $ at time \emph{t}, then
\[
H_i(t) \leftarrow H_i(t^-) \cup \{h\}.
\]
Clearly, if $ h \in H_i(t^-) $ then the transfer event has no effect.
\end{myprop}

Starting from a single set $ H(-\infty) = \{\emptyset\} $, the process $ H(t) $ evolves as a continuous-time Markov chain through a combination of branching~(T\ref{tp:branch}) and trait~(T\ref{tp:birth}--\ref{tp:transfer}) events to yield the diverse set of taxa $ H(0) = \{H_i(0) : i \in V_L\} $ that we observe at time 0. When the lateral transfer rate $ \beta = 0 $, we recover the binary Stochastic Dollo process of \citet{nicholls08}.

\section{Likelihood calculation}
\label{sec:like}

We may calculate the likelihood of a given trait history in terms of independent holding times and jumps between states (T\ref{tp:branch}--\ref{tp:transfer}). However, trait histories are nuisance parameters here as we are interested in the overall phylogeny so we must integrate them out of the model likelihood. Furthermore, we must account for the histories of traits born on the tree which did not survive into the taxa. In order to describe how to simultaneously integrate over all possible trait histories on the tree under our model, we now recast the trait process in terms of evolving patterns of presence and absence across branches.

\subsection{Pattern evolution}
\label{sec:pattern}

If we cut through the tree at time \emph{t}, each trait in $ \cHt $ displays a \emph{pattern} of presence and absence across the $ \Lt $ extant species $ \kt = (\kit : i \in [\Lt]) $, where $ [\Lt] = \{1, \dotsc, \Lt\} $. These patterns of presence and absence evolve over time as new branches arise and instances of traits die and transfer. The pattern displayed by trait $ h \in \cHt $ at time \emph{t} is $ \pht = (\phit : i \in [\Lt]) $, where
\[
\phit =
\left\{
\begin{array}{ll}
1, \quad & h \in H_{\kit}(t), \\
0, \quad & \text{otherwise},
\end{array}
\right.
\]
indicates the presence or absence of trait \emph{h} on lineage $ \kit $ at time \emph{t}.

The space of binary patterns of trait presence and absence across $ \Lt $ lineages is $ \cPt = \{0, 1\}^{\Lt} \setminus \{\zeros\} $, where $ \zeros $ denotes an $ \Lt $-tuple of zeros. Trait labels are exchangeable and there are $ \Npt = |\{h \in \cHt : \pht = \bp\}| $ traits displaying pattern $ \bp \in \cPt $ at time \emph{t}. The dynamics of the pattern frequency process $ \bNt = (\Npt : \bp \in \cPt) $ follow directly from Properties~T\ref{tp:branch}--\ref{tp:transfer} of the trait process in Section~\ref{sec:model}.

\subsubsection{Patterns at branching events}
\label{sec:patternsAtBranching}

At a branching event, patterns gain an entry and the space of patterns increases accordingly. The tuple $ \kt $ of branch labels is consistent across speciation events in the sense that when lineage $ k_i^{(t_j^-)} $ branches at time $ t_j $,
\[
\bk^{(t_j^-)}
	\rightarrow \bk^{(t_j)}
	= \left(k_1^{(t_j^-)}, \dotsc, k_{i - 1}^{(t_j^-)}, k_i^{(t_j)}, k_{i + 1}^{(t_j)}, k_{i + 1}^{(t_j^-)}, \dotsc, k_{L^{(t_j^-)}}^{(t_j^-)}\right),
\]
where species $ k_i^{(t_j)} $ and $ k_{i + 1}^{(t_j)} $ are the offspring of species $ k_i^{(t_j^-)} $ (T{\ref{tp:branch}). It follows that each trait $ h \in \cH^{(t_j^-)} $ transitions to display a pattern $ \ph(t_j) $ with entries $ p^h_i(t_j) = p^h_{i + 1}(t_j) \leftarrow p^h_i(t_j^-) $. For example, reading from top to bottom in Figure~\ref{fig:phyTree},
\begin{gather*}
\begin{aligned}
\bk^{(t_4^-)} &= (8, 4, 7, 15), &\quad \bk^{(t_4)} &= (8, 6, 5, 7, 15), \\
\bp^h(t_4^-) &= (1, 0, 0, 0), &\quad \bp^h(t_4) &= (1, 0, 0, 0, 0),
\end{aligned}
\end{gather*}
as a result of the speciation event at node 4.

A pattern $ \bp \in \cP^{(t_j)} $ with entries $ p_i = p_{i + 1} $ is consistent with the branching event on lineage $ k_i^{(t_j^-)} $ as it may be formed by duplicating the \emph{i}th entries of a pattern in $ \cP^{(t_j^-)} $. On the other hand, the trait process cannot generate a pattern $ \bp \in \cPt $ with $ p_i \neq p_{i + 1} $ at time $ t_j $ by definition~(T\ref{tp:branch}). We denote by $ \bT^{(j)} : \bN(t_j^-) \rightarrow \bN(t_j) $ the operation which initialises the pattern frequencies $ \bN(t_j) $ with entries of $ \bN(t_j^-) $ for patterns consistent with the branching event, and zeros otherwise. We return to this initialisation operation when we compute the expected pattern frequencies in Section~\ref{sec:epm}.

\subsubsection{Patterns between branching events}
\label{sec:patternsBetweenBranching}

In order to formally describe the Markovian evolution of the pattern frequencies $ \bNt $ between branching events, we first define how patterns relate to each other. The Hamming distance between patterns $ \bp $ and $ \bq \in \cPt $ is $ d(\bp, \bq) = |\{i \in [\Lt] : p_i \neq q_i\}| $ and $ s(\bp) = d(\bp, \zeros) $ is the Hamming weight of $ \bp $. A trait displaying pattern $ \bp $ at time \emph{t} \emph{communicates} with patterns in the sets
\begin{align*}
\Spm &= \{\bq \in \cPt : s(\bq) = s(\bp) - 1,\, d(\bp, \bq) = 1\}, \\
\Spp &= \{\bq \in \cPt : s(\bq) = s(\bp) + 1,\, d(\bp, \bq) = 1\},
\end{align*}
the patterns which differ from $ \bp $ through a single trait death~(T\ref{tp:death}) or transfer~(T\ref{tp:transfer}) event respectively. Figure~\ref{fig:patternTransitions} describes the transition rates between pattern states $ \bp $ and $ \bq \in \Spm \cup \Spp $. New traits displaying patterns of Hamming weight 1 arise on each branch through trait birth events~(T\ref{tp:birth}). For example, reading from top to bottom in Figure~\ref{fig:phyTree}, a copy of trait \emph{h} transfers at time \emph{t} from branch $ k_1^{(t^-)} = 1 $ to $ k_3^{(t)} = 11 $, so
\begin{align*}
\bp^h(t^-) &= (1, 0, 0, 0, 0, 0), &\quad \bp^h(t) &= (1, 0, 1, 0, 0, 0) \in S_{100000}^+, \\
N_{100000}(t) &= N_{100000}(t^-) - 1, &\quad N_{101000}(t) &= N_{101000}(t^-) + 1.
\end{align*}

\begin{figure}[t]
\centering
\includegraphics[width=\textwidth]{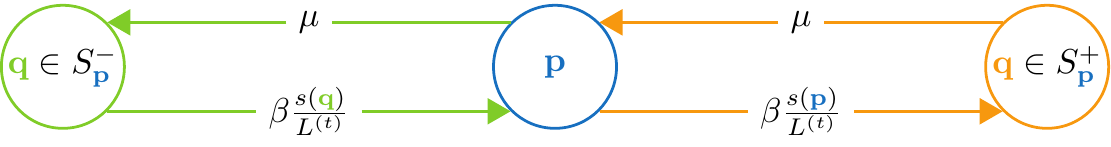}
\caption{Transition rates between pattern states $ \bp \in \cPt $ and $ \bq \in \Spm \cup \Spp $.}
\label{fig:patternTransitions}
\end{figure}

\subsection{Expected pattern frequencies}
\label{sec:epm}

Instances of the same trait evolve independently of each other and of other traits. If we sum over the rates in Figure~\ref{fig:patternTransitions} for each trait displaying a given pattern $ \bp \in \cPt $, then on a short interval of length $ \dt $ between branching events, by a standard argument for Markov chains,
\begin{equation}
\label{eq:transProb}
\begin{split}
&\PP[N_{\bp}(t + \dt) - \Npt = k | g, \lambda, \mu, \beta] \\
&\quad =
\left\{
\begin{array}{ll}
s(\bp) \left[ \mu + \beta \left(1 - \tfrac{s(\bp)}{\Lt}\right) \right] \Npt \dt + o(\dt), \quad &	k = -1, \vspace{4pt} \\
\left[\lambda  \ind{s(\bp) = 1} + \beta \sum_{\bq \in \Spm} \tfrac{s(\bq)}{\Lt} N_{\bq}(t) \right. \\
	\qquad \left. + \mu \sum_{\bq \in \Spp} N_{\bq}(t) \right] \dt + o(\dt), \quad & k = 1.
\end{array}
\right.
\end{split}
\end{equation}
Let $ \xpt = \xp(t; g, \lambda, \mu, \beta) = \EE[\Npt | g, \lambda, \mu, \beta] $, the expected number of traits in $ \cHt $ displaying pattern $ \bp \in \cPt $ at time \emph{t}. From Equation~\ref{eq:transProb}, $ \xpt $ evolves according to the following differential equation:
\begin{equation}
\label{eq:dxp}
\begin{split}
\dot{x}_{\bp}(t)
	=& \lim_{\dt \rightarrow 0} \frac{ \EE\left[N_{\bp}(t + \dt) - \Npt | g, \lambda, \mu, \beta\right] }{\dt} \\
	=& - s(\bp) \left[ \mu + \beta \left(1 - \tfrac{s(\bp)}{\Lt}\right) \right] \xpt + \lambda  \ind{s(\bp) = 1} \\
		 &\qquad + \beta \sum_{\bq \in \Spm} \tfrac{s(\bq)}{\Lt} \xqt + \mu \sum_{\bq \in \Spp} \xqt.
\end{split}
\end{equation}
There are $ |\cPt| = 2^{\Lt} - 1 $ coupled differential equations \eqref{eq:dxp} describing the expected evolution of the pattern frequencies $ \bNt $. We may write these equations as $ \dot{\bx}(t) = \bAt \bxt + \bbt $ where: $ \bxt = (\xpt : \bp \in \cPt) $ is the vector of expected pattern frequencies at time \emph{t}, and the sparse matrix $ \bAt $ and vector $ \bbt $ respectively describe the flow between patterns from trait death and transfer events and the flow into patterns of Hamming weight 1 through trait birth events.

In Section~\ref{sec:model}, we state that a branch of infinite length connects the Adam and root nodes. As a result, the pattern frequency process $ \bNt $ is in equilibrium just before the first branching event at time $ t_1 $, with the result that $ \bx(t_1^-) = x_1(t_1^-) = \lambda / \mu $ and $ N_1(t_1^-) \sim \Pois{\lambda / \mu} $. With this initial condition at the root, we can write the expected pattern frequencies at the leaves, $ \bx(0) $, recursively as a sequence of initial value problems between branching events: for each interval $ i = 1, \dotsc, L - 1 $, solve
\begin{equation}
\label{eq:ivp}
\dot{\bx}(t) = \bAt \bxt + \bbt \quad \text{for} \quad t \in [t_i, t_{i + 1}) \quad \text{where} \quad \bx(t_i) = \bT^{(i)} \bx(t_i^-),
\end{equation}
and we recall from Section~\ref{sec:patternsAtBranching} the operator $ \bT^{(i)} $ which propagates $ \bNtm $ and $ \bxtm $ across the \emph{i}th branching event. We illustrate this procedure graphically in Figure~\ref{fig:patternMeanTransfer}.

\begin{figure}[t]
\centering
\includegraphics[width=\textwidth]{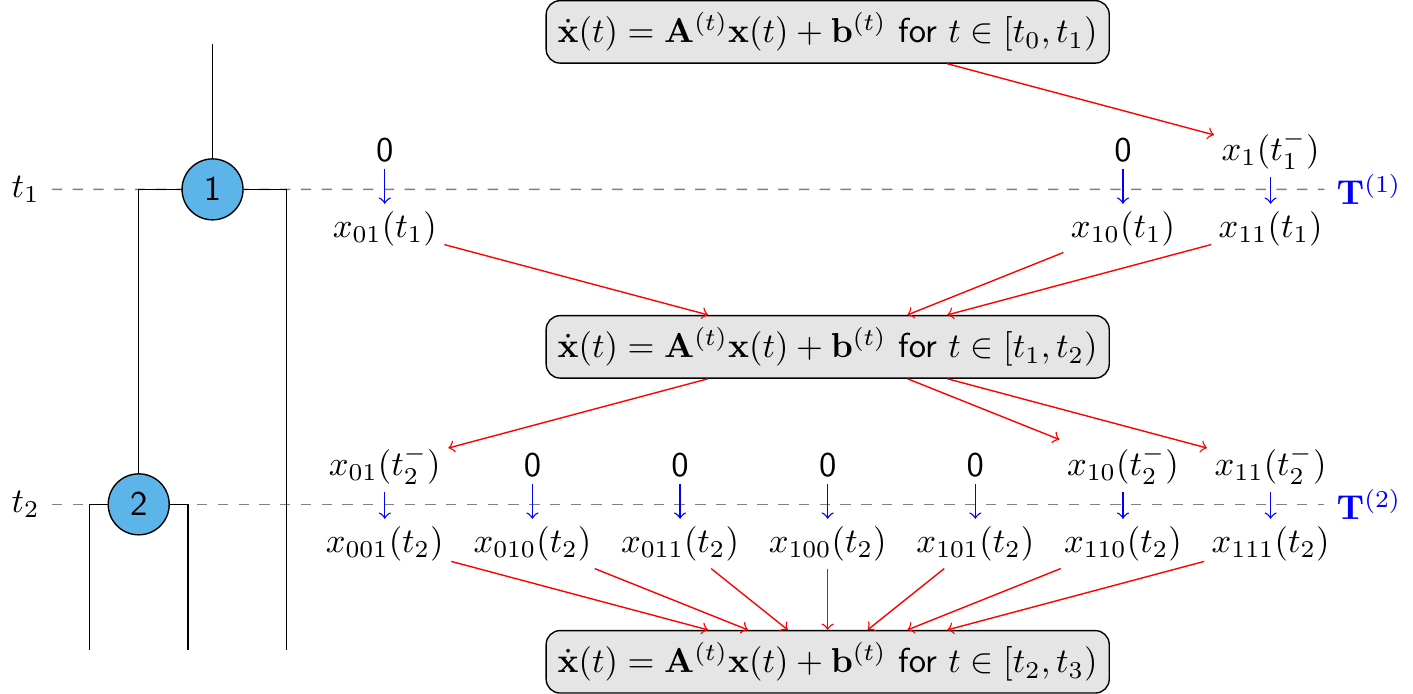}
\caption{Computing the expected pattern frequencies $ \bxt $ as a sequence of initial value problems~\eqref{eq:ivp} on a given tree. The initialisation operation $ \bx(t_i) = \bT^{(i)} \bx(t_i^-) $ from Section~\ref{sec:patternsAtBranching} provides the initial condition at the start of the \emph{i}th interval between branching events.}
\label{fig:patternMeanTransfer}
\end{figure}

\subsection{Likelihood}
\label{sec:dist}

Theorem~\ref{thm:bdl} describes the distribution of the pattern frequencies. We prove this result in Appendix~\ref{app:like}.

\begin{mytheorem}[Binary data distribution]
\label{thm:bdl}
The components of the vector of pattern frequencies $ \bNt = (\Npt : \bp \in \cPt) $ are independent Poisson random variable with corresponding rate parameters $ \bx(t; g, \lambda, \mu, \beta) $ given by the solution of the sequence of initial value problems in Equation~\ref{eq:ivp}.
\end{mytheorem}

\section{Model extensions}
\label{sec:modExt}

We now extend the model and likelihood calculation to allow for rate variation, missing data, offset leaves and the systematic removal of patterns from the data.

\subsection{Rate heterogeneity}
\label{sec:cat}

We introduce spikes of evolutionary activity in the form of \emph{catastrophes} \citep{ryder11}. Catastrophes, illustrated in Figure~\ref{fig:phyTree}, occur at rate $ \rho $ along each branch of the tree. A catastrophe advances the trait process along a branch by $ \delta = -\mu^{-1} \log(1 - \kappa) $ units of time relative to the other branches. In the model of \citet{ryder11}, this is equivalent to killing each trait on the branch independently with probability $ \kappa $ and adding a $ \Pois{\lambda \kappa / \mu} $ number of new traits. To ensure that catastrophes are identifiable with respect to the underlying trait process, we enforce a minimum \emph{catastrophe severity} $ \kappa \geq 0.25 $.

A branch may acquire traits through birth and transfer events and lose traits to death events during a catastrophe. The trait process at a catastrophe is equivalent to thinning the overall trait process to events on a single branch so we account for a catastrophe at time \emph{t} on branch $ \kit $ in the expected pattern frequency calculation~\eqref{eq:ivp} with the update
\begin{gather*}
\begin{aligned}
\xpt
	&\leftarrow e^{-\mu  \delta} \xptm + (1 - e^{-\mu  \delta}) \frac{\lambda}{\mu}
		&\quad &\begin{array}{l} \bp \in \cPt \\ s(\bp) = 1, p_i = 1, \end{array} \\
\begin{bmatrix} \xqt \\ \xrt \end{bmatrix}
	&\leftarrow
		\exp\left[
			\begin{pmatrix}
				-\beta \frac{s(\bq)}{\Lt} & \phantom{-}\mu \\
				\phantom{-}\beta \frac{s(\bq)}{\Lt} & -\mu
			\end{pmatrix} \delta
		\right]
	\begin{bmatrix} \xqtm \\ \xrtm \end{bmatrix}
		&\quad &\begin{array}{l} \bq, \br \in \cPt, d(\bq, \br) = 1 \\ q_i = 0, r_i = 1, \end{array}
\end{aligned}
\end{gather*}
where we exploit the property that each pattern communicates with at most one other during a catastrophe.

\subsection{Missing data}
\label{sec:md}

We allow for \emph{missing-at-random} data. Following \citet{ryder11}, the true binary state of trait \emph{h} at taxon $ i \in V_L $ is recorded with probability $ \xi_i = \PP(d_i^h \in \{0, 1\}) $ independently of the other traits and taxa. Let $ \Xi = (\xi_i : i \in V_L) $ denote the set of true-state observation probabilities. The space of observable site-patterns with missing data across the $ L $ taxa at time 0 is $ \cQ = \{0, 1, ?\}^{L} \setminus \{\zeros\} $. The set of binary patterns consistent with pattern $ \bq \in \cQ $ is $ u(\bq) = \{ \bp \in \cP^{(0)} : p_i = q_i \text{ \emph{if} } q_i \neq\ ?,\, i \in [L]\} $. From Theorem~\ref{thm:bdl} and the restriction and superposition properties of Poisson processes \citep{kingman92}, the frequency of traits displaying pattern $ \bq $ is an independent Poisson random variable with mean
\[
\bx_q(0; g, \lambda, \mu, \beta, \Xi)
	= \sum_{\mathclap{\bp \in u(\bq)}} \xp(0; g, \lambda, \mu, \beta) \prod_{i = 1}^{L} \xi_{k_i^{(0)}}^{\ind{q_i \in \{0, 1\}}} \left(1 - \xi_{k_i^{(0)}}\right)^{\ind{q_i = ?}}.
\]

\subsection{Non-isochronous data}
\label{sec:nonIso}

Non-isochronous data arise when taxa are sampled at different times. The corresponding taxa appear as \emph{offset} leaves in the phylogeny; nodes 12 and 15 in Figure~\ref{fig:phyTree}, for example. Similar to catastrophes, the trait process is frozen on offset leaves and a pattern may now only communicate with those patterns which are identical to it on the extinct lineages and differ at a single entry on the extant lineages.

The $ \Lt $ extinct and evolving lineages at time \emph{t}, of which $ \hat{L}^{(t)} $ are extant, are labelled $ \kt = (i \in E : t_{\pa(i)} \leq t < t_i \ind{i \in V_A}) $. The Hamming distance between patterns $ \bp $ and $ \bq \in \cPt $ across the extant lineages only is $ \hat{d}(\bp, \bq) = |\{i \in [\Lt] : p_i \neq q_i, \, t < t_{\kit}\}| $, and the corresponding Hamming weight of $ \bp $ across the extant lineages is $ \hat{s}(\bp) = \hat{d}(\bp, \zeros) $. Recalling $ \Spm $ and $ \Spp $ from Section~\ref{sec:patternsBetweenBranching}, pattern $ \bp \in \cPt $ communicates with patterns in the sets
\begin{align*}
\hat{S}_{\bp}^- &= \{\bq \in \Spm : \hat{s}(\bq) = \hat{s}(\bp) - 1,\, \hat{d}(\bp, \bq) = 1\}, \\
\hat{S}_{\bp}^+ &= \{\bq \in \Spp : \hat{s}(\bq) = \hat{s}(\bp) + 1,\, \hat{d}(\bp, \bq) = 1\},
\end{align*}
and its expected frequency evolves as
\begin{align*}
\dot{x}_{\bp}(t)
	=& -\hat{s}(\bp) \left[ \mu + \beta \left(1 - \tfrac{\hat{s}(\bp)}{\hat{L}^{(t)}}\right) \right] \xpt + \lambda  \ind{s(\bp)=\hat{s}(\bp) = 1} \\
		 &\qquad + \beta \sum_{\bq \in \hat{S}_{\bp}^-} \tfrac{\hat{s}(\bq)}{\hat{L}^{(t)}} \xqt + \mu \sum_{\bq \in \hat{S}_{\bp}^+} \xqt.
\end{align*}
We allow for offset leaves in our goodness-of-fit tests in Section~\ref{sec:applications}.

\subsection{Data registration}
\label{sec:reg}

Patterns which may be uninformative or unreliable with respect to the model are typically removed from the data. Given a registration rule \emph{R}, which may be a composition of other simpler rules such as those in Table~\ref{tab:reg}, we discard the columns in the data array \textbf{D} not satisfying \emph{R}, leaving registered data $ R(\bD) $, and restrict our analyses to patterns in $ R(\cQ) $. In Section~\ref{sec:applications}, we discard traits not marked present in a single taxon.

\begin{table}[t]
\centering
\caption{Registration rules of \citet{alekseyenko08} and \citet{ryder11}.}
\label{tab:reg}
\begin{tabular}{@{}ll@{}} \toprule
Unregistered traits
	& Unregistered patterns $ \cQ \setminus R(\cQ) $ \\ \midrule
Absent in taxon $ k_i^{(0)} $
	& $ \{\bq \in \cQ : q_i = 0\} $ \\
Observed in $ j $ taxa or fewer
	& $ \{\bq \in \cQ : |\{i \in [L] : q_i = 1\}| \leq j\} $ \\
Observed in $ j $ or more taxa
	& $ \{\bq \in \cQ : |\{i \in [L] : q_i = 1\}| \geq j\} $ \\
Potentially present in $ j $ taxa or greater
	& $ \{\bq \in \cQ : |\{i \in [L] : q_i \neq 0\}| \geq j\} $ \\ \bottomrule
\end{tabular}
\end{table}

\section{Bayesian inference}
\label{sec:inference}

In order to efficiently estimate both the node times and the rate parameters, we calibrate the space $ \Gamma $ of rooted phylogenetic trees on \emph{L} taxa with \emph{clade constraints}. The constraint $ \Gamma^{(0)} = \{g \in \Gamma : \ubar{t}_1 \leq t_1 < 0\} $ restricts the earliest admissible root time to $ \ubar{t}_1 $. Each additional constraint $ \Gamma^{(c)} $ places either time or ancestry constraints on the remaining nodes. We denote by $ \Gamma^C = \bigcap_c \Gamma^{(c)} $ the space of phylogenies satisfying the clade constraints.

\citet{nicholls11} describe a prior distribution on trees with the property that the root time $ t_ 1 $ is approximately uniformly distributed across a specified interval $ [\ubar{t}_1, \bar{t}_1] $. For a given tree $ g = (V, E, T, C) $, there are $ Z(g) $ possible time orderings of the nodes amongst the admissible node times $ T(g) = \{T' : (V, E, T', C) \in \Gamma^C\} $. For each node $ i \in V $, $ \ubar{t}_i = \inf_{T \in T(g)} t_i $ and $ \bar{t}_i = \sup_{T \in T(g)} t_i $ are the earliest and most recent times that \emph{i} may achieve in an admissible tree with topology $ (V, E) $. If $ S(g) = \{i \in V : \ubar{t}_i = \ubar{t}_1\} $ denotes the set of \emph{free} internal nodes with times bounded below by $ \ubar{t}_1 $, the prior with density
\[
f_G(g) \prop \frac{\ind{g \in \Gamma^C}}{Z(g)} \prod_{i \in S(g)}  \frac{\ubar{t}_1 - \bar{t}_i}{t_1 - \bar{t}_i},
\]
is approximately uniform across topologies and root times provided that $ \ubar{t}_1 \ll \min_{i \in V \setminus S} \ubar{t}_i $ \citep{ryder11}. Uniform priors on offset leaf times completes our prior specification on the tree. \citet{heled12} describe an exact method for computing uniform calibrated tree priors but we do not pursue that approach here. Table~\ref{tab:priors} lists the prior distributions on the remaining parameters.

\begin{table}[t]
\centering
\caption{Prior distributions on parameters in the Stochastic Dollo and Stochastic Dollo with Lateral Transfer models.}
\label{tab:priors}
\begin{tabular}{@{}lll@{}} \toprule
Parameter				&	Prior										&	Reasoning \\ \midrule
Trait birth rate		&	$ \lambda \sim 1 / \lambda $				&	Improper, scale invariant \\
Trait death rate		&	$ \mu \sim \Gamma(10^{-3}, 10^{-3}) $		&	Approximately $ 1 / \mu $ \\
Trait transfer rate		&	$ \beta \sim \Gamma(10^{-3}, 10^{-3}) $		&	Approximately $ 1 / \beta $ \\
Catastrophe rate		&	$ \rho \sim \Gamma(1.5, 5 \times 10^3) $	&	$ \EE[\rho^{-1}] = 10^4 $ years \\
Catastrophe severity	&	$ \kappa \sim \mathrm{U}[0.25, 1] $			&	$ \EE[\delta|\mu] = \mu^{-1} [1 - \log(0.75)] $ years \\
Observation probabilities	&	$ \Xi \sim \mathrm{U}[0, 1]^{L} $		&	Independent, uniform \\ \bottomrule
\end{tabular}
\end{table}

Inspecting the solution of the expected pattern frequency calculation~\eqref{eq:ivp} with initial condition $ \mathbf{x}(t_1^-) = \lambda / \mu $ at the root, we see that $ \bx(t; g, \lambda, \ldots) = \lambda \bx(t; g, 1, \ldots) $. We can integrate $ \lambda $ out of the Poisson likelihood in Theorem~\ref{thm:bdl} with respect to its prior in Table~\ref{tab:priors} to obtain a multinomial likelihood whereby a pattern $ \bp \in R(\cQ) $ is observed with probability proportional to its expected frequency. Furthermore, we may integrate the catastrophe rate $ \rho $ out of the Poisson prior on the number of catastrophes $ |C| $ to obtain a Negative Binomial prior instead. We describe these steps in detail in Appendix~\ref{app:inference}.

Let $ n_{\bp} = |\{h \in \cH(0) : \bp = \bd^h \in R(\bD)\}| $ denote the frequency of traits in the registered data displaying pattern $ \bp \in R(\cQ) $. Putting everything together, the posterior distribution is
\begin{equation}
\label{eq:intPost}
\pi(g, \mu, \beta, \kappa, \Xi | R(\bD))
	\propto f_G(g) f_M(\mu) f_B(\beta) \prod_{\bp \in R(\cQ)} \left(\frac{\xp}{\sum_{\bq \in R(\cQ)} \xq}\right)^{\np},
\end{equation}
where the expected pattern frequencies $ \bx \equiv \bx(0; g, 1, \mu, \beta, \kappa, \Xi) $~\eqref{eq:ivp} account for catastrophes, missing data and offset leaves where necessary. This completes the specification of the Stochastic Dollo with Lateral Transfer (SDLT) model.

The posterior distribution~\eqref{eq:intPost} is intractable but may be explored using standard Markov chain Monte Carlo (MCMC) sampling schemes for phylogenetic trees and Stochastic Dollo models \citep{nicholls08, ryder11}. We describe the MCMC transition kernels for moves particular to the SDLT model in Appendix~\ref{app:inference}.

\subsubsection*{Implementation}

Code to implement the SDLT model in the software package \texttt{TraitLab} \citep{nicholls13} is available from the authors.

\section{Method testing}
\label{sec:testing}

We describe a number of tests to validate our model and inference scheme in Appendix~\ref{app:testing}. We compare the exact and empirical distributions of synthetic data to validate our implementation of the expected pattern frequency calculation~\eqref{eq:ivp}. We test the identifiability of the SDLT model, its consistency with the SD model when the lateral transfer rate $ \beta = 0 $, and its robustness to a common form of model misspecification whereby recently transferred traits are discarded from the data. In each case, we obtain a satisfactory fit to the data and recover the true parameters.

\section{Application}
\label{sec:applications}

The order and timing of human settlement in Eastern Polynesia is a matter of debate. In the standard subgrouping of the Eastern Polynesian languages, Rapanui diverges first, followed by the split leading to the Marquesic (Hawaiian, Mangarevan, Marquesan) and Tahitic (Manihiki, Maori, Penrhyn, Rarotongan, Rurutuan, Tahitian, Tuamotuan) language subgroups \citep{marck00}. Recent linguistic and archaeological evidence has challenged this theory. In an implicit phylogenetic network study of lexical traits, \citet{gray10} detect non-tree-like signals in the data; furthermore, the Tahitic and Marquesic languages do not form clean clusters in their study. In a meta-analysis of radiocarbon-dated samples from archaeological sites in the archipelago, \citet{wilmshurst11} claim that Eastern Polynesia was settled in two distinct phases: the Society Islands between 900 and 1000 years before the present (BP) and the remainder between 700 and 900 years BP. These dates, much later than those reported by \citet{spriggs93}, for example, do not allow much time for the development of the Eastern Polynesian language subgroups. \citet{conte14} present evidence of human settlement in the Marquesas Islands approximately 1100 years BP. On the basis of the above and further evidence of lateral transfer in primary source material, \citet{walworth14} disputes Marquesic and Tahitic as distinct subgroups.

To add to this debate, we compare the SDLT and SD models on a data set of lexical traits in eleven Eastern Polynesian languages drawn from the approximately 1200 languages in the Austronesian Basic Vocabulary Database \citep{greenhill08}. The data is a subset of the Polynesian language data set in the study of \citet{gray10}. We analyse the 968 traits marked present in at least one of the eleven languages, hereafter referred to as \texttt{POLY-0}. The data are isochronous. Consistent with \citet{gray09}, the sole clade constraint limits the root of the tree to lie between 1150 and 1800 years BP.

We plot samples from the marginal tree posterior under the SDLT and SD models in Figure~\ref{tree:poly}. We summarise these distributions with \emph{majority rule consensus trees} in the Appendix~\ref{app:applications}. In agreement with \citet{gray10} and \citet{walworth14}, the standard subgroupings do not appear as subtrees in either model. Rapanui does not form an outgroup in either of our analyses. There is little evidence in the tree posteriors to support the claim of \citet{wilmshurst11}, however, as the posterior distributions of the root time, $ t_1 $, resemble its approximately uniform prior distribution on the range [1150, 1800] years BP.

The majority of the uncertainty under the SDLT model is in the topology of the subtree containing Rarotongan, Penrhyn, Tuamotu, Rapanui, Mangareva and Marquesan. This subtree also has 100\% posterior support under the SD model, but most of the uncertainty here is in relationships further up the tree. We use \texttt{BEAST} \citep{drummond12} to obtain the 95\% highest posterior probability sets for the tree topologies under the respective models. This set comprises 135 topologies for the SDLT model and 19 for the SD model. This level of confidence in relatively few topologies is likely a result of the SD model's misspecification on the laterally transferred traits.

\begin{figure}[t]
\centering
\subfloat[SDLT model.]{\includegraphics[width=0.49\textwidth]{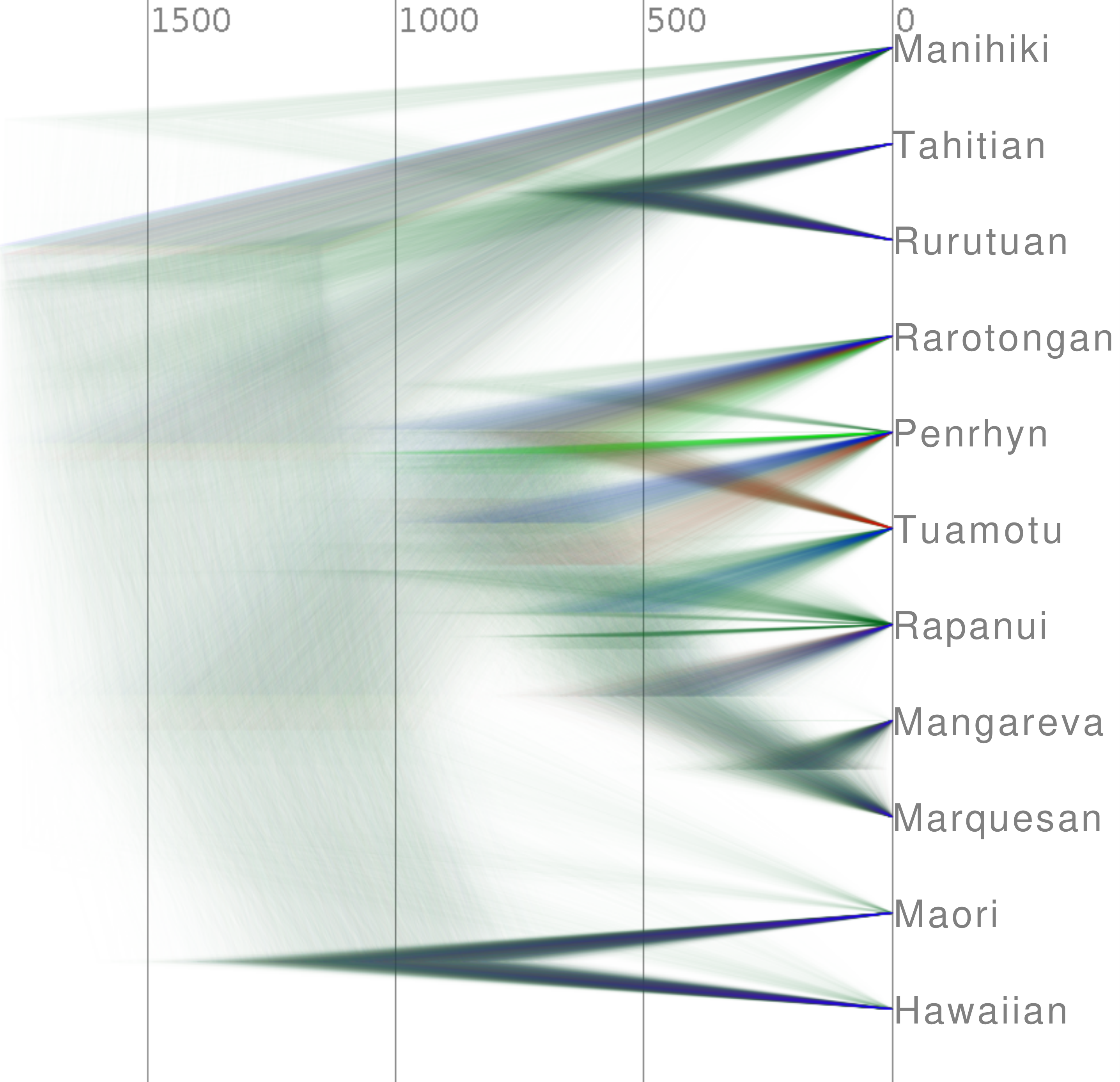} \label{tree:polySDLT}}
\subfloat[SD model.]{\includegraphics[width=0.49\textwidth]{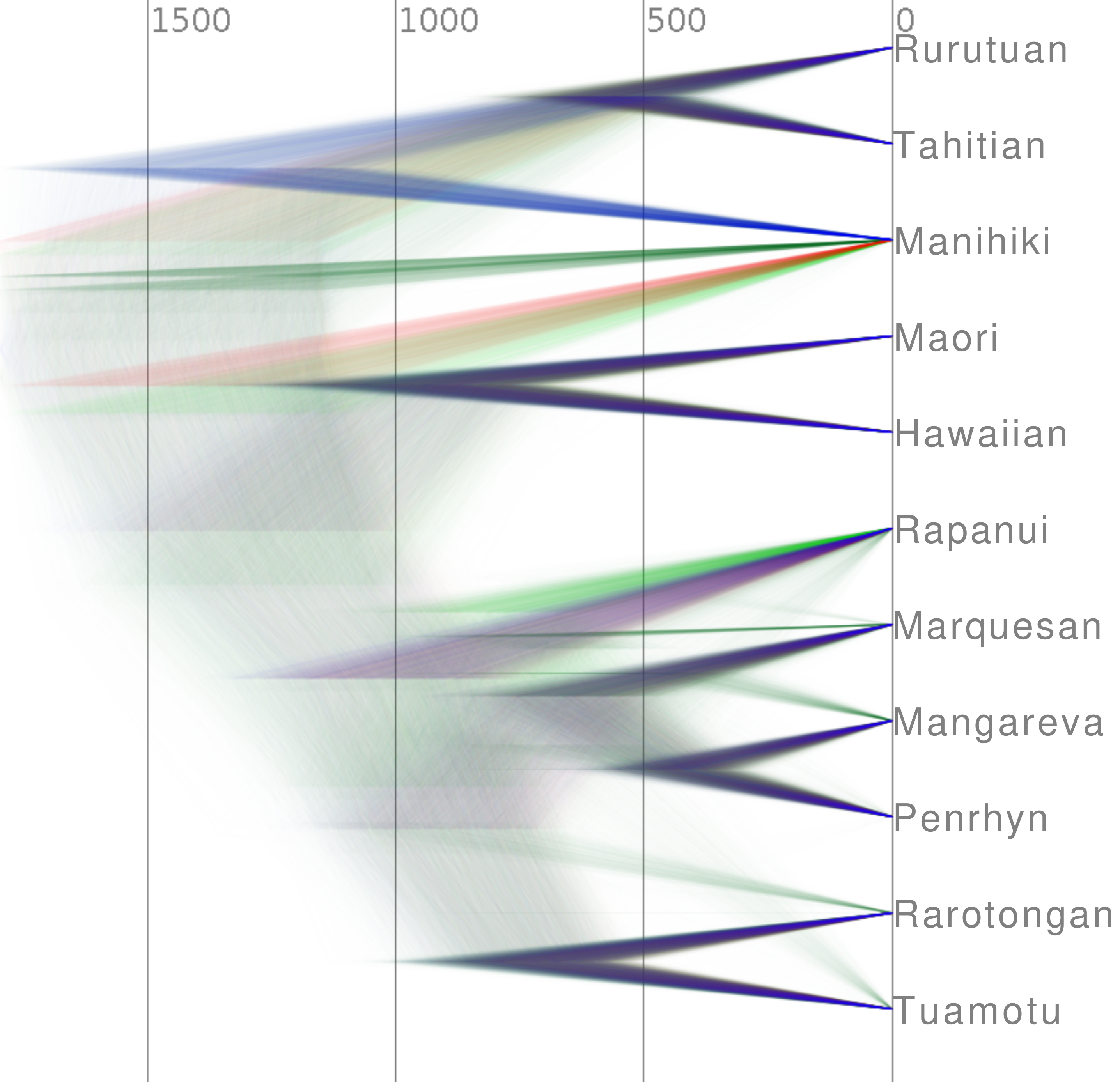} \label{tree:polySD}}
\caption{\texttt{DensiTree} \protect\citep{bouckaert14} plots of samples from the marginal tree posterior under the SDLT and SD models fit to \texttt{POLY-0}. Heavier lines indicate higher posterior support. Time is in units of years before the present.}
\label{tree:poly}
\end{figure}

The effect of the laterally transferred traits in the data is also evident in the histograms in Figure~\ref{pars:poly}. The death rate $ \mu $ is approximately $ 50\% $ higher under the SD model as traits must be born further up the tree and killed off at a higher rate to explain the variation in the data due to lateral transfer. The relative transfer rate $ \beta / \mu $ is the expected number of times that a single instance of a trait transfers before dying out; its posterior distribution under the SDLT model is centred on 1.35. In contrast, on the basis of simulation studies, both \citet{nicholls08} and \citet{greenhill09} consider a relative transfer rate of 0.5 high. We report histograms for the remaining parameters as well as the trace and autocorrelation plots we use to diagnose the convergence of our Markov chains \citep{geyer92} in Appendix~ \ref{app:applications}.

\begin{figure}[t]
\centering
\begin{tabular}{@{}c@{}@{}c@{}@{}c@{}}
\includegraphics[width=0.5\textwidth, trim = 0.05cm 0cm 0.6cm 0.25cm, clip]{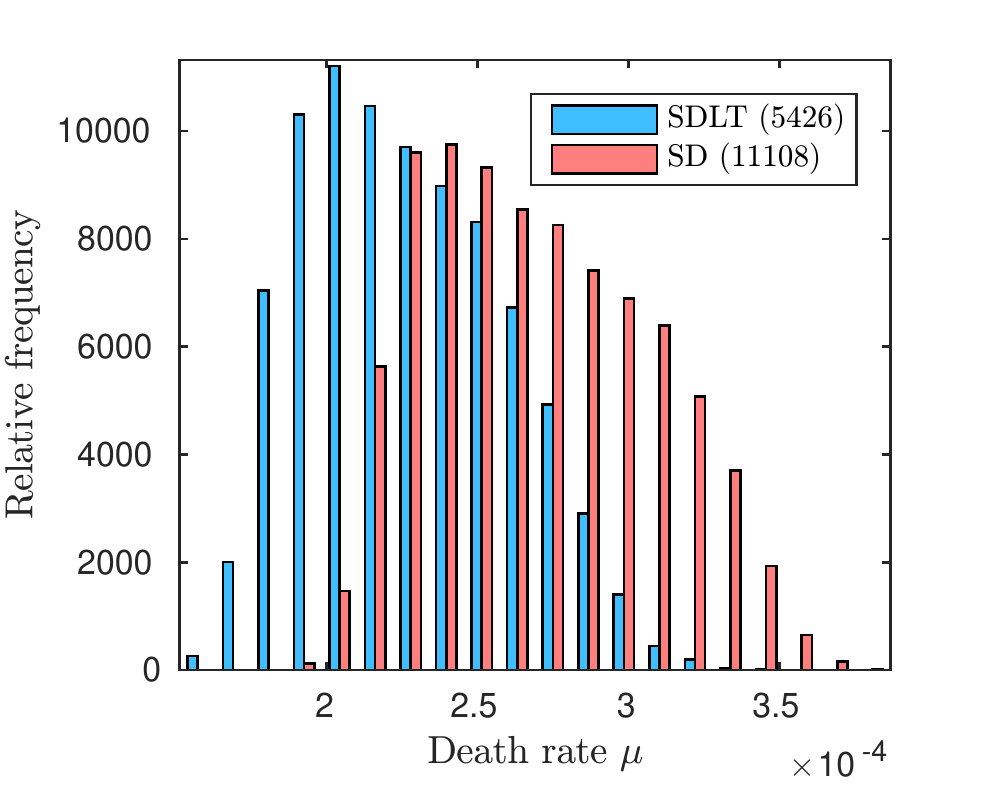} &
\includegraphics[width=0.5\textwidth, trim = 0.05cm 0cm 0.6cm 0.25cm, clip]{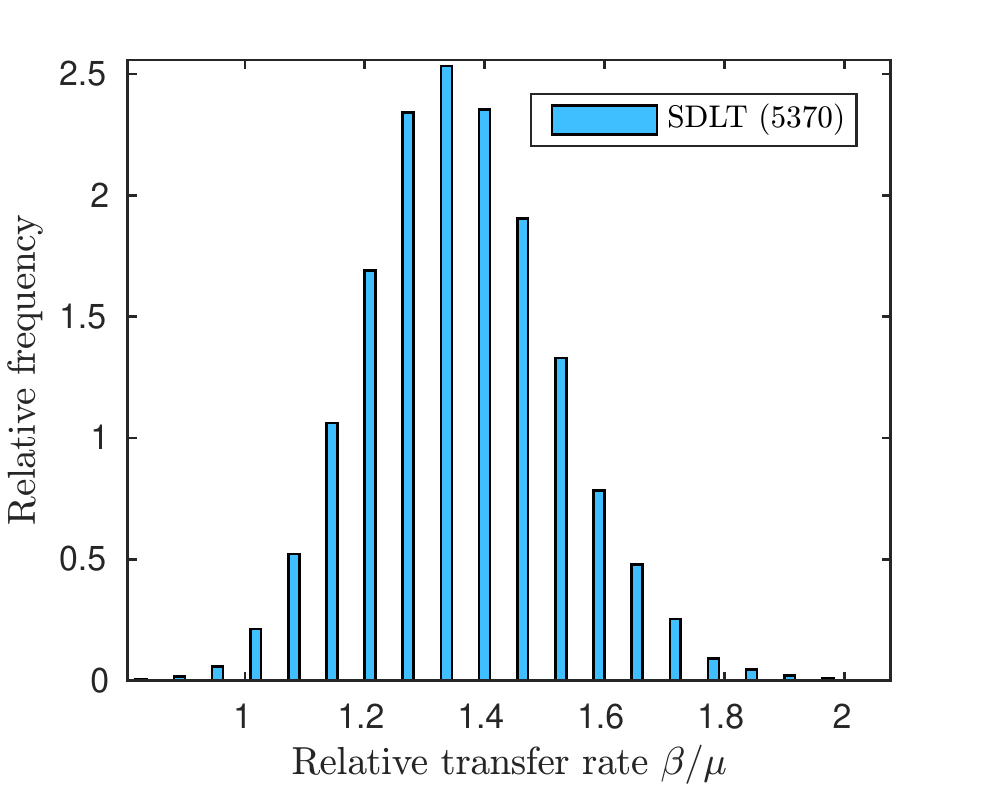}
\end{tabular}
\caption{Marginal parameter posterior distributions under the SDLT and SD models fit to the Eastern Polynesian data set \texttt{POLY-0}. Effective sample sizes are in parentheses.}
\label{pars:poly}
\end{figure}

With the above concerns about the SD model in mind, we now assess the validity of our analyses. To assess goodness-of-fit, we relax the constraints on each leaf time and attempt to reconstruct them. The constraint $ \Gamma^{(i)} = \{g \in \Gamma : t_i = 0\} $ fixes leaf $ i \in V_L $ at time 0 and $ \Gamma^{(i')} = \{g \in \Gamma : -10^3 \leq t_i \leq 10^4\} $ denotes its relaxation to a wide interval either side of time 0. We denote by $ \Gamma^{C'} $ the calibrated space of phylogenies with $ \Gamma^{(i)} $ replaced by $ \Gamma^{(i')} $. The constraint $ \Gamma^C \subset \Gamma^{C'} $ so the Bayes factor comparing the relaxed and constrained models is
\begin{align}
\label{eq:SDR}
B_{i', i}
	&= \frac{\pi(R(\bD) | g \in \Gamma^{C'})}{\pi(R(\bD) | g \in \Gamma^C)} \nonumber \\
	&= \frac{\pi(R(\bD) | g \in \Gamma^{C'})}{\pi(R(\bD) | g \in \Gamma^C \cap \Gamma^{C'})} \nonumber \\
	&= \frac{\pi(g \in \Gamma^C | g \in \Gamma^{C'})}{\pi(g \in \Gamma^C | R(\bD), g \in \Gamma^{C'})},
\end{align}
a Savage--Dickey ratio of the marginal prior and posterior densities that the constraint $ \Gamma^{(i)} $ is satisfied in the relaxed model. A large Bayes factor here indicates a lack of support for the leaf constraint and is a sign of model misspecification.

We cannot compute the Savage--Dickey ratio in Equation~\ref{eq:SDR} in closed form, so in practice we estimate the densities by the proportions of sampled leaf times in the range $ [-50, 50] $ years around time 0. We report log-Savage--Dickey ratios in Figure~\ref{fig:polyBF} and histograms of the marginal leaf ages in Appendix~\ref{app:applications}. There are clear signs that the SD model is misspecified here. In particular, the SD model rejects the constraints on Manihiki and Marquesan, so we report lower bounds on the corresponding Bayes factors. The large Bayes factor for the constraint on Rapanui provides `positive' evidence of misspecification on the scale of \citet{kass95}.

\begin{figure}[t]
\centering
\includegraphics[width=\textwidth, trim = 0.5cm 0cm 1.25cm 0.25cm, clip]{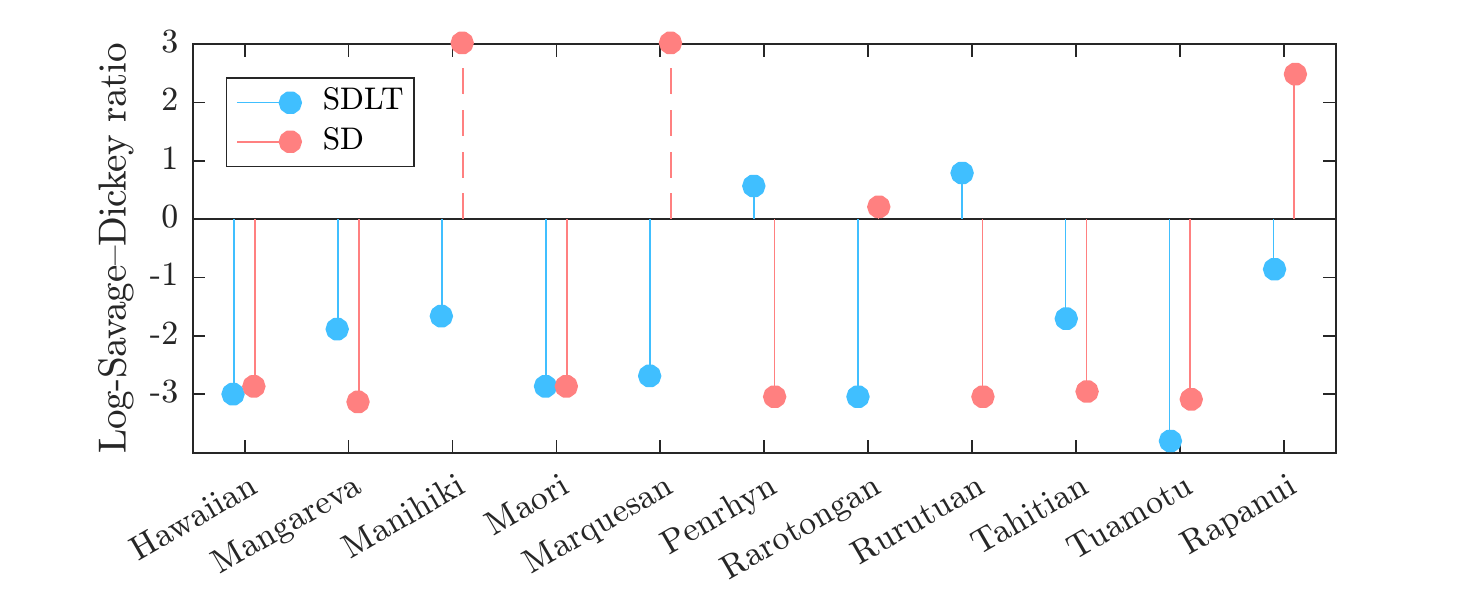}
\caption{Bayes factors comparing the support for the leaf constraints in the SDLT and SD models fit to \texttt{POLY-0}. We estimate lower bounds on the Bayes factors for the constraints on Manihiki and Marquesan under the SD model.}
\label{fig:polyBF}
\end{figure}

We assess the predictive performance of each model on a random splitting of the registered data $ R(\bD) $ into evenly sized training and test sets labelled $ \bDtr $ and $ \bDte $ respectively. \citet{madigan94} propose to score each model by its log-posterior predictive probability, $ \log \pi(\bDte | \bDtr) $, where $ \pi(\bDte | \bDtr) = \int \pi(\bDte | x) \pi(x | \bDtr) \ud x $, with $ x = (g, \mu, \beta, \kappa, \Xi) $ for the SDLT model and $ x = (g, \mu, \kappa, \Xi) $ for the SD model. The difference in scores is a log-Bayes factor measuring the relative success of the models at predicting the test data \citep{kass95}. The results in Table~\ref{tab:ppp} strongly support the superior fit of the SDLT model to \texttt{POLY-0}.

\begin{table}[t]
\centering
\caption{Posterior predictive model assessment.}
\label{tab:ppp}
\begin{tabular}{@{}llll@{}} \toprule
Data set		&	SDLT score	&	SD score 	&	Log-Bayes factor \\ \midrule
\texttt{POLY-0}	&	-3058.2		&	-3105.8		&	47.6	\\
\texttt{POLY-1}	&	-1401.2		&	-1481.1 	&	79.9	\\ \bottomrule
\end{tabular}
\end{table}

Traits marked present in a single language are often deemed unreliable and removed in the registration process. To address this concern, we repeat our analyses on the data set \texttt{POLY-1} which we form by removing the \emph{singleton} patterns from \texttt{POLY-0}. Although the outcome of the predictive model selection in Table~\ref{tab:ppp} is unchanged, these singleton patterns play an important role in SDLT model inference and parameter credible intervals are affected by their removal.

\section{Concluding remarks}
\label{sec:conclusions}

Lateral transfer is an important problem, but practitioners lack the tools to perform fully likelihood-based inference for dated phylogenies in this setting. We address this issue with a novel model for species diversification which extends the Stochastic Dollo model for lateral transfer in trait presence/absence data. To our knowledge, the method we describe is the first fully likelihood-based approach to control for lateral transfer in reconstructing a rooted phylogenetic tree. The second major contribution of this paper is the inference procedure whereby we integrate out the locations of the trait birth, death and transfer events through a sequence of initial value problems.

In the application we consider, accounting for lateral transfer results in an improved fit over the regular Stochastic Dollo model but comes at a significant computational cost. The sequence of initial value problems to compute the likelihood parameters in the lateral transfer model is easy to state but difficult to solve in practice. On a tree with \emph{L} leaves, we can exploit symmetry in the differential systems to compute the expected pattern frequencies exactly in $ \cO(2^{2 L}) $ operations. In practice, we use an ordinary differential equation (ODE) solver to approximate their values within an error tolerance dominated by the Monte Carlo error. This approach requires $ \cO(L 2^L C(L)) $ operations, where $ C(L) $ is the number of matrix-vector multiplications required by the ODE solver; for example, with the \texttt{Matlab} \texttt{ODE45} solver and typical choices of parameters, we observe $ C(10) \in [80, 90] $ and $ C(20) \in [95, 100] $. This approach is feasible for approximately $ L = 20 $ leaves on readily available hardware. As we must evaluate the likelihood many times over the course of an MCMC analysis, this computational burden is a major stumbling block towards applying our model to data sets with more taxa or multiple character states, and is the focus of ongoing research [\citet{kelly16Thesis}, Chapter 4].

The model as described is not \emph{projective} in the sense that we cannot marginalise out the effect of unobserved lineages, which in our analyses correspond to the many Polynesian languages not included in our data set. Consequently, the probability that a trait transfers between sampled lineages decreases as the number of unobserved lineages increases. Similarly, a trait which previously died out on the sampled lineages may transfer back into the system from an unobserved lineage. One possible solution to this problem is to introduce \emph{ghost} lineages \citep{szollosi12, szollosi13} to allow for lateral transfer between sampled and unsampled taxa at the expense of an increase in computational cost. There are many other avenues for future work on the model. For example, one could: partition the data across a mixture of models and trees, relax the global lateral transfer regime or the assumption that traits are independent, model multiple character states \citep{alekseyenko08}, allow individual catastrophes to vary in their effect, jointly model sequence and trait presence/absence data \citep{cybis15} and account for other types of missing data.

There are many open problems which have been ignored due to the expense of fitting models that account for lateral transfer. One such example occurs in the model of \citep{chang15} whereby ancestral nodes may have data. Stochastic Dollo without lateral transfer cannot be used to model the observation process here as traits absent in an ancestral state but present in both descendent and non-descendent leaves violate the Dollo parsimony assumption. Our method provides a model-based solution to this problem and many others.

\section*{Acknowledgements}

The authors wish to thank Robin Ryder and Simon Greenhill for their assistance with this project, and acknowledge the feedback of the associate editor and two anonymous reviewers.

\bibliographystyle{plainnat}
\bibliography{SDLT}

\clearpage
\appendices

\section{Likelihood calculation}
\label{app:like}

Theorem~\ref{thm:bdl} states that under the SDLT model, the pattern frequencies $ \bNt = (\Npt : \bp \in \cPt) $ at time \emph{t} on a tree \emph{g} is a vector of independent Poisson-distributed random variables with rate parameters $ \bxt = (\xpt : \bp \in \cPt) = \EE[\bNt | g, \lambda, \mu, \beta] $ given by the sequence of initial value problems in Equation~\ref{eq:ivp}.

\begin{proof}[Proof of Theorem~\ref{thm:bdl}]

Starting at the Adam node at time $ t_0 = -\infty $, the trait process evolves forwards in time along the branches of the tree. The pure birth-death trait process is in equilibrium when it reaches the root at time $ t_1 $ by construction, so $ N_1(t_1^-) \sim \Pois{x_1(t_1^-)} $ where $ x_1(t_1^-) = \lambda / \mu $. Following Property~T\ref{tp:branch} in Section~\ref{sec:model},
\begin{itemize}
\item The pattern $ (1, 1) $ is consistent with the branching event at time $ t_1 $ so $ N_{11}(t_1) \equiv N_1(t_1^-) $ and $ x_{11}(t_1) \equiv x_1(t_1^-) $
\item The patterns $ (0, 1) $ and $ (1, 0) $ are inconsistent with the branching event at the root so $ N_{01}(t_1) \equiv N_{10}(t_1) \equiv 0 $ and $ x_{01}(t_1) \equiv x_{10}(t_1) \equiv 0 $.
\end{itemize}
This provides us with the initial condition for the start of the next interval $ [t_1, t_2^-) $ between branching events. We can calculate the expected pattern frequencies at any time \emph{t} by alternatively solving the initial value problems in Equation~\ref{eq:ivp} and computing initial conditions using the initialisation operators $ \bT^{(1)}, \, \bT^{(2)}, \dotsc, \bT^{(L - 1)} $.

To complete the proof, we derive the Kolmogorov forward equation describing the temporal evolution of $ p_{\bn}(t) = \PP(\bNt = \bn | g, \lambda, \mu, \beta) $ for an integer vector $ \bn = (\np : \bp \in \cPt) $ and show that it is equivalent to the time-derivative of the hypothesised Poisson probability mass function
\begin{equation}
\label{eq:pn}
\pi_{\bn}(t) =
\prod_{\bp \in \cPt} \frac{\xpt^{\np} e^{-\xpt}}{\np!}.
\end{equation}

For patterns $ \bp $ and $ \bq \in \cPt $, we require the operators $ \Upz $, $ \Upq $ and $ \Uzq $ which applied to $ \bn $ yield
\begin{alignat*}{4}
\Upz \bn &= (\dotsc, n_{\bp - 1}, \np - 1, n_{\bp + 1}, &\dotsc&				   &), \\
\Upq \bn &= (\dotsc, n_{\bp - 1}, \np - 1, n_{\bp + 1}, &\dotsc&, n_{\bq - 1}, \nq + 1, n_{\bq + 1}, \dotsc&), \\
\Uzq \bn &= (				  &\dotsc&, n_{\bq - 1}, \nq + 1, n_{\bq + 1}, \dotsc&),
\end{alignat*}
where we have abused notation and used $ \bp - 1 $ and $ \bp + 1 $ to index the entries either side of $ \np $ in $ \bn $. These operators respectively correspond to the change in $ \bn $ observed if: a trait displaying pattern $ \bp $ becomes extinct~(T\ref{tp:death}), a trait which displayed pattern $ \bp $ transitions to display pattern $ \bq $ through either a death~(T\ref{tp:death}) or transfer~(T\ref{tp:transfer}) event, and a trait is born displaying pattern $ \bq $~(T\ref{tp:birth}). Of course, these transitions may only occur if the patterns communicate. If $ \rho(\bn, \bn') $ denotes the transition rate from state $ \bNt = \bn $ to $ \bn' $, then from Section~\ref{sec:patternsBetweenBranching},
\begin{alignat}{3}
\label{eq:lpq}
\rho(\bn, \Upz \bn)
	&= \np \lpz \quad &\text{where} \quad
	\lpz &=
	\left\{ \begin{array}{ll}
		\mu, & s(\bp) = 1, \\
		0, \qquad & \text{otherwise},
	\end{array} \right. \nonumber \\
\rho(\bn, \Upq \bn)
	&= \np \lpq \quad &\text{where} \quad
	\lpq &=
	\left\{ \begin{array}{ll}
		\mu, & \bq \in \Spm, \\
		\beta  \frac{s(\bp)}{L}, & \bq \in \Spp, \\
		0, \qquad & \text{otherwise},
	\end{array} \right. \\
\rho(\bn, \Uzq \bn)
	&= \lzq \quad &\text{where} \quad
	\lzq &=
	\left\{ \begin{array}{ll}
		\lambda, & s(\bq) = 1, \\
		0, \qquad & \text{otherwise}.
	\end{array} \right. \nonumber
\end{alignat}

\paragraph{The forward equation}

We derive the forward equation from first principles. For a short interval of length $ \dt $, from Equations~\ref{eq:transProb}~and~\ref{eq:lpq}, we obtain
\begin{align}
\label{eq:ptdt}
p_{\bn}(t + \dt)
	&= p_{\bn}(t) \left[ 1 - \left( \sum_{\bp \in \cPt} \sum_{\bq \in \cPt} \rho(\bn, \Upq \bn) + \sum_{\bp \in \cPt} \rho(\bn, \Upz \bn) \right. \right. \nonumber\\
	&\qquad \left. \left. + \sum_{\bq \in \cPt} \rho(\bn, \Uzq \bn) \right) \dt + o(\dt) \right] \nonumber\\
	&\quad + \sum_{\bp \in \cPt} \sum_{\bq \in \cPt} p_{\Upq \bn}(t) \left[ \rho(\Upq \bn, \bn) \dt + o(\dt) \right] \nonumber\\
	&\quad + \sum_{\bp \in \cPt} p_{\Upz \bn}(t) \left[ \rho(\Upz \bn, \bn) \dt + o(\dt) \right] \nonumber\\
	&\quad + \sum_{\bq \in \cPt} p_{\Uzq \bn}(t) \left[ \rho(\Uzq \bn, \bn) \dt + o(\dt) \right] \nonumber\\
	&= p_{\bn}(t) \left[ 1 - \left( \sum_{\bp \in \cPt} \sum_{\bq \in \cPt} \np \lpq + \sum_{\bp \in \cPt} \np \lpz \right. \right. \\
	&\qquad \left. \left. + \sum_{\bq \in \cPt} \lzq \right) \dt + o(\dt) \right] \nonumber\\
	&\quad + \sum_{\bp \in \cPt} \sum_{\bq \in \cPt} p_{\Upq \bn}(t) \left[ (\nq + 1) \lqp \dt + o(\dt) \right] \nonumber\\
	&\quad + \sum_{\bp \in \cPt} p_{\Upz \bn}(t) \left[\lzp \dt + o(\dt) \right] \nonumber\\
	&\quad + \sum_{\bq \in \cPt} p_{\Uzq \bn}(t) \left[(\nq + 1) \lqz \dt + o(\dt) \right]. \nonumber
\end{align}
We subtract $ p_{\bn}(t) $ from both sides of Equation~\ref{eq:ptdt}, then divide by $ \dt $ and take the limit as $ \dt \downarrow 0 $ to obtain
\begin{equation}
\begin{split}
\label{eq:pdotn1}
\dot{p}_{\bn}(t)
	&= - p_{\bn}(t) \left[\sum_{\bp \in \cPt} \sum_{\bq \in \cPt} \np \lpq  + \sum_{\bp \in \cPt} \np \lpz + \sum_{\bq \in \cPt} \lzq \right] \\
	&\quad + \sum_{\bp \in \cPt} \sum_{\bq \in \cPt} p_{\Upq \bn}(t) (\nq + 1) \lqp \\
	&\quad + \sum_{\bp \in \cPt} p_{\Upz \bn}(t) \lzp + \sum_{\bq \in \cPt} p_{\Uzq \bn}(t) (\nq + 1) \lqz.
\end{split}
\end{equation}
We shall drop the dependence on \emph{t} in our notation for the remainder of this section. From the hypothesised probability mass function~\eqref{eq:pn}, we see that
\begin{align*}
\pi_{\Upz \bn} &= \pi_{\bn}  \frac{ \np }{ \xp }, \\
\pi_{\Upq \bn} &= \pi_{\bn}  \frac{\np}{\xp}  \frac{\xq}{\nq + 1}, \\
\pi_{\Uzq \bn} &= \pi_{\bn}  \frac{ \xq }{\nq + 1 },
\end{align*}
and to simplify our argument later, we suppose that $ p_{\bn}(t) = \pi_{\bn}(t) $ and substitute these identities into Equation~\ref{eq:pdotn1} to obtain
\begin{align}
\label{eq:pdotn2}
\dot{p}_{\bn}
	&= - p_{\bn} \left[ \sum_{\bp \in \cP} \sum_{\bq \in \cP} \np \lpq + \sum_{\bp \in \cP} \np \lpz + \sum_{\bq \in \cP} \lzq \right] \nonumber \\
		&\quad + p_{\bn} \sum_{\bp \in \cP} \sum_{\bq \in \cP} \frac{\np}{\xp}  \frac{\xq}{\nq + 1} (\nq + 1) \lqp + p_{\bn} \sum_{\bp \in \cP} \frac{\np}{\xp}  \lzp \nonumber \\
		&\quad + p_{\bn} \sum_{\bq \in \cP} \frac{\xq}{\nq + 1} (\nq + 1) \lqz \nonumber \\
	&= -p_{\bn} \left[ \sum_{\bp \in \cP} \sum_{\bq \in \cP} \np \lpq + \sum_{\bp \in \cP} \np \lpz + \sum_{\bq \in \cP} \lzq \right] \nonumber \\
		&\quad + p_{\bn} \left[ \sum_{\bp \in \cP} \sum_{\bq \in \cP} \np \frac{\xq}{\xp} \lqp
			+ \sum_{\bp \in \cP} \frac{\np}{\xp}  \lzp + \sum_{\bq \in \cP} \xq \lqz \right] \nonumber \\
	&= p_{\bn} \left[ \sum_{\bp \in \cP} \sum_{\bq \in \cP} \np \left( -\lpq + \frac{\xq}{\xp}  \lqp \right) + \sum_{\bp \in \cP} \np \left( - \lpz + \frac{1}{\xp}  \lzp \right) \right. \\
		&\qquad\left. + \sum_{\bq \in \cP} \left( - \lzq + \xq \lqz \right) \right]. \nonumber
\end{align}
Equation~\ref{eq:pdotn2} is the forward equation for $ \bNt $.

\paragraph{Time derivative of the probability mass function}

Equation~\ref{eq:dxp} describes the temporal evolution of $ \xpt $, the expected number of traits displaying pattern $ \bp \in \cPt $ at time \emph{t}, which we can rewrite as
\begin{equation}
\label{eq:xpdot1}
\dot{x}_{\bp}(t)
	= - \xpt \left( \lpz + \sum_{\bq \in \cPt} \lpq \right) + \lzp + \sum_{\bq \in \cPt} \xqt \lqp,
\end{equation}
using the identities in Equation~\ref{eq:lpq} and the fact that $ s(\bp) = |\Spm| + \ind{s(\bp) = 1} $ and $ \Lt - s(\bp) = |\Spp| $, where we recall the sets $ \Spm $ and $ \Spp $ from Section~\ref{sec:patternsBetweenBranching}. Differentiating the hypothesised probability mass function~\eqref{eq:pn} with respect to time \emph{t}, we obtain
\begin{equation}
\label{eq:pdash1}
\dot{\pi}_{\bn}(t) = \pi_{\bn}(t) \frac{\mathrm{d}}{\dt} \log(\pi_{\bn}(t)) = \pi_{\bn}(t) \sum_{\bp \in \cPt} \frac{\np}{\xpt} \dot{x}_{\bp}(t) - \pi_{\bn}(t) \sum_{\bp \in \cPt} \dot{x}_{\bp}(t).
\end{equation}
Dropping the dependence on time \emph{t} from our notation and substituting Equation~\ref{eq:xpdot1} into Equation~\ref{eq:pdash1} yields
\begin{align}
\label{eq:pdash2}
\dot{\pi}_\bn
	&= \pi_{\bn} \sum_{\bp \in \cP} \np \left[ - \left( \lpz + \sum_{\bq \in \cP} \lpq \right)
		+ \frac{1}{\xp} \left( \lzp + \sum_{\bq \in \cP} \xq \lqp \right) \right] \nonumber \\
	&\quad + \pi_{\bn} \sum_{\bp \in \cP} \left[ \xp \left( \lpz + \sum_{\bq \in \cP} \lpq \right) - \lzp - \sum_{\bq \in \cP} \xq \lqp \right] \nonumber \\
	&= \pi_{\bn} \sum_{\bp \in \cP} \np \left[ \sum_{\bq \in \cP} \left( - \lpq + \frac{\xq}{\xp}  \lqp \right) - \lpz + \frac{1}{\xp} \lzp \right] \nonumber \\
		&\quad + \pi_{\bn} \left[ \sum_{\bp \in \cP} \left( -\lzp + \xp \lpz \right) \right] + \pi_{\bn} \sum_{\bp \in \cP} \left[ \xp \sum_{\bq \in \cP} \lpq - \sum_{\bq \in \cP} \xq \lqp \right] \nonumber \\
	&= \pi_{\bn} \left[ \sum_{\bp \in \cP} \sum_{\bq \in \cP} \np \left( - \lpq + \frac{\xq}{\xp}  \lqp \right) + \sum_{\bp \in \cP} \np \left( -\lpz + \frac{1}{\xp} \lzp \right) \right. \\
	&\quad \qquad \left. + \sum_{\bp \in \cP} \left( -\lzp + \xp \lpz \right) \right] + \pi_{\bn} \left[ \sum_{\bp \in \cP} \xp \sum_{\bq \in \cP} \lpq - \sum_{\bp \in \cP} \sum_{\bq \in \cP} \xq \lqp \right]. \nonumber
\end{align}
The final term in Equation~\ref{eq:pdash2} is 0, so it matches the forward equation in Equation~\eqref{eq:pdotn2}. We conclude that Equation~\ref{eq:pn} with parameters given by Equation~\ref{eq:ivp} correctly describes the distribution of the pattern frequencies.

\end{proof}

\section{Bayesian inference}
\label{app:inference}

\subsection{Prior and posterior distributions}

We demonstrate how to integrate the birth rate $ \lambda $ out of the likelihood in Theorem~\ref{thm:bdl}. Let $ \by(t) = \bx(t; g, \lambda = 1, \mu, \beta, \ldots) $ with entry $ y_{\bp}(t) $ for pattern $ \bp \in \cPt $. Immediately after the branching event at the root,
\begin{equation*}
\begin{split}
\dot{\bx}(t_1; g, \lambda, \ldots) &= \bA^{(t_1)} \bx(t_1; g, \lambda, \ldots) + \bb^{(t_1)} \\
	&= \bA^{(t_1)} \begin{bmatrix} 0 \\ 0 \\ \lambda / \mu \end{bmatrix} + \begin{bmatrix} \lambda \\ \lambda \\ 0 \end{bmatrix} \\
	&= \lambda \dot{\by}(t_1),
\end{split}
\end{equation*}
by construction, so $ \bx(t_2^-) = \lambda \by(t_2^-) $. The initial value problems~\eqref{eq:ivp} and initialisation operations $ \bT^{(i)} $ are linear so we can repeat the above argument to see that $ \bx(t; g, \lambda, \ldots) = \lambda  \by(t) $ for any time \emph{t}.

Dropping the dependence on time from our notation, we now integrate $ \lambda $ out of the Poisson likelihood in Theorem~\ref{thm:bdl} with respect to its improper prior in Table~\ref{tab:priors},
\begin{align*}
\pi(\bD | g, \mu, \beta, \dotsc)
	&= \int_0^{\infty} \pi(\bD | g, \lambda, \mu, \beta, \dotsc) \pi_{\Lambda}(\lambda) \dl \nonumber \\
	&\propto \int_0^{\infty} \frac{1}{\lambda} \prod_{\bp} (\lambda  y_{\bp})^{n_{\bp}} e^{-\lambda  y_{\bp}} \dl \nonumber \\
	&= \left(\prod_{\bp} y_{\bp}^{n_{\bp}}\right) \int_0^{\infty} \lambda^{\left(\sum_{\bp} n_{\bp}\right) - 1} e^{-\lambda \sum_{\bp} y_{\bp}} \dl \nonumber \\
	&= \left(\prod_{\bp} y_{\bp}^{n_{\bp}}\right) \left(\sum_{\bp} y_{\bp}\right)^{-\sum_\bp \np} \Gamma\left(\sum_{\bp} \np\right) \nonumber \\
	&\propto \prod_{\bp} \left(\frac{y_{\bp}}{\sum_{\bq} y_{\bq}}\right)^{\np},
\end{align*}
which we recognise as the multinomial likelihood term in the posterior~\eqref{eq:intPost}.

Catastrophes occur according to a $ \Pois{\rho} $ process along the branches of the tree. Let $ n^{(i)} $ denote the number of catastrophes on branch $ i \in E \setminus \{1\} $ of length $ \Delta_i = t_{\pa(i)} - t_i $. Let \emph{n} denote the total number of catastrophes on the tree and $ \Delta $ the length of the tree below the root. Conditional on the catastrophe rate $ \rho $, the prior distribution on the number of catastrophes and their locations is
\begin{equation}
\label{eq:priorCatRho}
\pi_{C | R}(C | \rho)
	= \prod_{i \in E \setminus \{1\}} \frac{(\rho \Delta_i)^{n^{(i)}} e^{-\rho \Delta_i}}{n^{(i)}!} \frac{n^{(i)}!}{\Delta^{n^{(i)}}}
	= \rho^n e^{-\rho \Delta}.
\end{equation}
where we recall that conditional on their number, catastrophes are uniformly distributed across a branch. The factorial terms in the numerator in Equation~\ref{eq:priorCatRho} account for the fact that the set of catastrophes on each branch is invariant to relabelling.

The prior on $ \rho $ in Table~\ref{tab:priors} is $ \Gamma(a, b) $ where $ a = 1.5 $ and $ b = 5 \times 10^3 $. We now integrate $ \rho $ out of Equation~\ref{eq:priorCatRho} with respect to its Gamma prior,
\begin{align*}
\pi_C(C)
	&= \int_0^{\infty} \pi_{C | R}(C | \rho) \pi_R(\rho) \mathrm{d}\rho \\
	&= \frac{b^a}{\Gamma(a)} \int_0^{\infty} \rho^{n + a - 1} e^{-\rho(\Delta + b)}  \mathrm{d}\rho \\
	&= \frac{b^a}{\Gamma(a)} \frac{\Gamma(n + a)}{(\Delta + b)^{n + a}} \\
	&= \frac{\Gamma(n + a)}{\Gamma(a) n!} \left(\frac{\Delta}{\Delta + b}\right)^n \left(\frac{b}{\Delta + b}\right)^a \frac{n!}{\Delta^n},
\end{align*}
to obtain a Negative Binomial distribution on the number of catastrophes, with catastrophe locations uniformly distributed across the tree.

\subsection{MCMC transition kernels}

We extend existing sampling algorithms for the Stochastic Dollo model \citep{nicholls08, ryder11} to construct a Markov chain whose invariant distribution is the posterior $ \pi(g, \mu, \beta, \kappa, \Xi | R(\bD)) $ in Equation~\ref{eq:intPost}. Let $ x = [(V, E, T, C), \mu, \beta, \kappa, \Xi] $ denote the current state of the chain, and a move to a new state $ \xs $ drawn from the proposal distribution $ Q(x, \cdot) $ is accepted with probability
\[
\min\left[1, \frac{\pixs}{\pix} \frac{\Qxsx}{\Qxxs}\right].
\]

We apply the same scaling update to the lateral transfer rate $ \beta $ and the death rate $ \mu $. If $ \xs = [(V, E, T, C), \mu, \beta^*, \kappa, \Xi] $ where $ \beta^* | \beta \sim \Unif{\varrho^{-1} \beta}{\varrho \beta} $ for some constant $ \varrho > 1 $, the Hastings ratio for this move is
\[
\frac{\Qxsx}{\Qxxs} = \frac{\beta}{\beta^*}.
\]

A catastrophe $ c = (b, u) \in C $ in state \emph{x} occurs on branch $ b \in E $ at time $ t_{b} + u (t_{\pa(b)} - t_{b}) $ where $ u \in (0, 1) $ is the relative location of the catastrophe along the branch. The location for a new catastrophe $ c^* = (b^*, u^*) $ is chosen uniformly at random across the branches of the tree to form the proposed state $ \xs $ with catastrophe set $ C \cup \{c^*\} $. We choose catastrophes uniformly at random for deletion in the reverse move, so
\[
\frac{\Qxsx}{\Qxxs} = \frac{p_{DC}}{p_{AC}} \frac{1}{|C| + 1} \sum_{i \in E \setminus \{1\}} (t_{\pa(i)} - t_i),
\]
where $ p_{AC} $ and $ p_{DC} $ denote the probabilities of proposing to add and delete a catastrophe respectively.

We chose catastrophe $ c = (b, u) $ uniformly from the catastrophe set \emph{C} to move to branch $ b^* $ chosen uniformly from the $ \deg(b) + \deg[\pa(b)] - 2 $ branches neighbouring branch \emph{b}, where $ \deg(b) $ denotes the degree of node \emph{b}. This is equivalent to deleting a randomly chosen catastrophe and adding it to a neighbouring branch, although we do not resample the relative location \emph{u}. If $ c^* = (b^*, u) $ replaces \emph{c} in the proposed state $ \xs $,
\[
\frac{\Qxsx}{\Qxxs} = \frac{\deg(b) + \deg[\pa(b)] - 2}{\deg(b^*) + \deg[\pa(b^*)] - 2} \frac{t_{\pa(b^*)} - t_{b^*}}{t_{\pa(b)} - t_{b}},
\]
where the Jacobian term $ (t_{\pa(b^*)} - t_{b^*}) / (t_{\pa(b)} - t_{b}) $ accounts for the change in sampling distribution of the catastrophe position due to the change in branch lengths. In fact, for every proposed move \emph{x} to $ \xs = [(V', E', T', C), \dots] $ which affects tree branch lengths, the Hastings ratio includes a Jacobian term
\[
\prod_{i \in E \setminus \{1\}} \frac{|C^{(i)}|!}{(t_{\pa(i)} - t_i)^{|C^{(i)}|}} \frac{(t_{\pa(i)}^* - t_i^*)^{|C^{*(i)}|}}{|C^{*(i)}|!}
\]
to account for the relative sampling densities for the catastrophe sets in each state, where $ C^{(i)} $ and $ C^{*(i)} $ respectively denote the catastrophe set on branch \emph{i} in the current and proposed states.

We construct subtree-prune-and-regraft moves on the tree in such a way that the total number of catastrophes on the tree remains constant and does not affect the ratio of proposal distributions outside the scaling term above. Let $ \edge{\pa(i)}{i} \in E $ denote a time-directed branch. From the current state \emph{x}, we choose a node $ i \in V \setminus \{0, 1\} $ below the root, prune the subtree beneath its parent $ \pa(i) $ and reattach it at a location chosen uniformly along a randomly chosen branch $ \edge{\pa(j)}{j} \in E $ to create state $ \xs $. Now, recall that catastrophes are indexed by the offspring node of the branch they lie on and suppose that vertices retain their labels in the move from state \emph{x} to state $ \xs $. There are three possible outcomes of the proposed move.
\begin{itemize}
\item If neither node \emph{i} or $ \pa(j) $ is the root in \emph{x} then catastrophes remain on their assigned branches in state $ \xs $. In more detail, if a catastrophe $ c = (b, u) $ is on branch $ \edge{\pa(b)}{b} $ in state \emph{x} then $ c = (b, u) $ in state $ \xs $ also, with the possible difference that node \emph{b}'s parent may change thereby affecting when the catastrophe occurs. We illustrate this in Figure~\ref{fig:catSPRx1}.
\item If $ \pa(i) $ is the root in state \emph{x} then \emph{i}'s sibling $ \sib(i) $ is the root in state $ \xs $. There are no catastrophes on $ \edge{0}{\pa(i)} $ in \emph{x} by definition so we move the catastrophes currently on $ \edge{\pa(i)}{\sib(i)} $ in \emph{x} to $ \edge{\pa(j)}{\pa(i)} $ in $ \xs $, and the catastrophes on $ \edge{\pa(j)}{j} $ in \emph{x} to $ \edge{\pa(i)}{j} $ in $ \xs $. We illustrate this move in Figure~\ref{fig:catSPRx2}.
\item Finally, if \emph{j} is the root in \emph{x} then $ \pa(i) $ becomes the root in $ \xs $ so we move the catastrophes on $ \edge{\pa(\pa(i))}{\pa(i)} $ in \emph{x} to $ \edge{\pa(i)}{j} $ in $ \xs $. This move, the reverse of the above, is illustrated in Figure~\ref{fig:catSPRx3}.
\end{itemize}

\begin{figure}[p!ht]
\centering
\subfloat[Neither node $ \pa(i) $ nor $ \pa(j) $ is the root in either state. Catastrophes remain on their assigned branches.]{\includegraphics[width=0.9\textwidth]{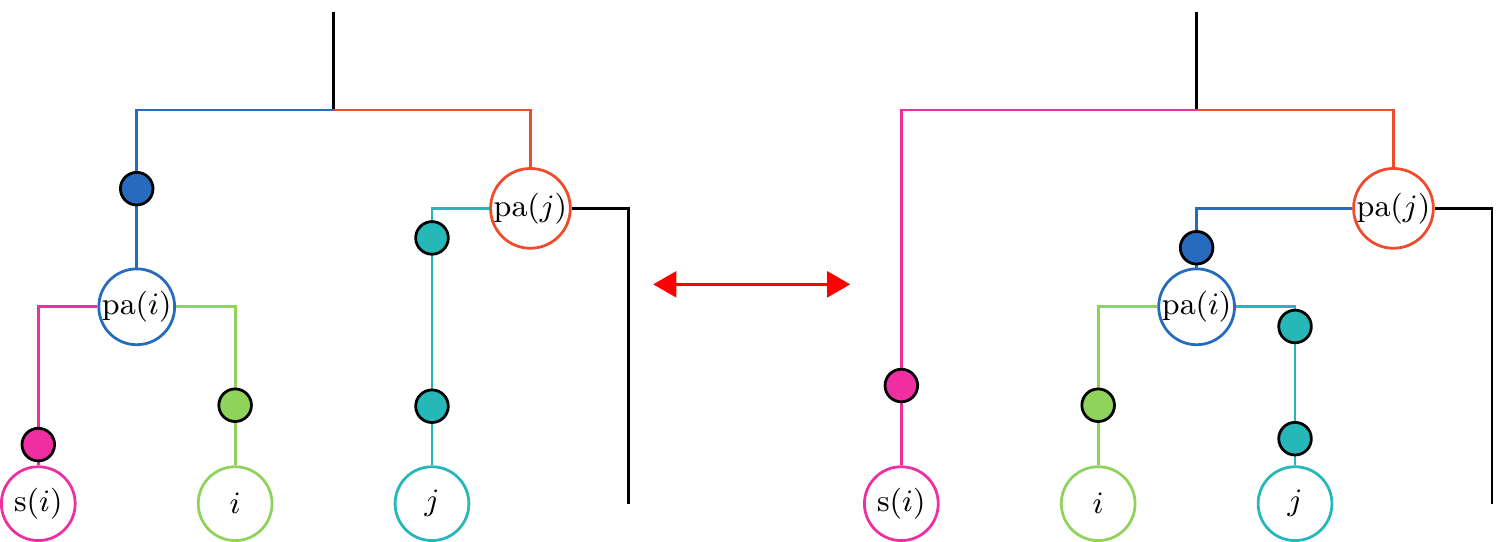} \label{fig:catSPRx1}}

\subfloat[Node $ \pa(i) $ is the root in the left-hand state and node $ \pa(j) $ is \emph{i}'s sibling. The catastrophes on branch $ \edge{\pa(i)}{\pa(j)} $ in the left-hand state are moved to edge $ \edge{\pa(j)}{\pa(i)} $ in the right-hand state when node \emph{j} becomes the root.]{\includegraphics[width=0.9\textwidth]{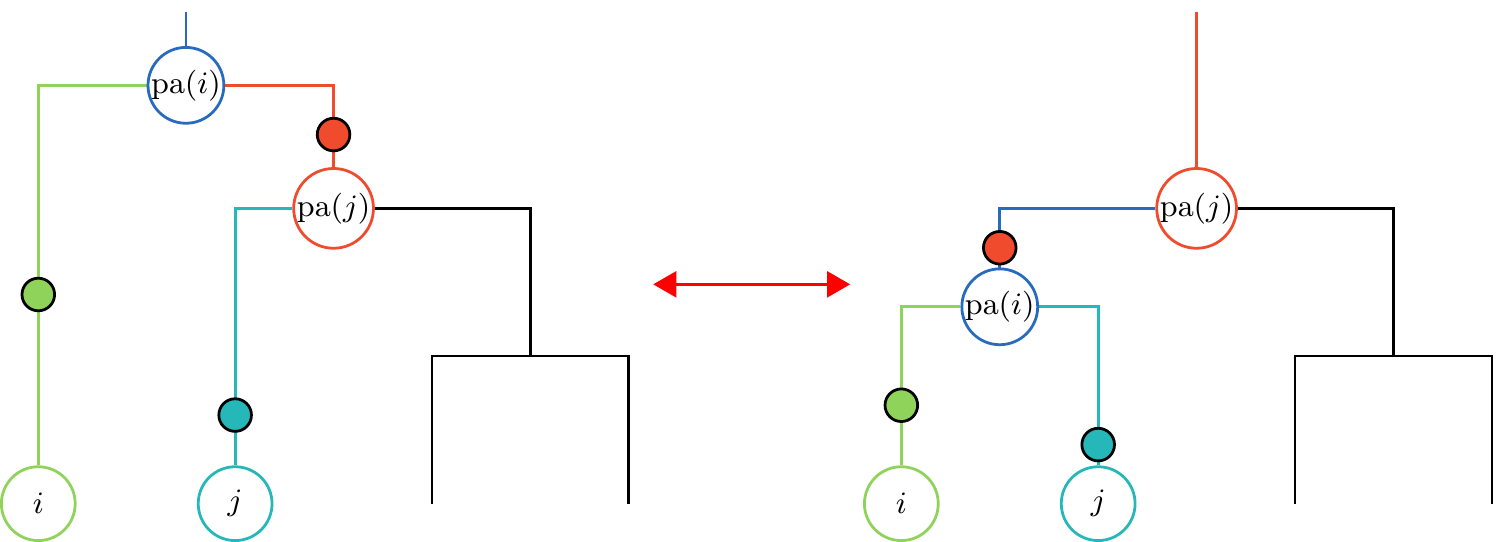} \label{fig:catSPRx2}}

\subfloat[Node \emph{j} is the root in the left-hand state. The catastrophes on branch $ \edge{j}{\pa(i)} $ in the left-hand state are moved to edge $ \edge{\pa(i)}{j} $ in the right-hand state.]{\includegraphics[width=0.9\textwidth]{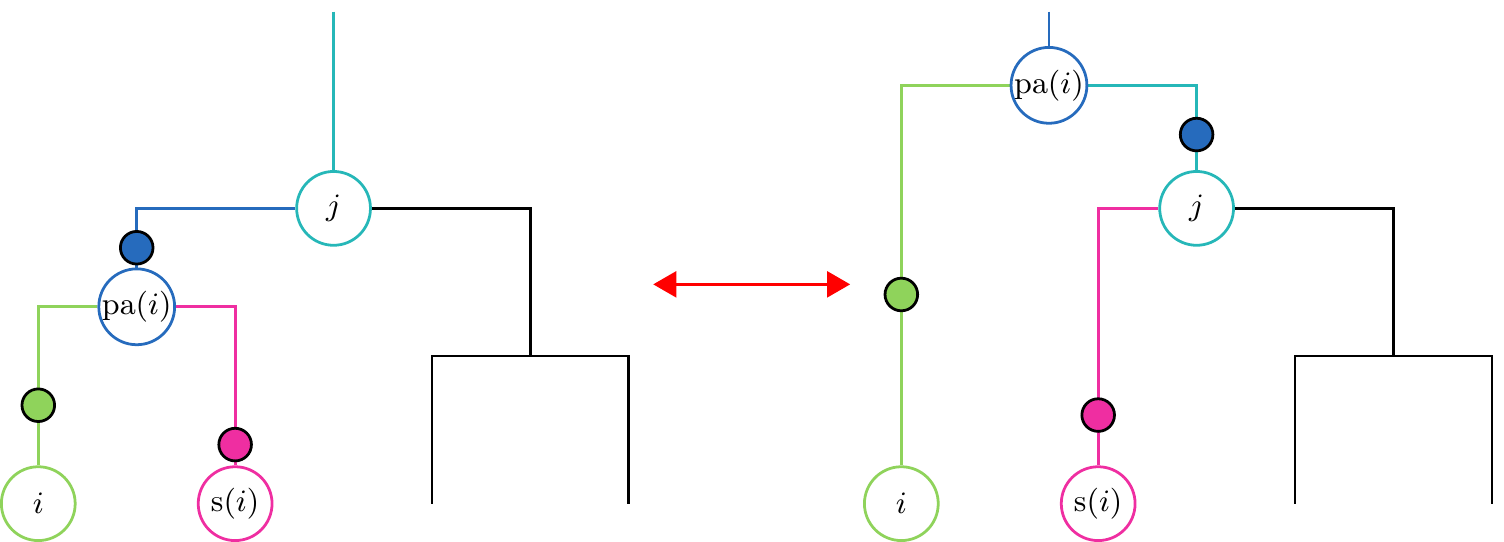} \label{fig:catSPRx3}}

\caption{Subtree-prune-and-regraft moves do not affect the number of catastrophes on the tree. Only the catastrophe branch index \emph{b} changes in a move; the location \emph{u} along the branch remains fixed.}
\label{fig:catSPR}
\end{figure}

\section{Method testing}
\label{app:testing}

We validate our model and computer implementation using three coupled synthetic data sets. The data set \texttt{SIM-B} is a draw from the SDLT process on the tree in Figure~\ref{tree:simTrue} with parameters in Table~\ref{tab:simPars}. We use \texttt{SIM-B} here to test the identifiability of the SDLT model. The relative transfer rate $ \beta / \mu $ is high and although the shortcomings of the SD model in this setting have already been established \citep{nicholls08, greenhill09}, we also fit it here to highlight the effect of properly controlling for lateral transfer.

\begin{table}[t]
\centering
\caption{The parameter settings we used to simulate data set \texttt{SIM-B}. The effective catastrophe duration $ \delta = -\mu^{-1} \log(1 - \kappa) = 500 $ years.}
\label{tab:simPars}
\begin{tabular}{@{}llll@{}} \toprule
Parameter			&	Value							&	Parameter					&	Value \\ \cmidrule(r){1-2} \cmidrule(l){3-4}
Trait birth rate	&	$ \lambda = 10^{-1} $			&	Root time 					&	$ t_1 = -10^3	$	\\
Trait death rate	&	$ \mu = 5 \times 10^{-4} $		&	Catastrophe severity		&	$ \kappa = 0.2212 $ \\
Trait transfer rate	&	$ \beta = 5 \times 10^{-4} $	&	Observation probabilities	&	$ \Xi \sim \beta(1, 1/3)^L $ \\ \bottomrule
\end{tabular}
\end{table}

From \texttt{SIM-B}, we create two additional data sets: \texttt{SIM-N} and \texttt{SIM-T}. To form \texttt{SIM-N}, we discard any trait copy derived from a lateral transfer event; that is, we do not discard all instances of a given trait, only the copies which transferred. This is equivalent to ignoring all the lateral transfer events in the generation of \texttt{SIM-B}, so \texttt{SIM-N} is a draw from the SD process coupled to \texttt{SIM-B}. The SD model is nested within the SDLT model and we use \texttt{SIM-N} to test the consistency of the two models when the lateral transfer rate $ \beta = 0 $.

Recently transferred traits are more readily identified and discarded in practice. This potential bias is a common source of model misspecification. To this end, we only discard instances of traits in \texttt{SIM-B} which transferred in the final 250 years to create \texttt{SIM-T}. We fit the SDLT and SD models here to test their robustness to this model misspecification. We summarise the synthetic data sets in Table~\ref{tab:simSum}.

\begin{table}[t]
\centering
\caption{Summary of synthetic data sets for model testing.}
\label{tab:simSum}
\begin{tabular}{@{}llll@{}} \toprule
Data set		&	Traits	&	True model	&	Purpose	\\ \midrule
\texttt{SIM-B}	&	678		&	SDLT		&	Identifiability \\
\texttt{SIM-N}	&	672		&	SD			&	Consistency	\\
\multirow{2}{*}{\texttt{SIM-T}}	&	\multirow{2}{*}{675}	&	\multirow{2}{0.275\textwidth}{SDLT before time~$ -250 $, SD thereafter}		&	\multirow{2}{*}{Robustness} \\ \\ \bottomrule
\end{tabular}
\end{table}

In Figure~\ref{fig:cdfSIM}, we compare the exact Poisson cumulative distribution function of each pattern in $ \cP $ under the SDLT model (Theorem~\ref{thm:bdl}) with empirical estimates based on $ 10^3 $ replicates of \texttt{SIM-B}. On the basis of these tests, we conclude that our Gillespie-style simulation algorithm \citep{gillespie77} and expected pattern frequency calculation~\eqref{eq:ivp} are correct.

\begin{figure}[t]
\includegraphics[width=\textwidth, trim = 1cm 0cm 1.5cm 0.3cm, clip]{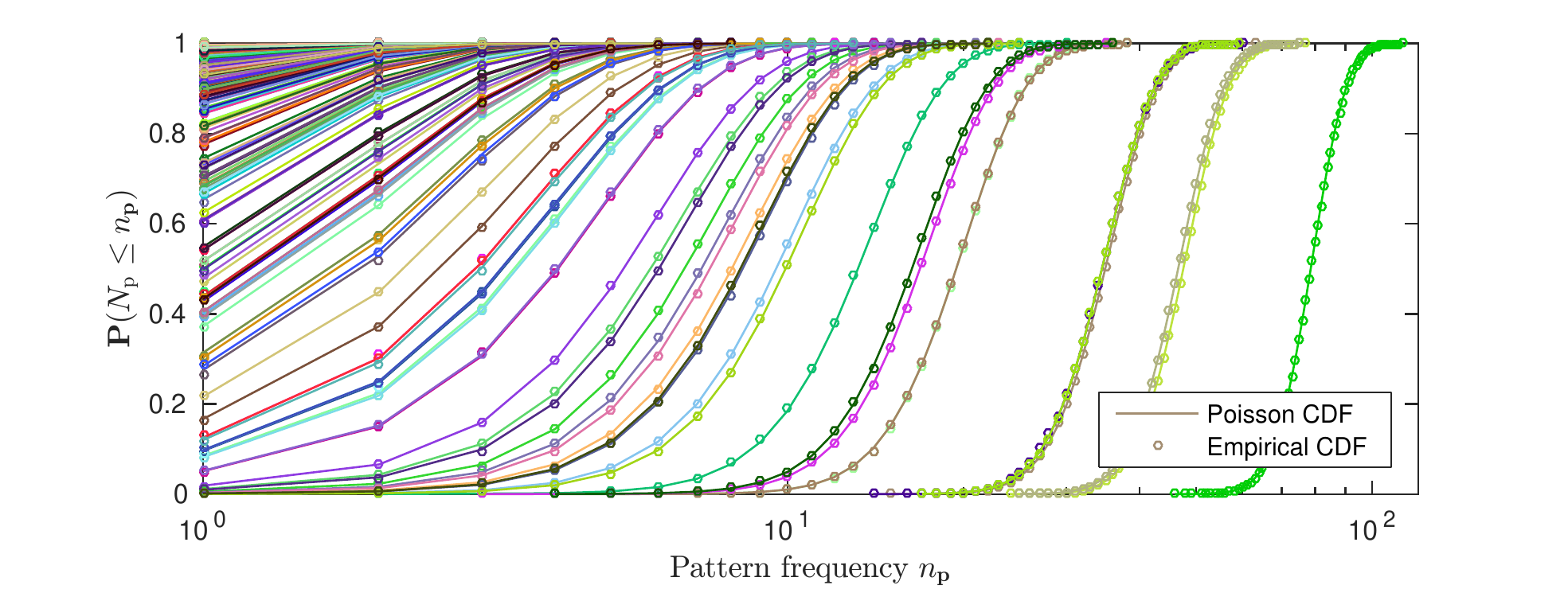}
\caption{Exact and empirical cumulative distribution functions (CDFs) of patterns in $ \cP $ for $ 10^3 $ replicates of \texttt{SIM-B}.}
\label{fig:cdfSIM}
\end{figure}

For the MCMC analyses, we discard traits not marked present in at least one taxon (Section~\ref{sec:reg}). In addition to the clade constraints depicted in Figure~\ref{tree:simTrue}, we enforce a minimum root time $ \ubar{t}_1 = -2000 $ years (equivalently, a maximum root age $ -t_1 $ of 2000 years). For the plots which follow, figures in parentheses denote effective sample sizes unless stated otherwise.

\begin{figure}[p!ht]
\centering
{\subfloat[True tree.]{\includegraphics[width=0.4\textwidth, trim = 0cm 1.7cm 4.5cm 0cm, clip]{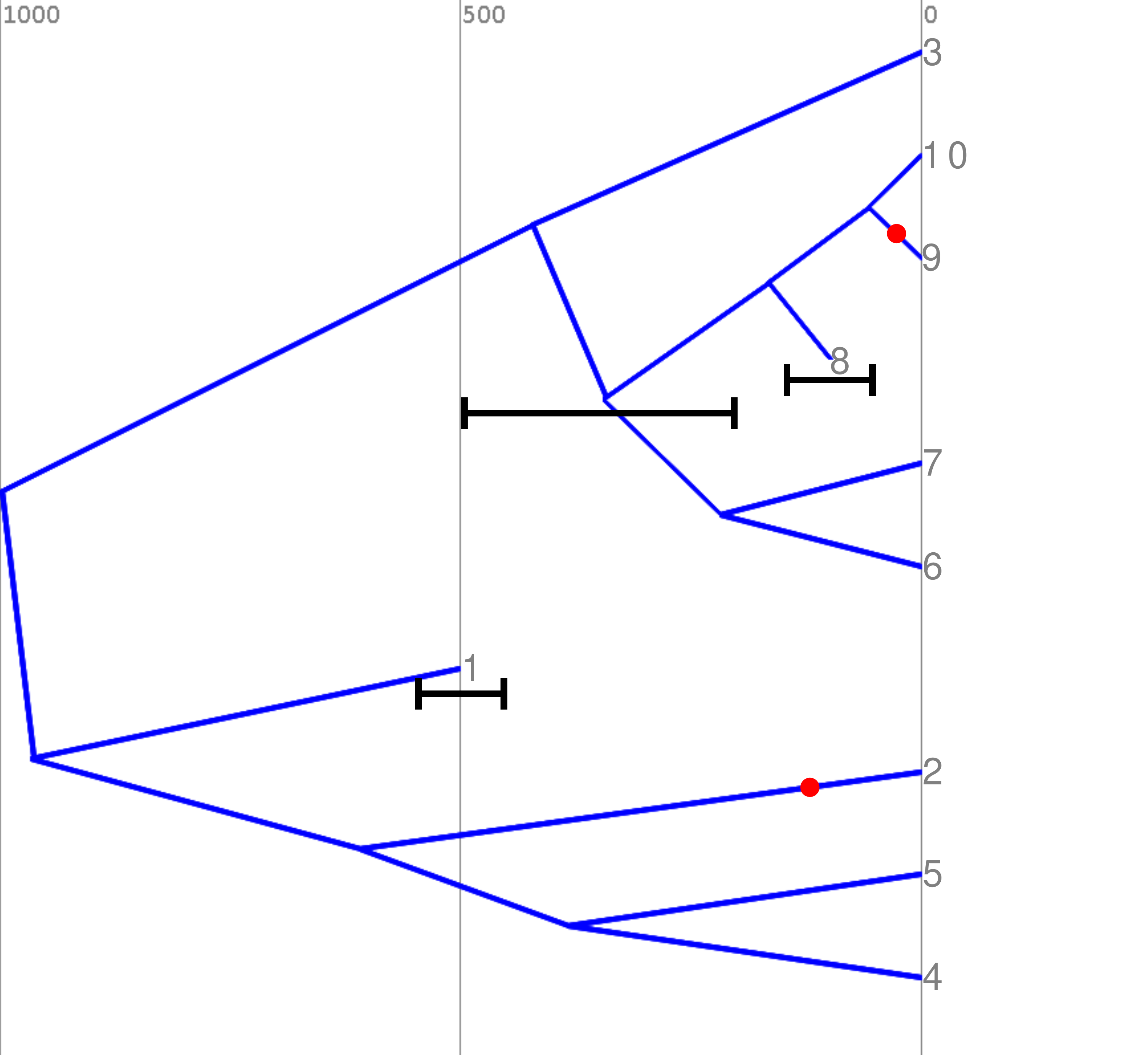} \label{tree:simTrue}}} ~
{
\begin{tabular}{@{}c@{}c@{}c@{}}
	\subfloat[SDLT on \texttt{SIM-B} ($ 31\% $).]{\includegraphics[width=0.325\textwidth, trim = 0cm 1.7cm 4.5cm 0cm, clip]{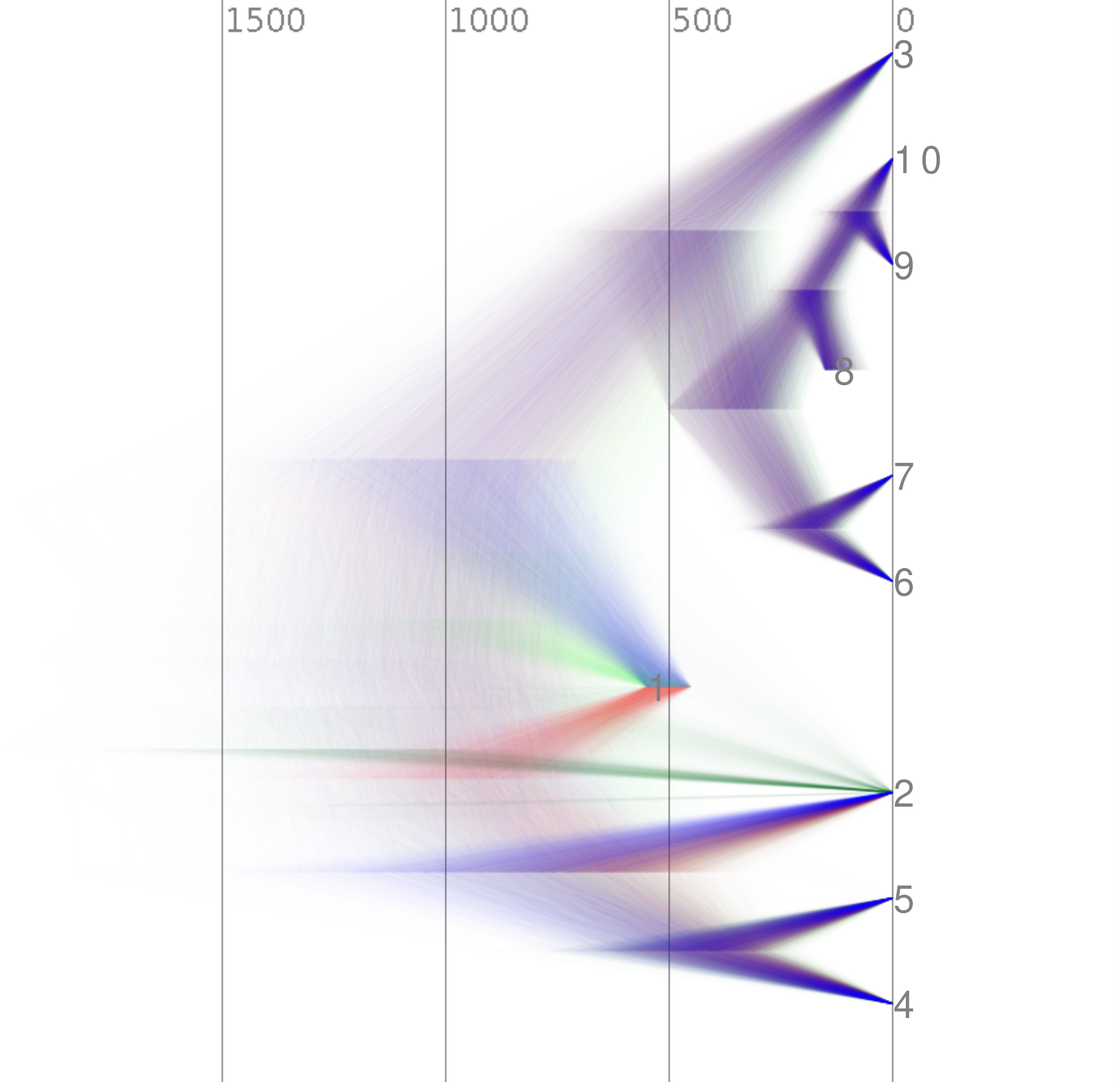} \label{tree:simBB}} &
	\subfloat[SDLT on \texttt{SIM-N} ($ 28\% $).]{\includegraphics[width=0.325\textwidth, trim = 0cm 1.7cm 4.5cm 0cm, clip]{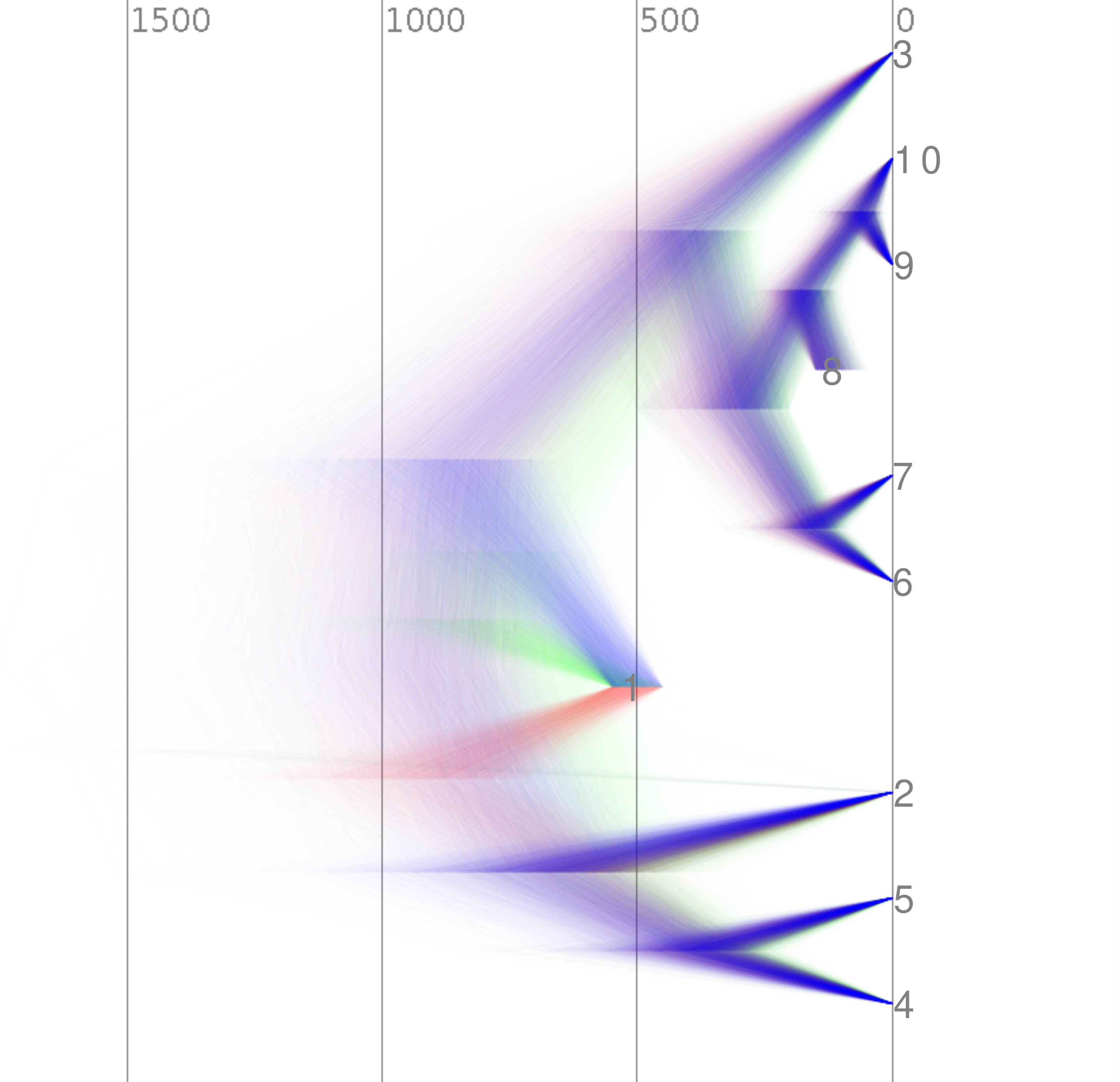} \label{tree:simNB}} &
	\subfloat[SDLT on \texttt{SIM-T} ($ 11\% $).]{\includegraphics[width=0.325\textwidth, trim = 0cm 1.7cm 4.5cm 0cm, clip]{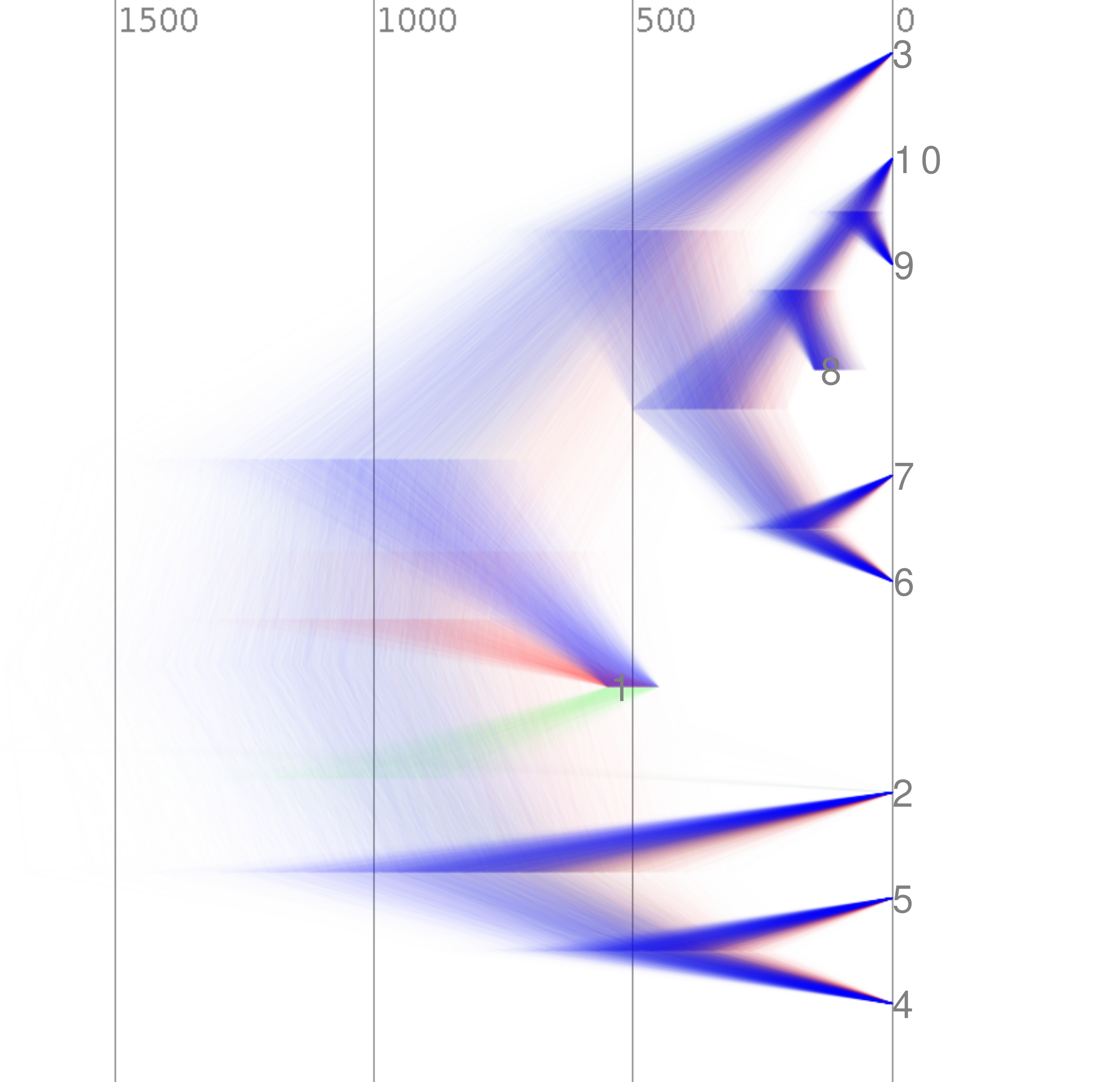} \label{tree:simTB}} \\
	\subfloat[SD on \texttt{SIM-B}	 ($ 0\% $).]{\includegraphics[width=0.325\textwidth, trim = 0cm 1.7cm 4.5cm 0cm, clip]{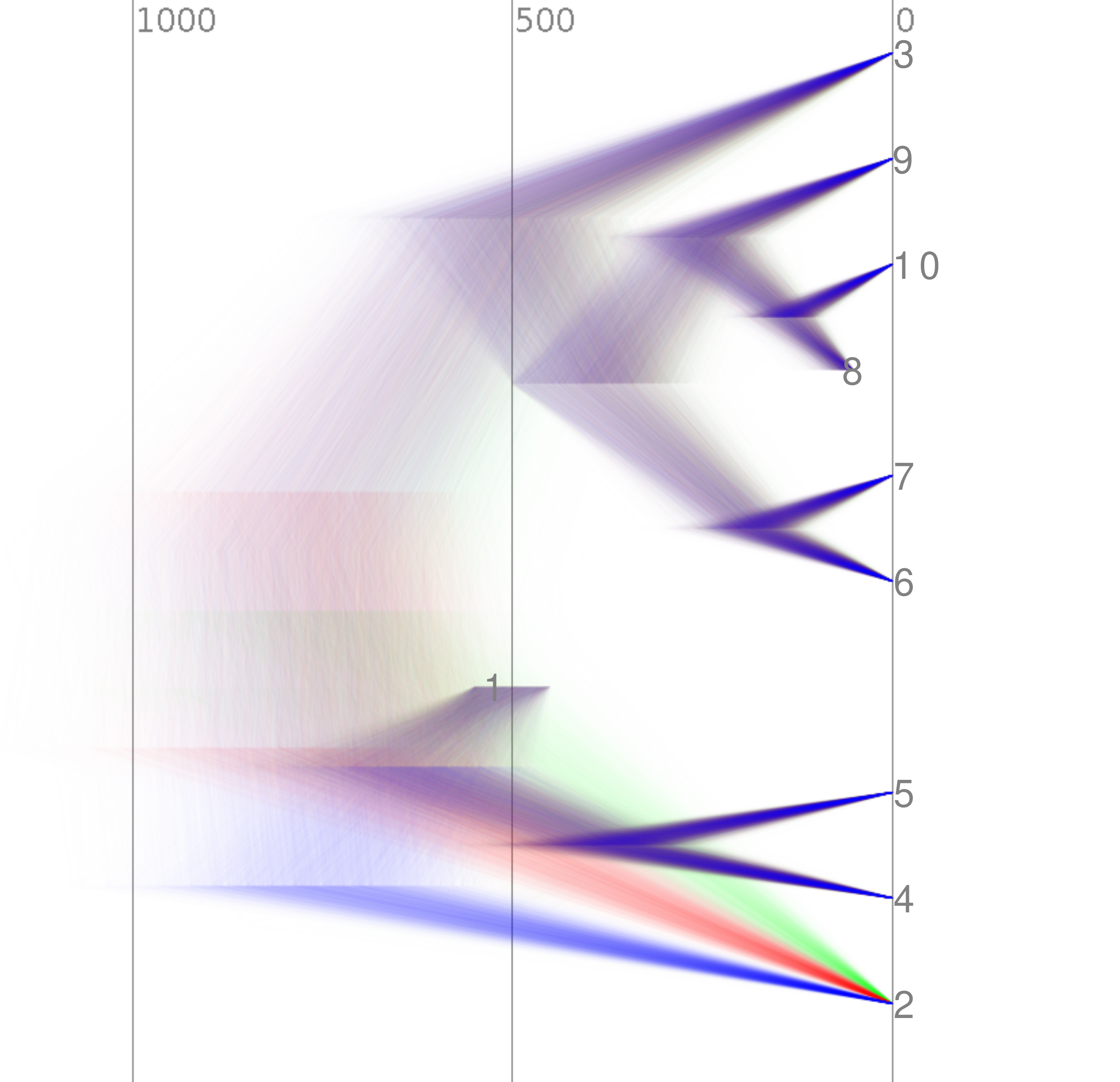} \label{tree:simBN}} &
	\subfloat[SD on \texttt{SIM-N} 	 ($ 27\% $).]{\includegraphics[width=0.325\textwidth, trim = 0cm 1.7cm 4.5cm 0cm, clip]{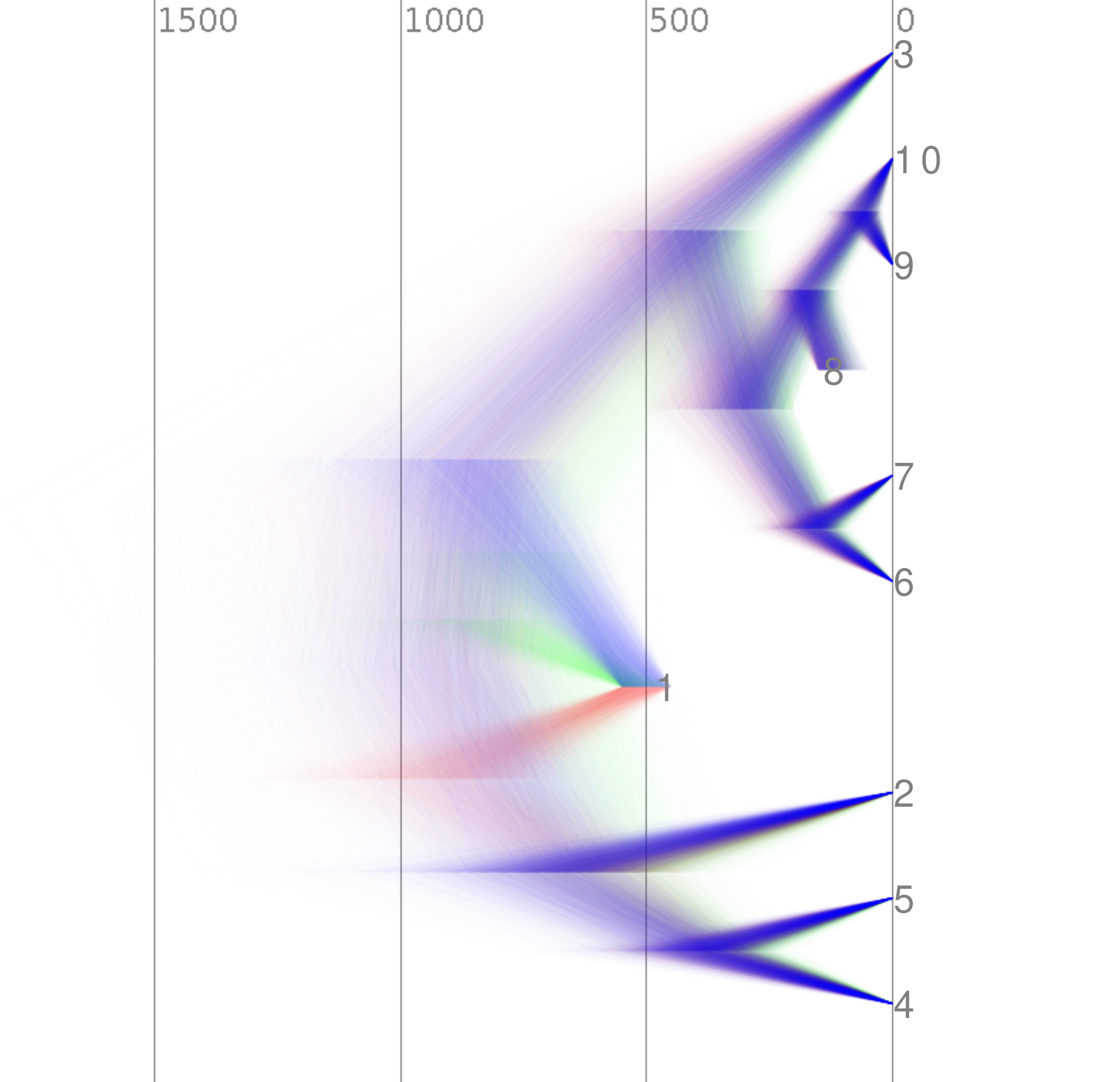} \label{tree:simNN}} &
	\subfloat[SD on \texttt{SIM-T} 	 ($ 10\% $).]{\includegraphics[width=0.325\textwidth, trim = 0cm 1.7cm 4.5cm 0cm, clip]{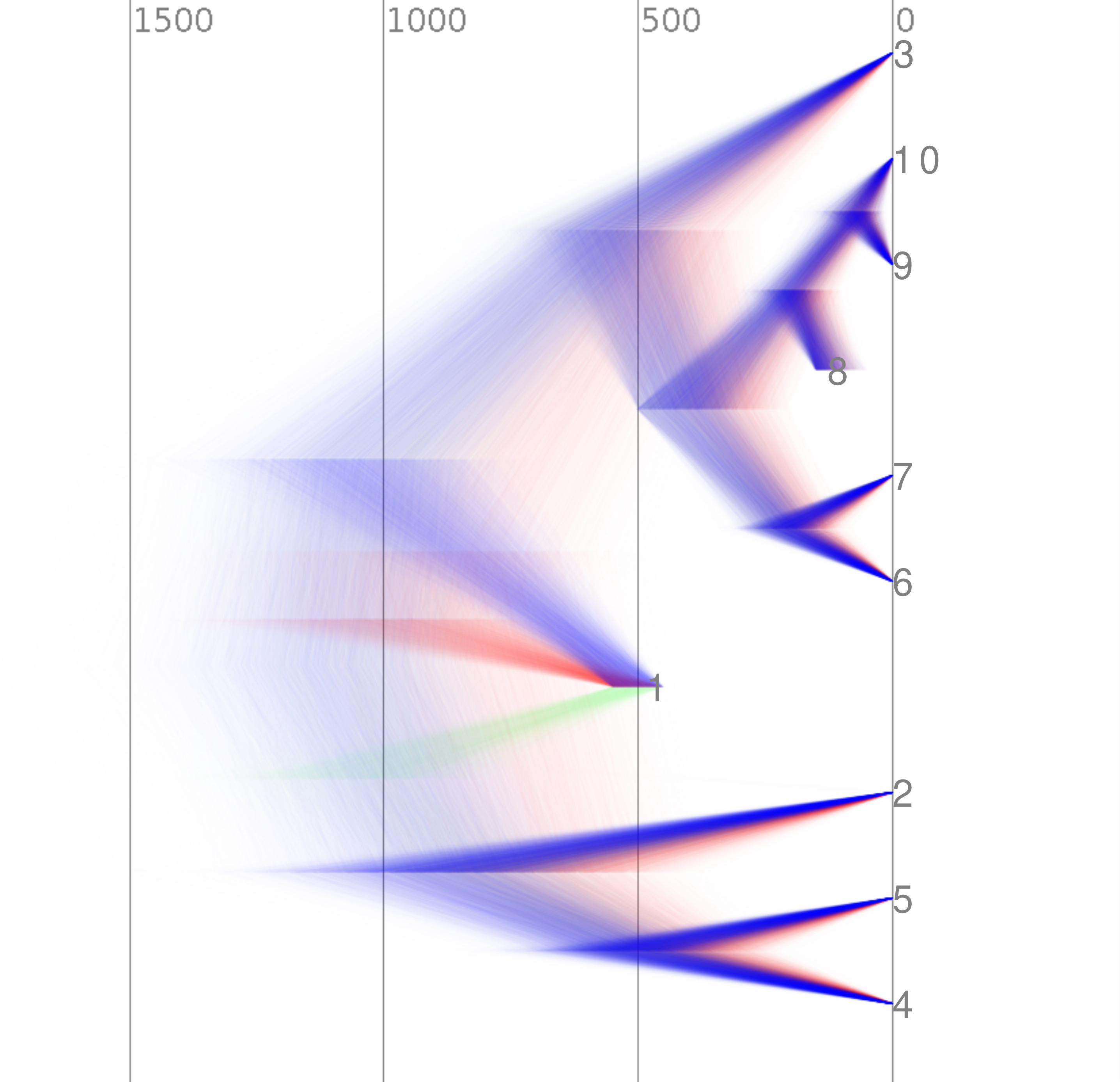} \label{tree:simTN}}
\end{tabular}
}
\caption{The tree \protect\subref{tree:simTrue} used to create the synthetic data sets. Catastrophe locations are marked in red and clade constraints for the MCMC analyses in black. \texttt{DensiTree} \protect\citep{bouckaert14} plots of the marginal tree posteriors for each synthetic dataset under the SDLT \protect\subref{tree:simBB}--\protect\subref{tree:simTB} and SD \protect\subref{tree:simBN}--\protect\subref{tree:simTN} models. Figures in parentheses denote posterior support for the true topology. The most frequently sampled topologies in each case are coloured blue, followed by red and green, with the remainder in dark green.}
\label{tree:sim}
\end{figure}

Of particular interest among the marginal tree posteriors in Figure~\ref{tree:sim} is the contrasting supports for the true topology in each case, particularly \texttt{SIM-B} where the SD model focuses on the wrong topology entirely. We return to this point in the goodness-of-fit analyses at the end of this section. There is little to distinguish here between the models fit to \texttt{SIM-N} and \texttt{SIM-T} so they are consistent in this respect.

We summarise parameter samples from the models in Figures~\ref{hist:simB}--\ref{hist:simT}. The SD model applied to \texttt{SIM-B} underestimates the root age $ -t_1 $ in Figure~\ref{hist:simB}. Similarly, the death rate $ \mu $ is inflated under the SD model as it attempts to account for variation due to lateral transfer through trait deaths instead. Posterior estimates of the relative transfer rate $ \beta / \mu $ are consistent with their true values for \texttt{SIM-B} and \texttt{SIM-N}. On \texttt{SIM-T}, the posterior resembles a mixture distribution, which is not surprising given the nature of the data. There is no cause for concern among the trace and autocorrelation plots in Figures~\ref{trace:simB}--\ref{autocorr:simT}, so it remains to assess the quality of our analyses.

\begin{figure}[p]
\centering
\begin{tabular}{@{}c@{}@{}c@{}}
	\includegraphics[width=0.5\textwidth, trim = 0.05cm 0cm 0.6cm 0cm, clip]{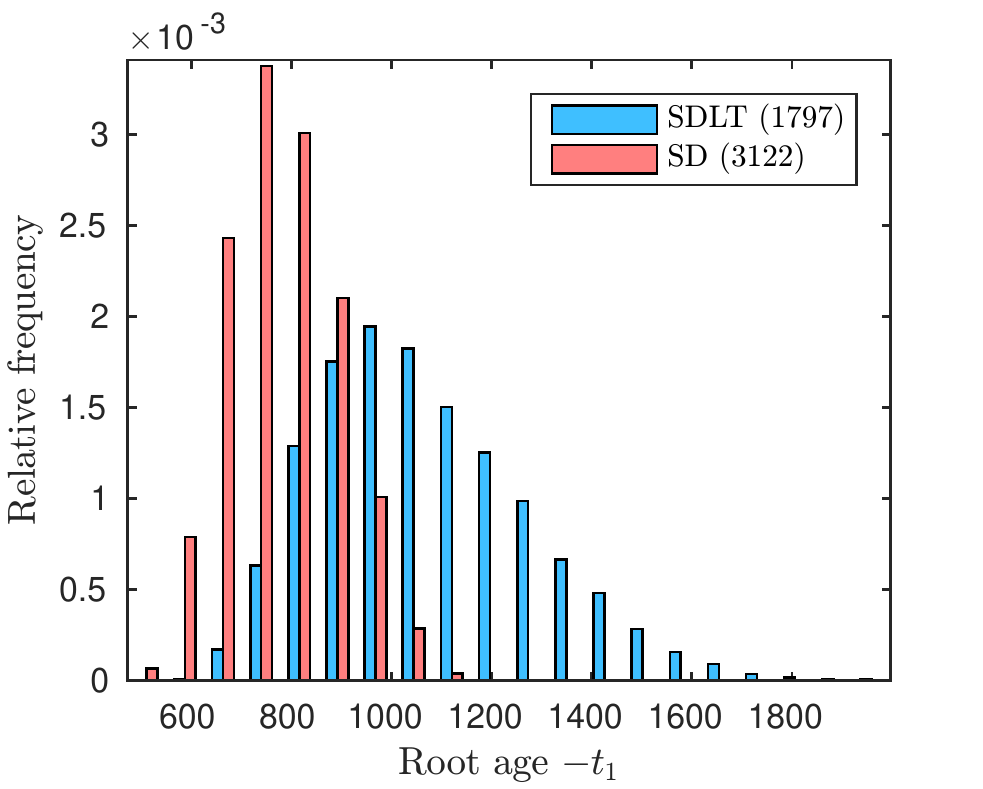} &
	\includegraphics[width=0.5\textwidth, trim = 0.05cm 0cm 0.6cm 0cm, clip]{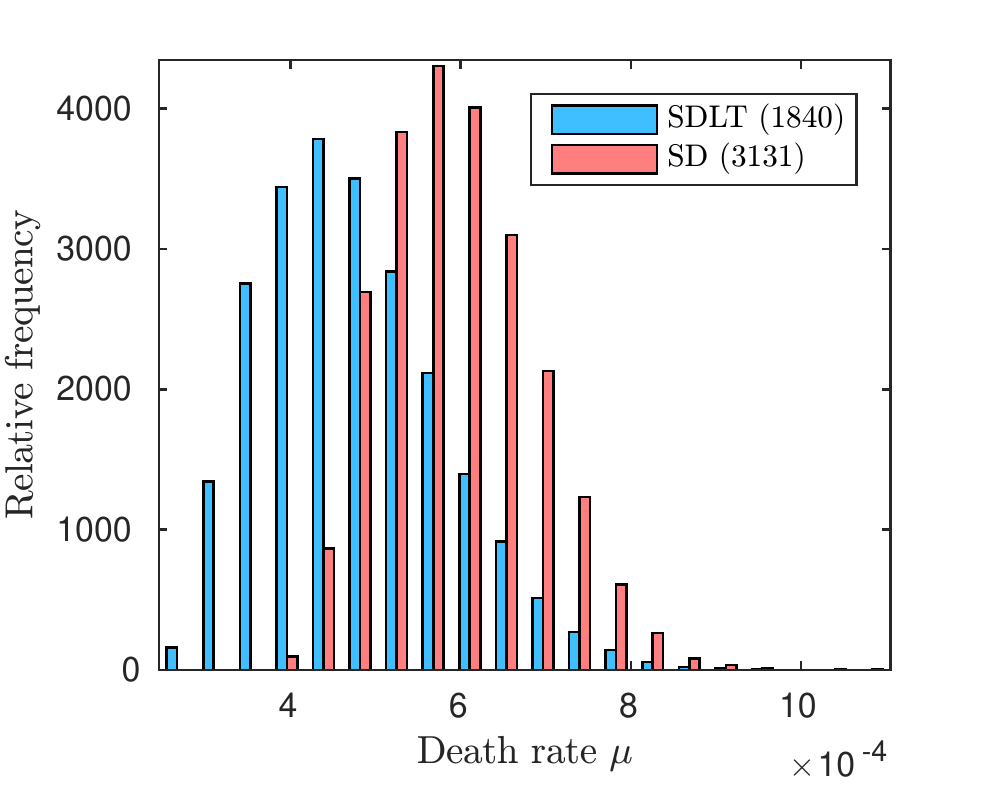} \\
	\includegraphics[width=0.5\textwidth, trim = 0.05cm 0cm 0.6cm 0cm, clip]{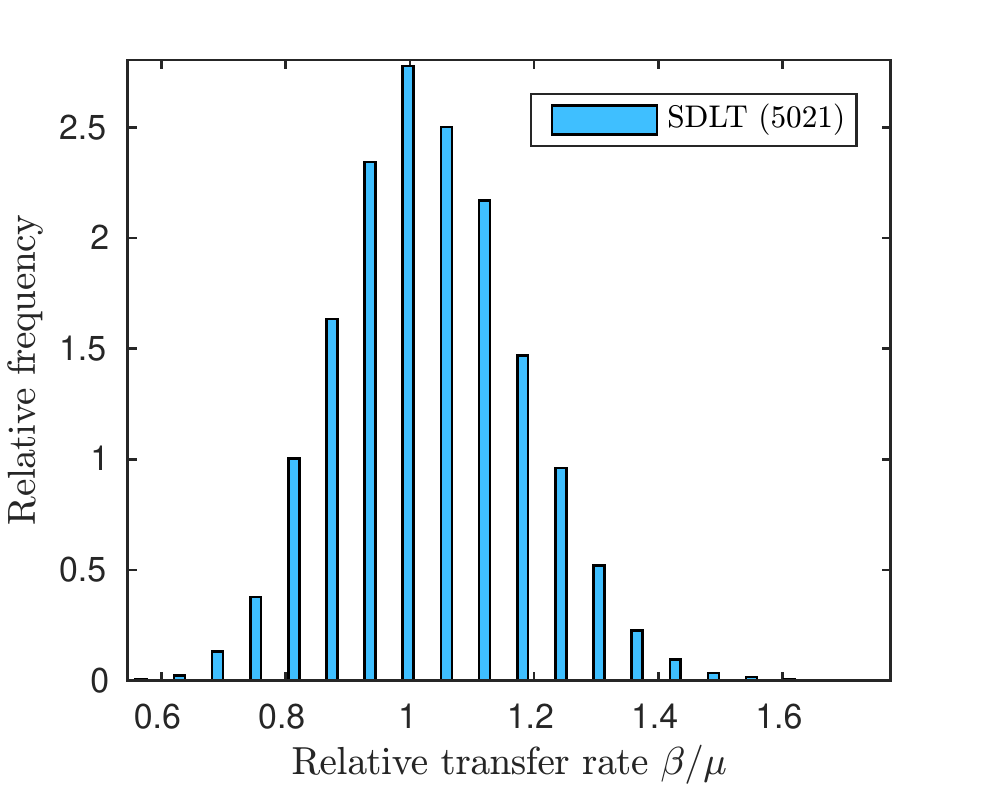} &
	\includegraphics[width=0.5\textwidth, trim = 0.05cm 0cm 0.6cm 0cm, clip]{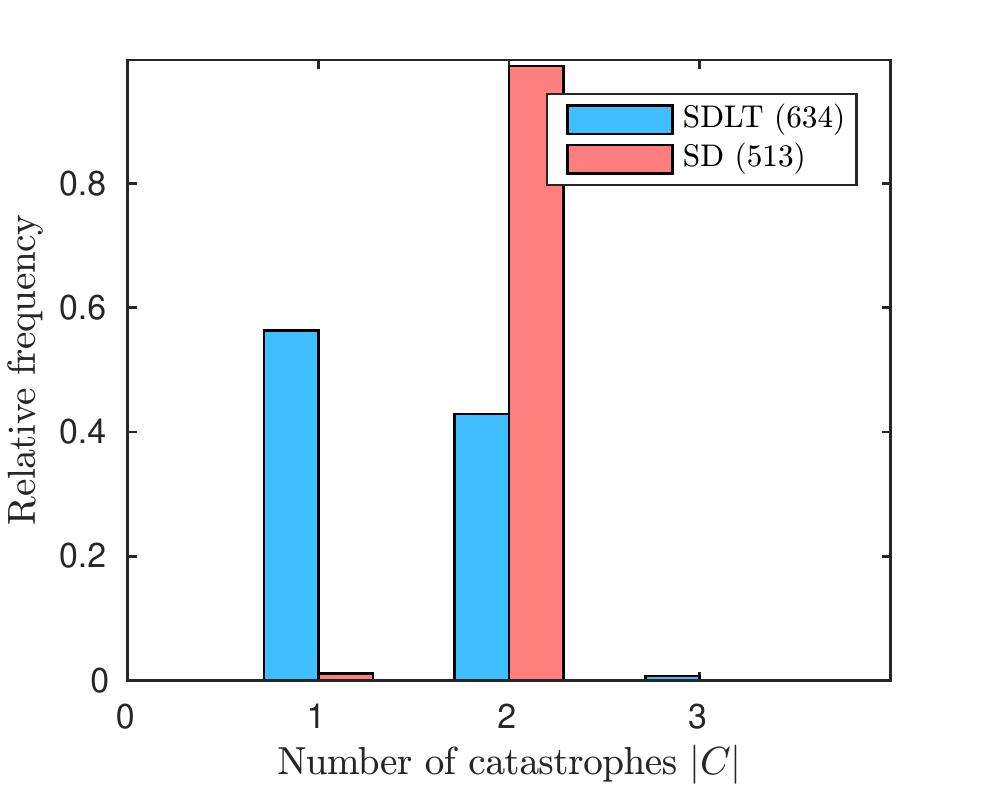} \\
	\includegraphics[width=0.5\textwidth, trim = 0.05cm 0cm 0.6cm 0cm, clip]{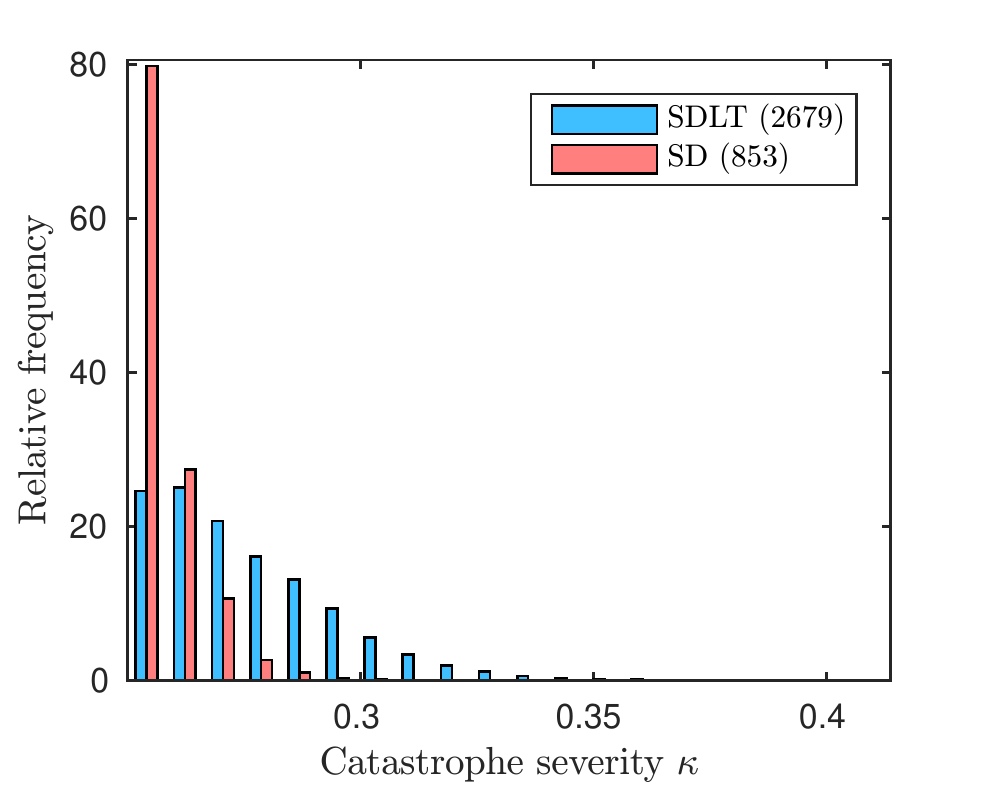} &
	\includegraphics[width=0.5\textwidth, trim = 0.05cm 0cm 0.6cm 0cm, clip]{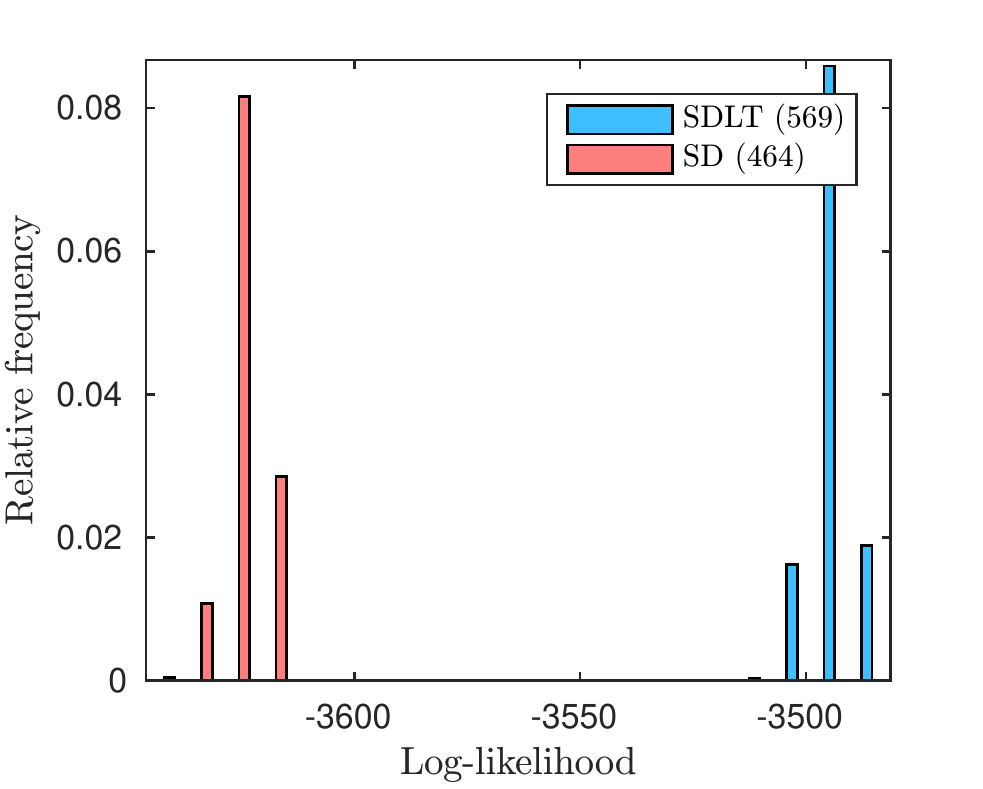}
\end{tabular}
\caption{Histograms of samples in our analyses of \texttt{SIM-B}.}
\label{hist:simB}
\end{figure}

\begin{figure}[p]
\centering
\begin{tabular}{@{}c@{}@{}c@{}}
	\includegraphics[width=0.5\textwidth, trim = 0.05cm 0cm 0.6cm 0cm, clip]{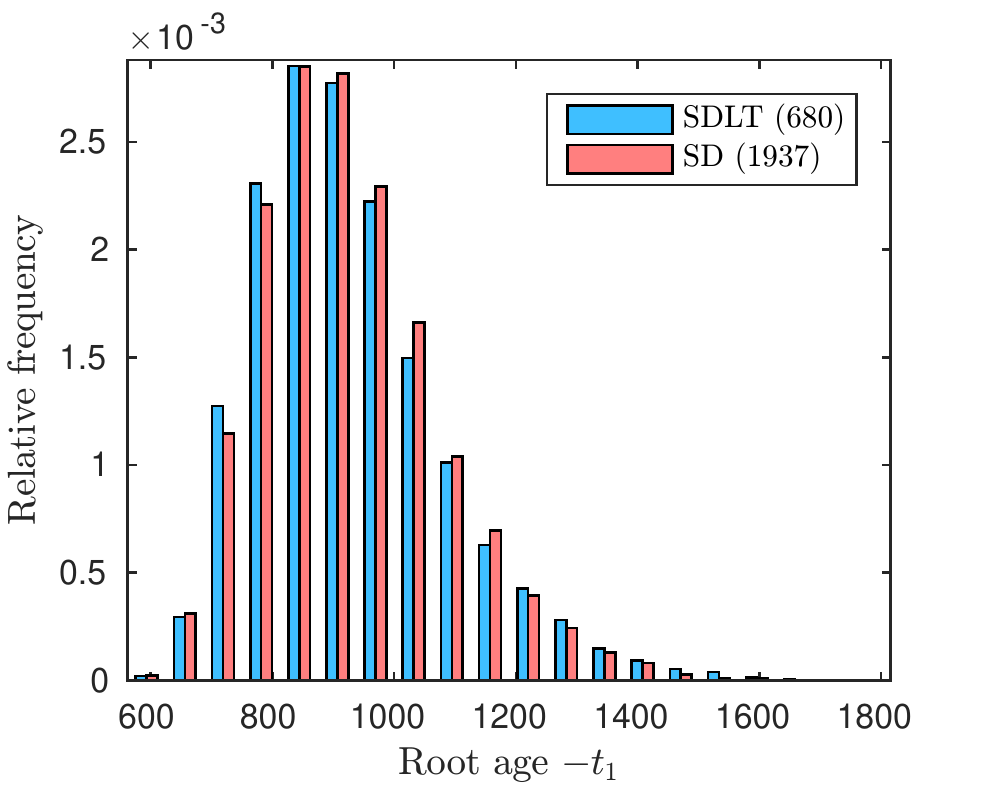} &
	\includegraphics[width=0.5\textwidth, trim = 0.05cm 0cm 0.6cm 0cm, clip]{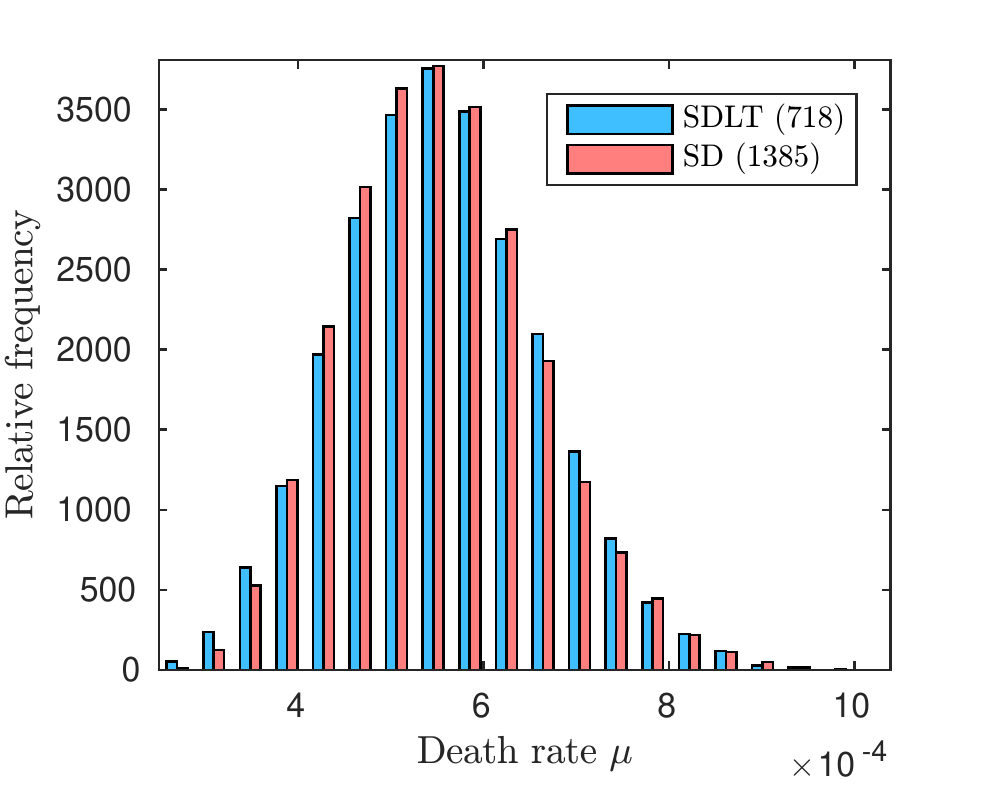} \\
	\includegraphics[width=0.5\textwidth, trim = 0.05cm 0cm 0.6cm 0cm, clip]{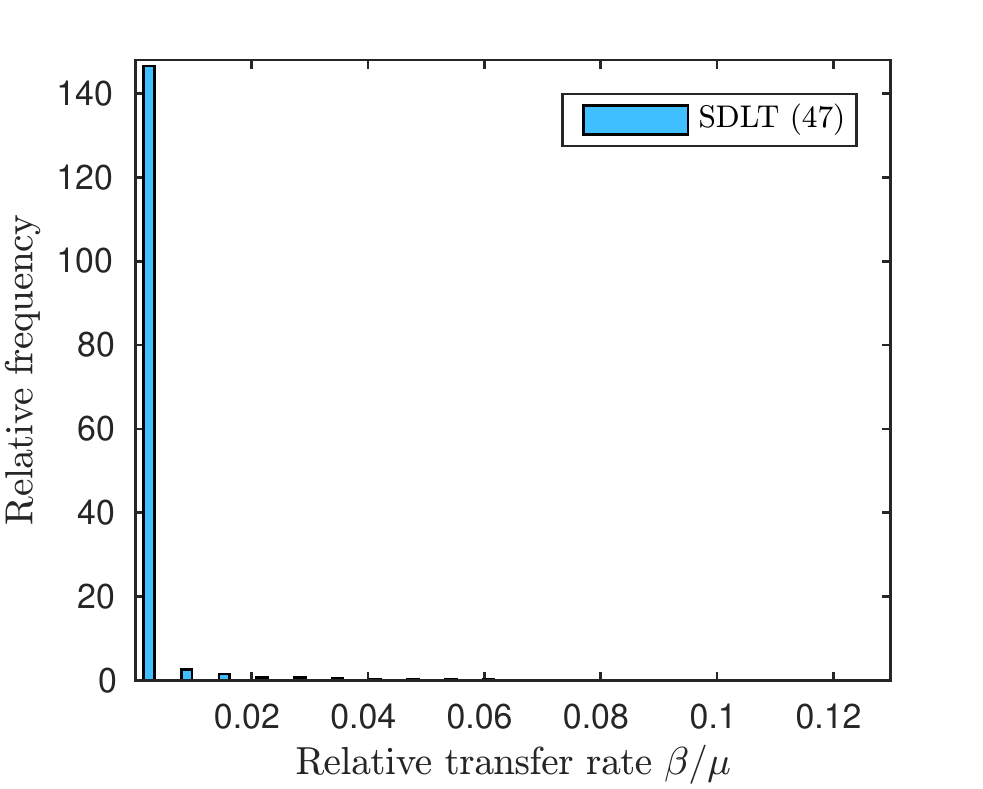} &
	\includegraphics[width=0.5\textwidth, trim = 0.05cm 0cm 0.6cm 0cm, clip]{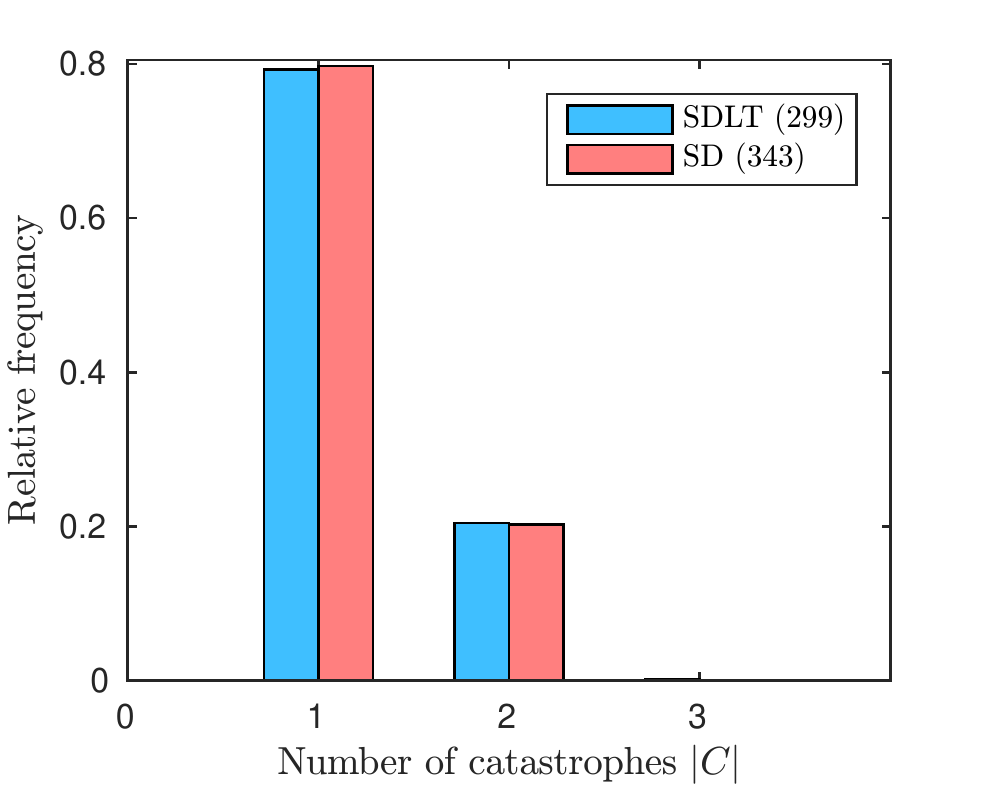} \\
	\includegraphics[width=0.5\textwidth, trim = 0.05cm 0cm 0.6cm 0cm, clip]{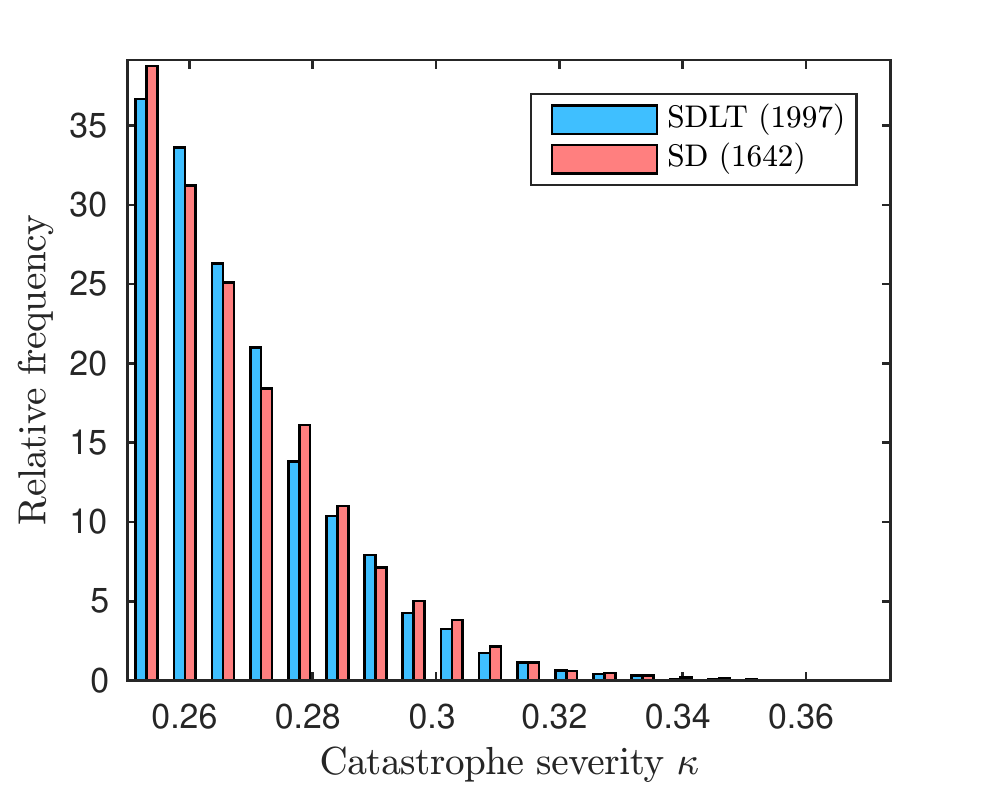} &
	\includegraphics[width=0.5\textwidth, trim = 0.05cm 0cm 0.6cm 0cm, clip]{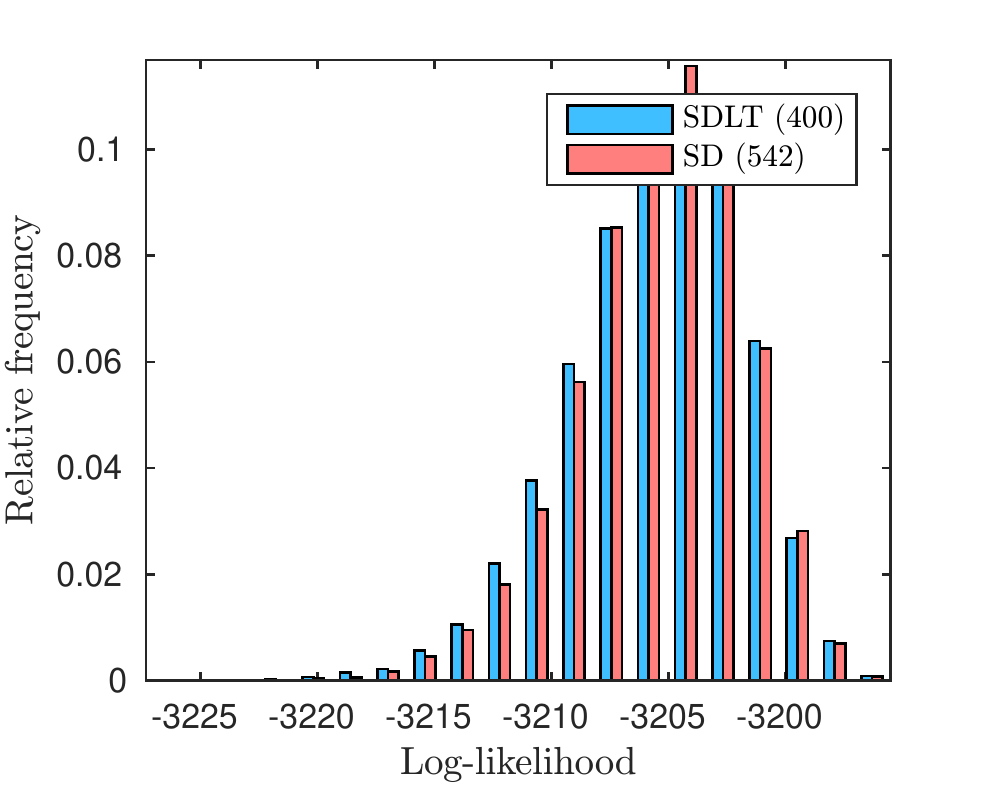}
\end{tabular}
\caption{Histograms of samples in our analyses of \texttt{SIM-N}.}
\label{hist:simN}
\end{figure}

\begin{figure}[p]
\centering
\begin{tabular}{@{}c@{}@{}c@{}}
	\includegraphics[width=0.5\textwidth, trim = 0.05cm 0cm 0.6cm 0cm, clip]{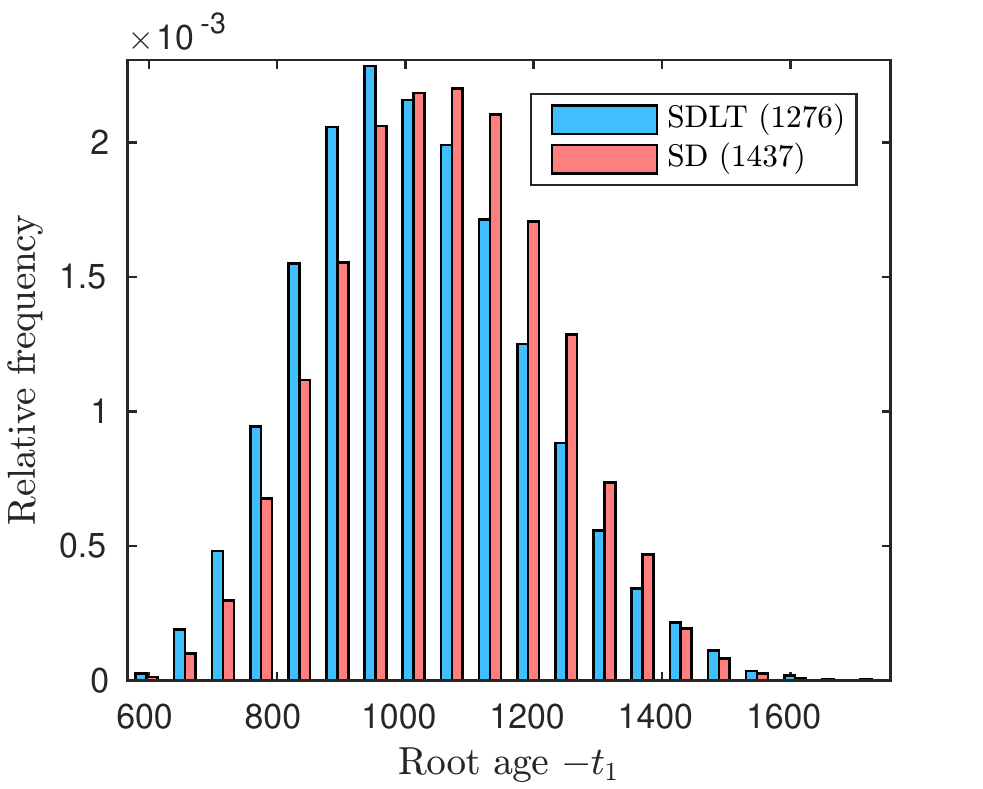} &
	\includegraphics[width=0.5\textwidth, trim = 0.05cm 0cm 0.6cm 0cm, clip]{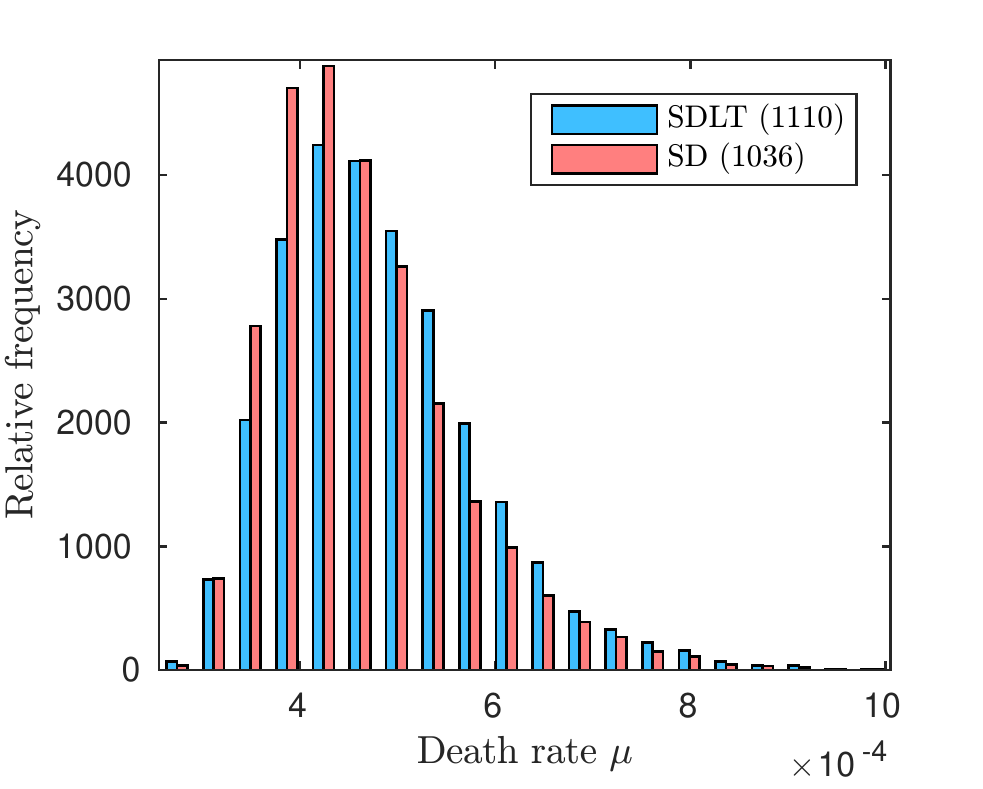} \\
	\includegraphics[width=0.5\textwidth, trim = 0.05cm 0cm 0.6cm 0cm, clip]{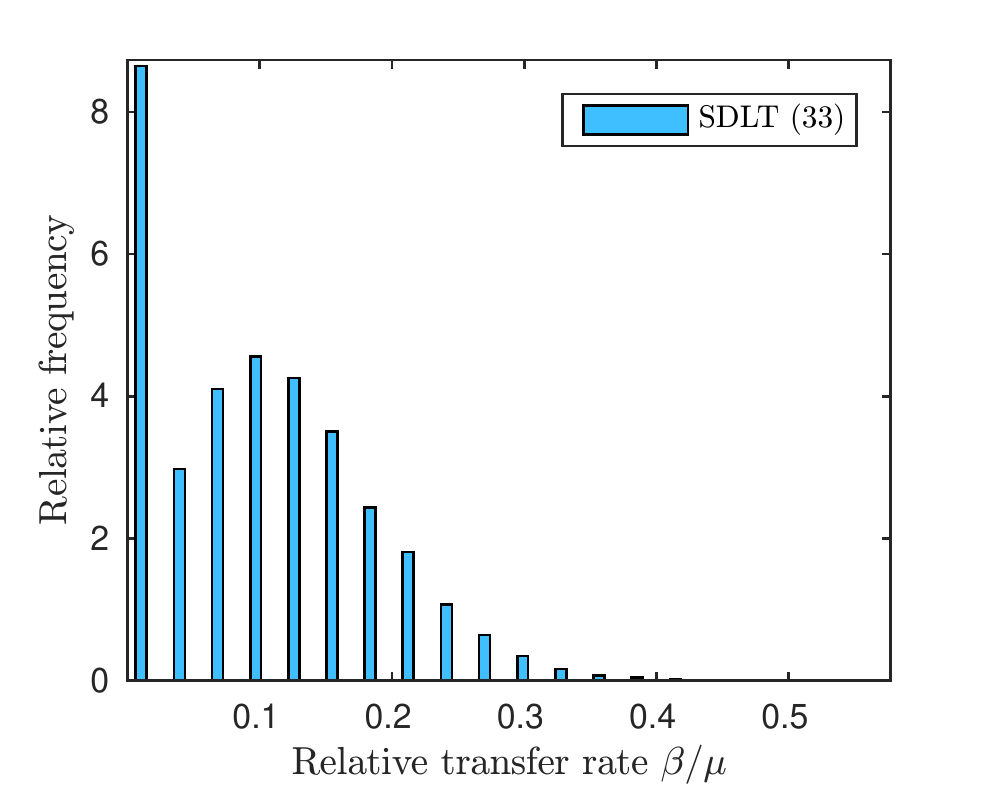} &
	\includegraphics[width=0.5\textwidth, trim = 0.05cm 0cm 0.6cm 0cm, clip]{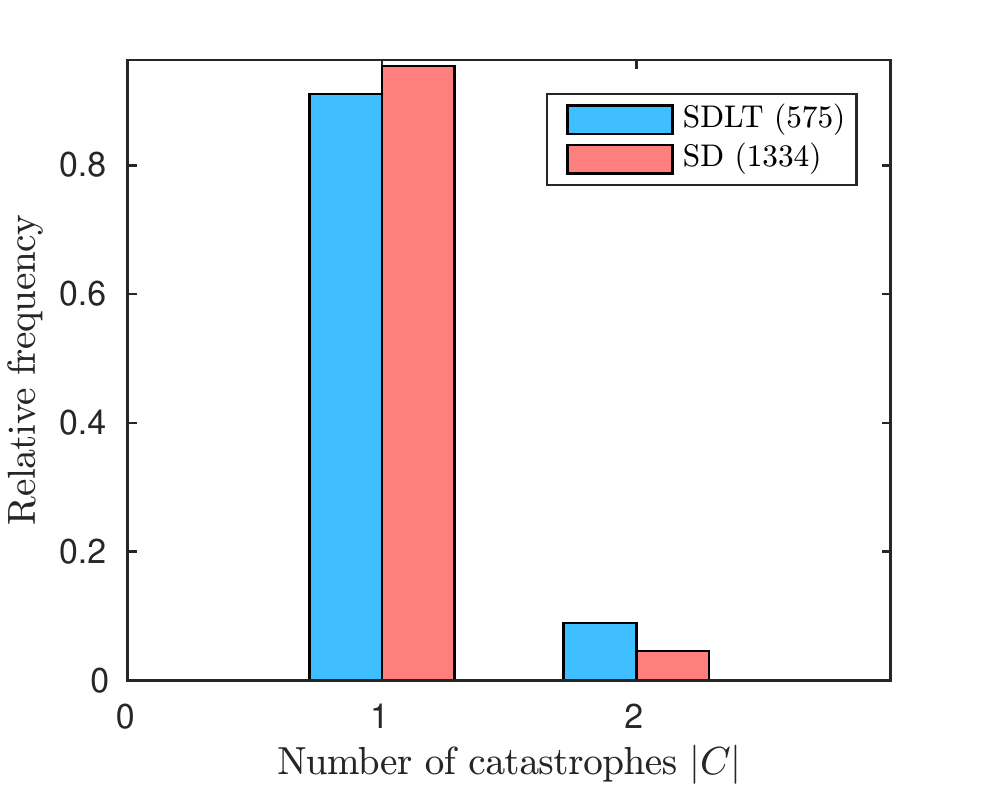} \\
	\includegraphics[width=0.5\textwidth, trim = 0.05cm 0cm 0.6cm 0cm, clip]{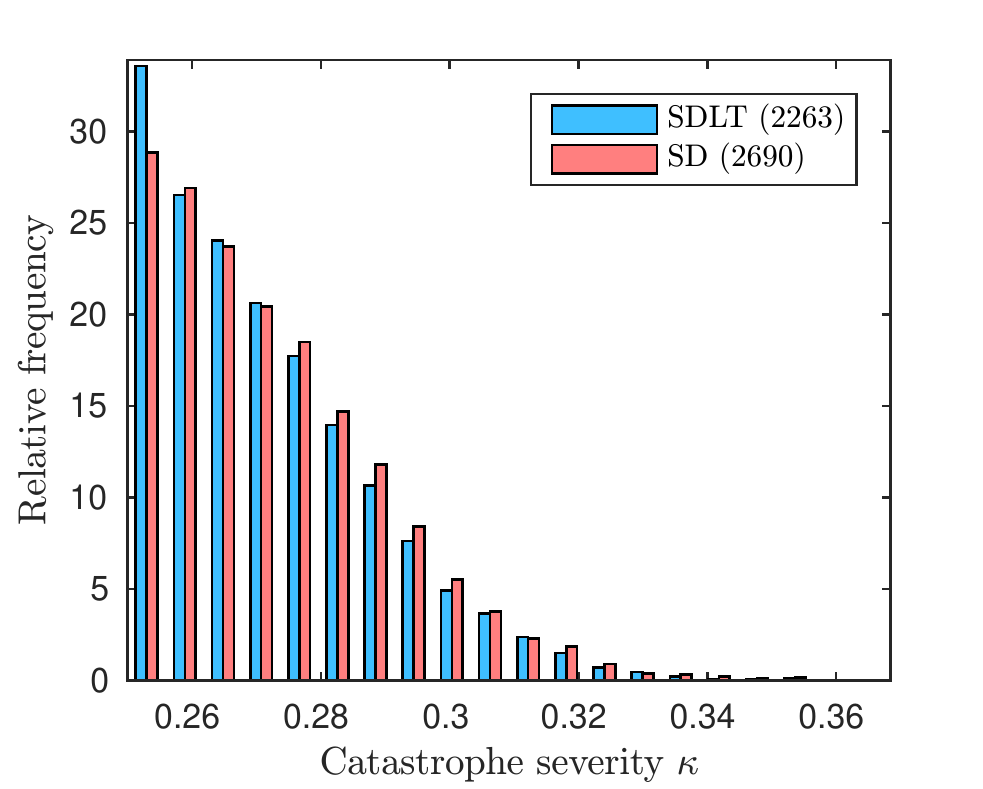} &
	\includegraphics[width=0.5\textwidth, trim = 0.05cm 0cm 0.6cm 0cm, clip]{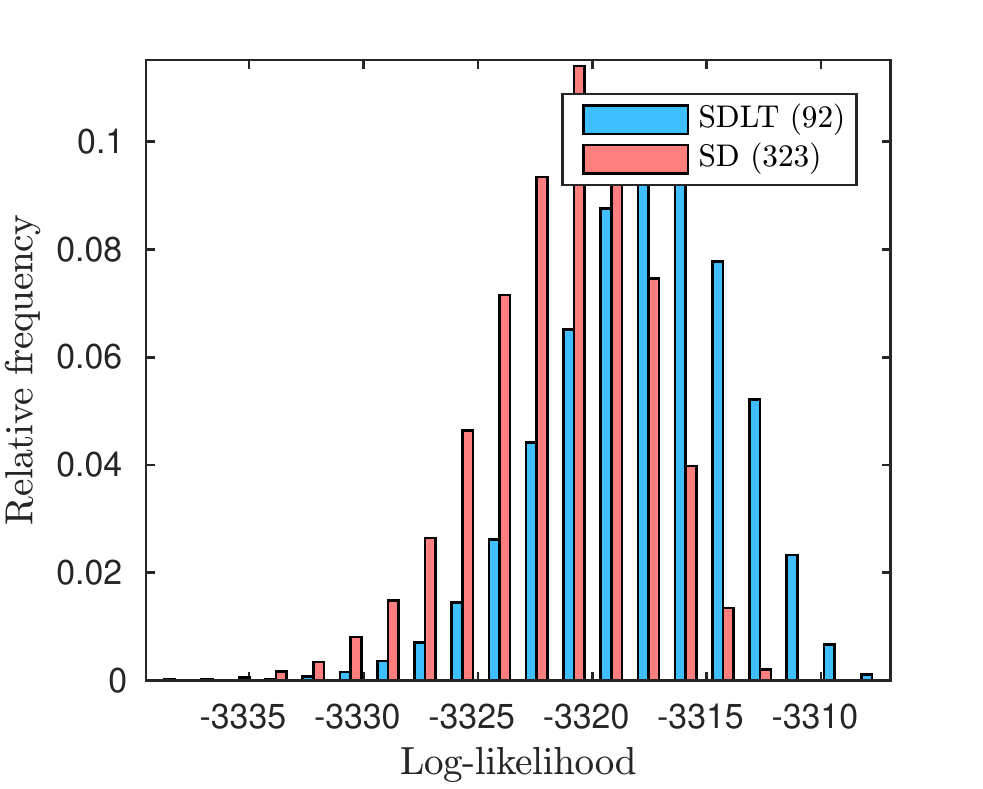}
	\end{tabular}
\caption{Histograms of samples in our analyses of \texttt{SIM-T}.}
\label{hist:simT}
\end{figure}

\begin{figure}[p]
\centering
\begin{tabular}{@{}c@{}@{}c@{}}
	\includegraphics[width=0.5\textwidth, trim = 0.05cm 0cm 0.6cm 0cm, clip]{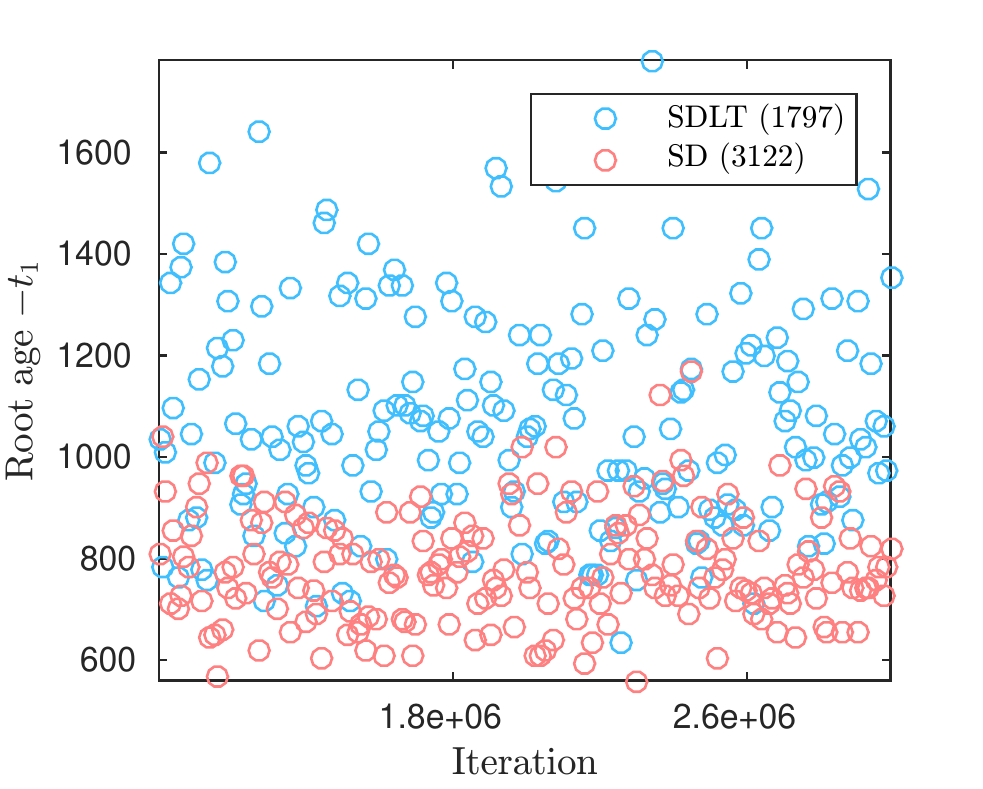} &
	\includegraphics[width=0.5\textwidth, trim = 0.05cm 0cm 0.6cm 0cm, clip]{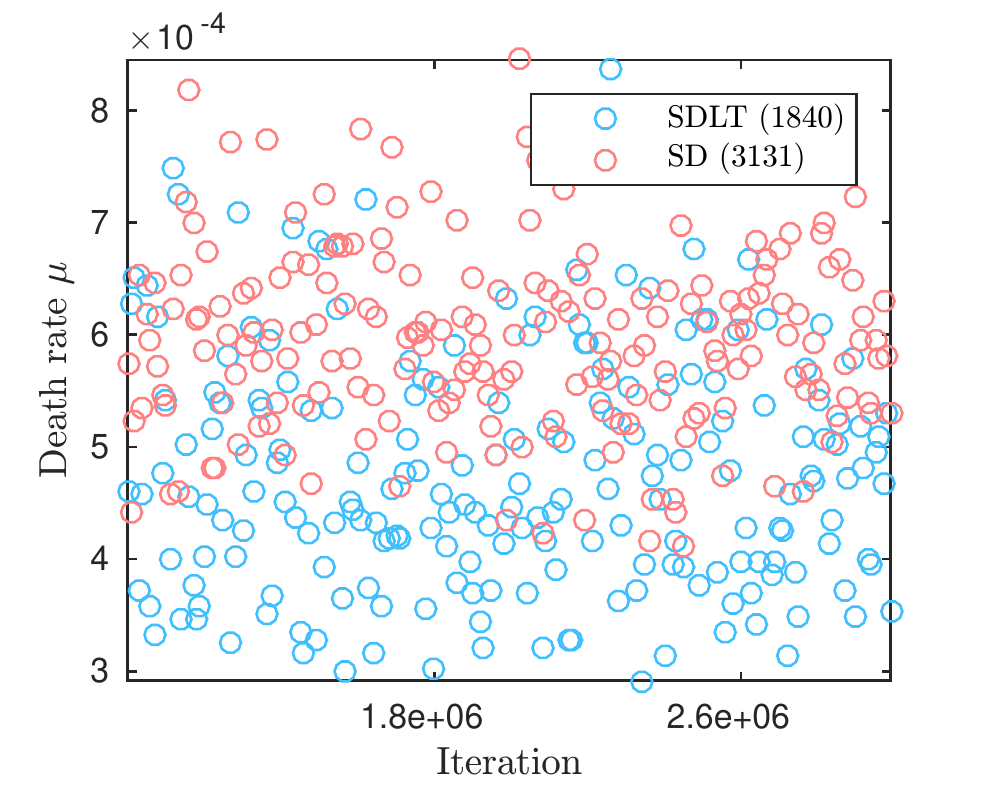} \\
	\includegraphics[width=0.5\textwidth, trim = 0.05cm 0cm 0.6cm 0cm, clip]{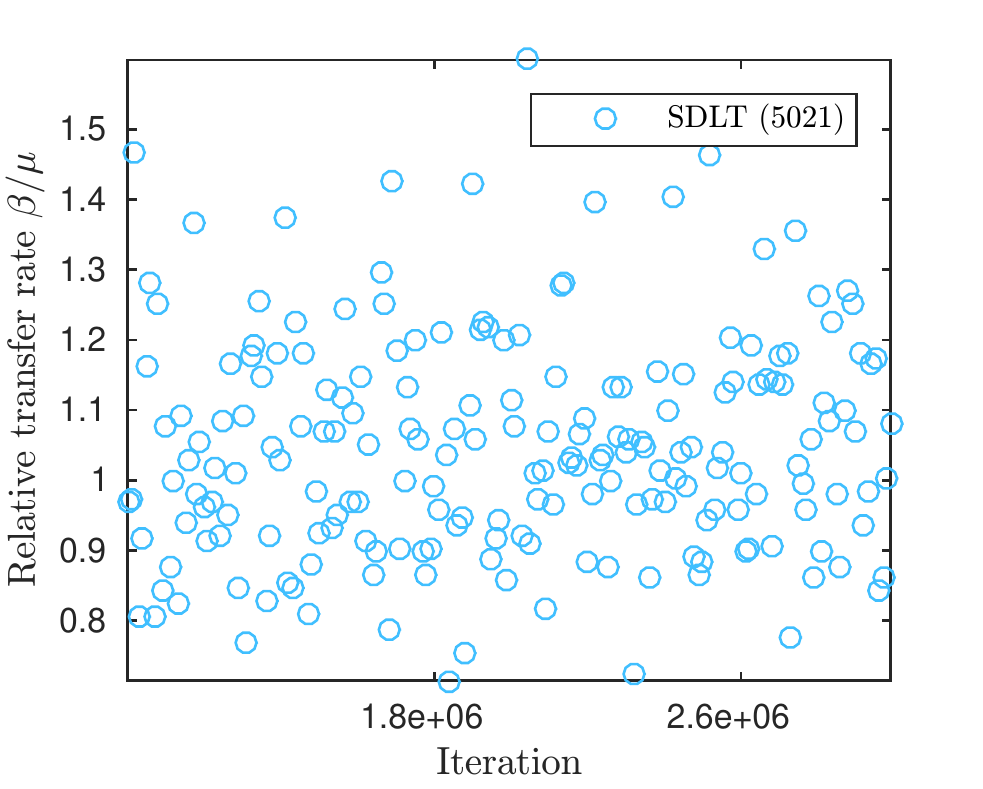} &
	\includegraphics[width=0.5\textwidth, trim = 0.05cm 0cm 0.6cm 0cm, clip]{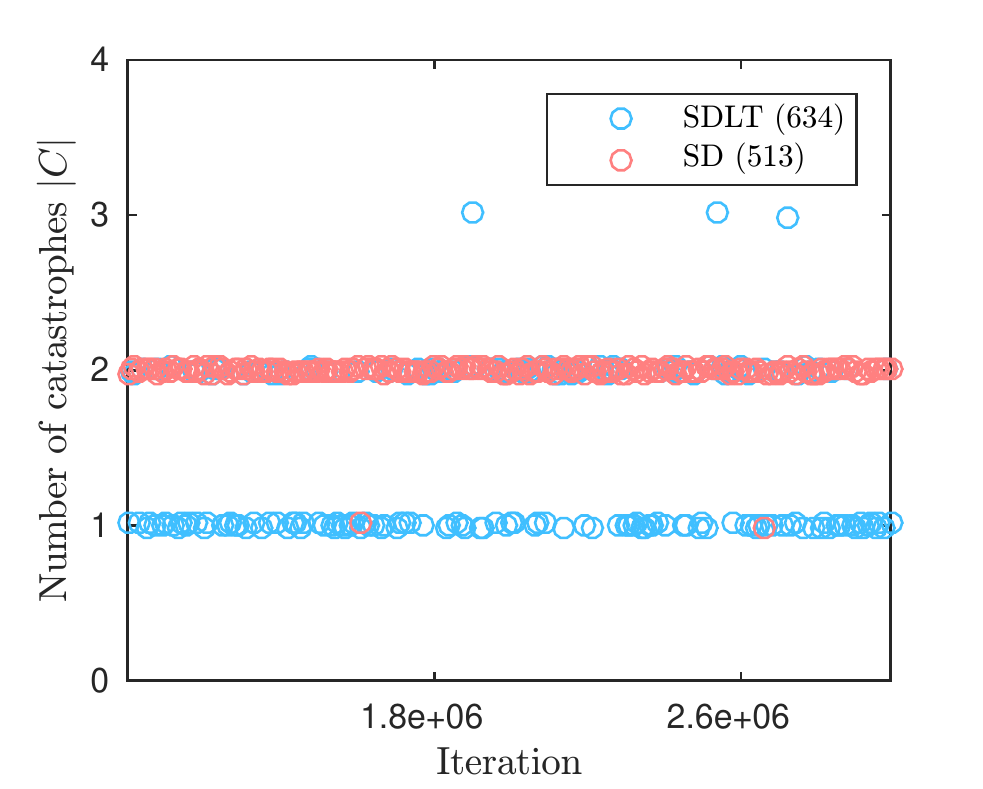} \\
	\includegraphics[width=0.5\textwidth, trim = 0.05cm 0cm 0.6cm 0cm, clip]{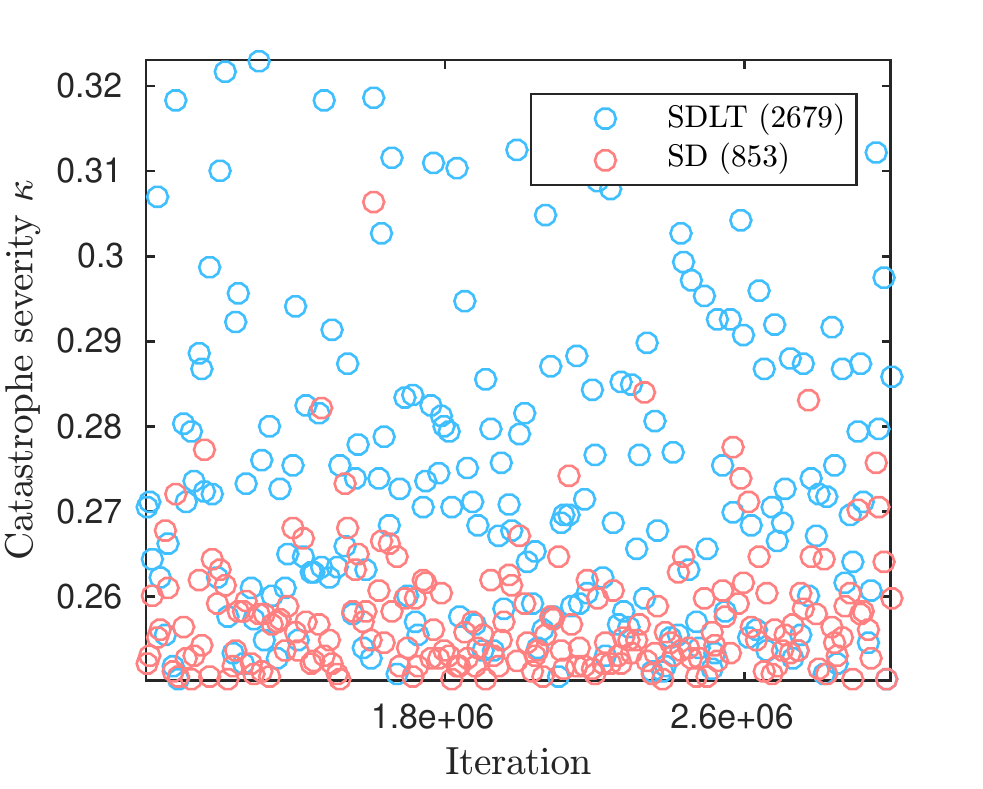} &
	\includegraphics[width=0.5\textwidth, trim = 0.05cm 0cm 0.6cm 0cm, clip]{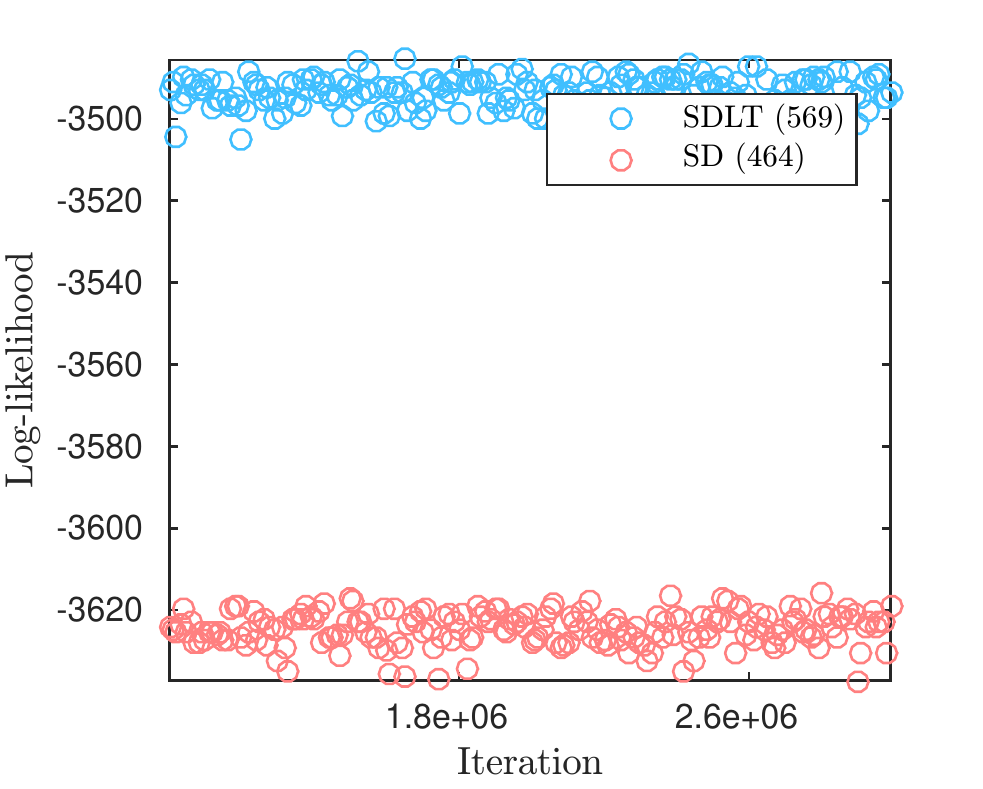}
\end{tabular}
\caption{Trace plots of samples in our analyses of \texttt{SIM-B}.}
\label{trace:simB}
\end{figure}

\begin{figure}[p]
\centering
\begin{tabular}{@{}c@{}@{}c@{}}
	\includegraphics[width=0.5\textwidth, trim = 0.05cm 0cm 0.6cm 0cm, clip]{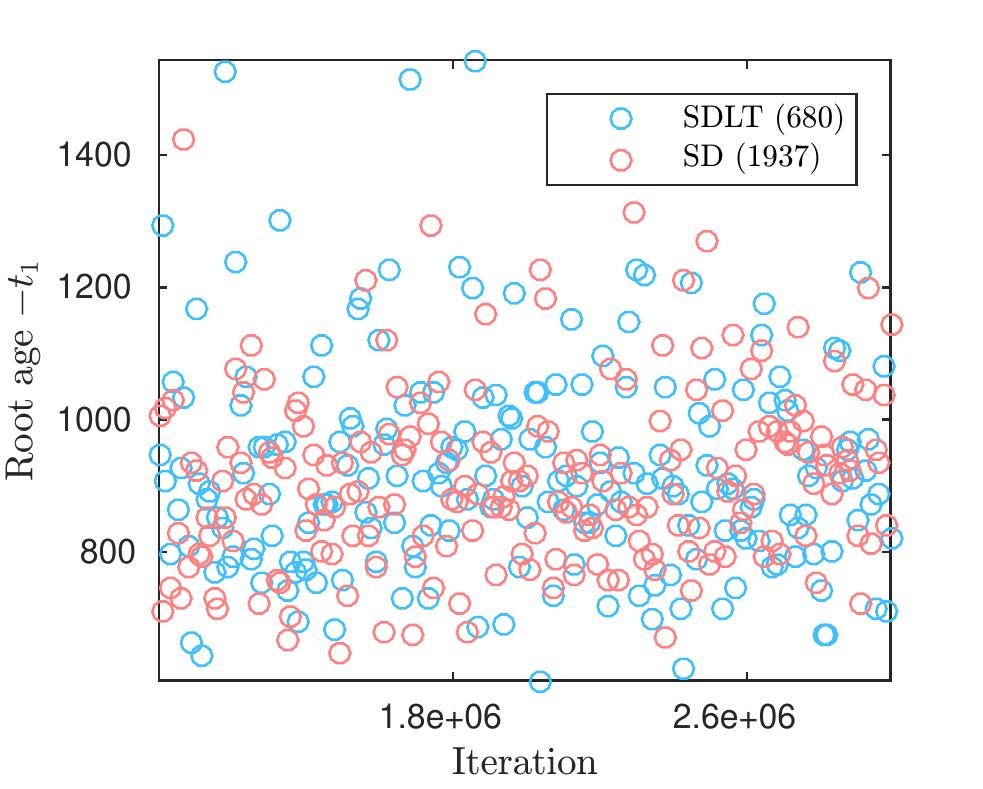} &
	\includegraphics[width=0.5\textwidth, trim = 0.05cm 0cm 0.6cm 0cm, clip]{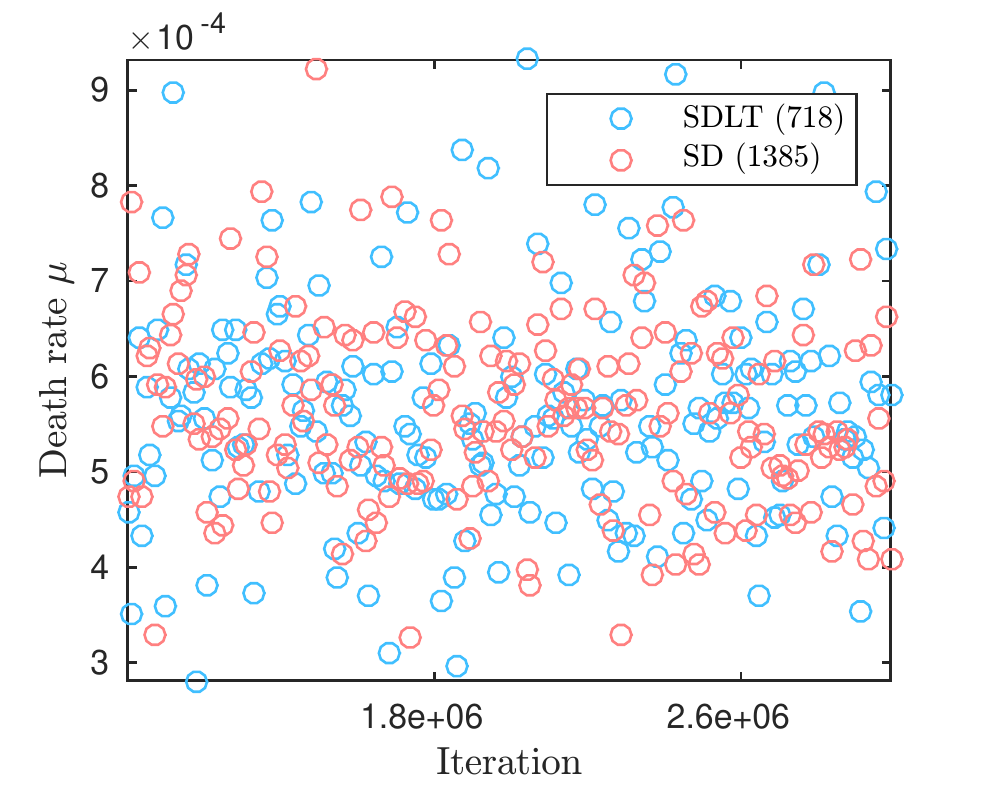} \\
	\includegraphics[width=0.5\textwidth, trim = 0.05cm 0cm 0.6cm 0cm, clip]{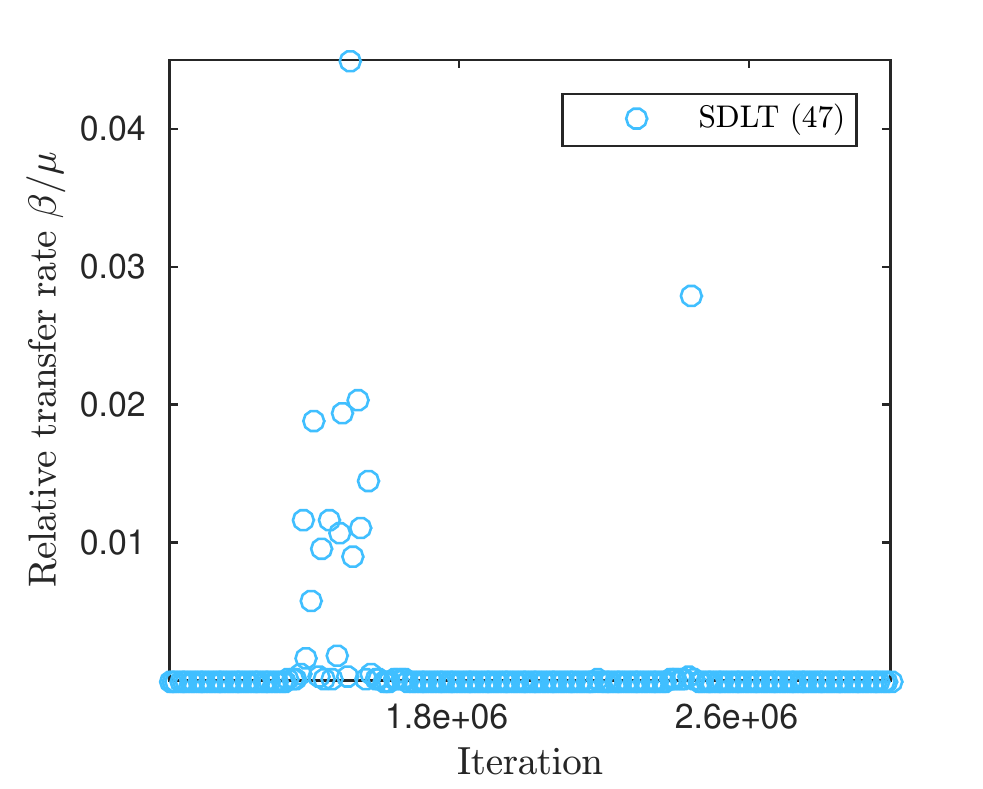} &
	\includegraphics[width=0.5\textwidth, trim = 0.05cm 0cm 0.6cm 0cm, clip]{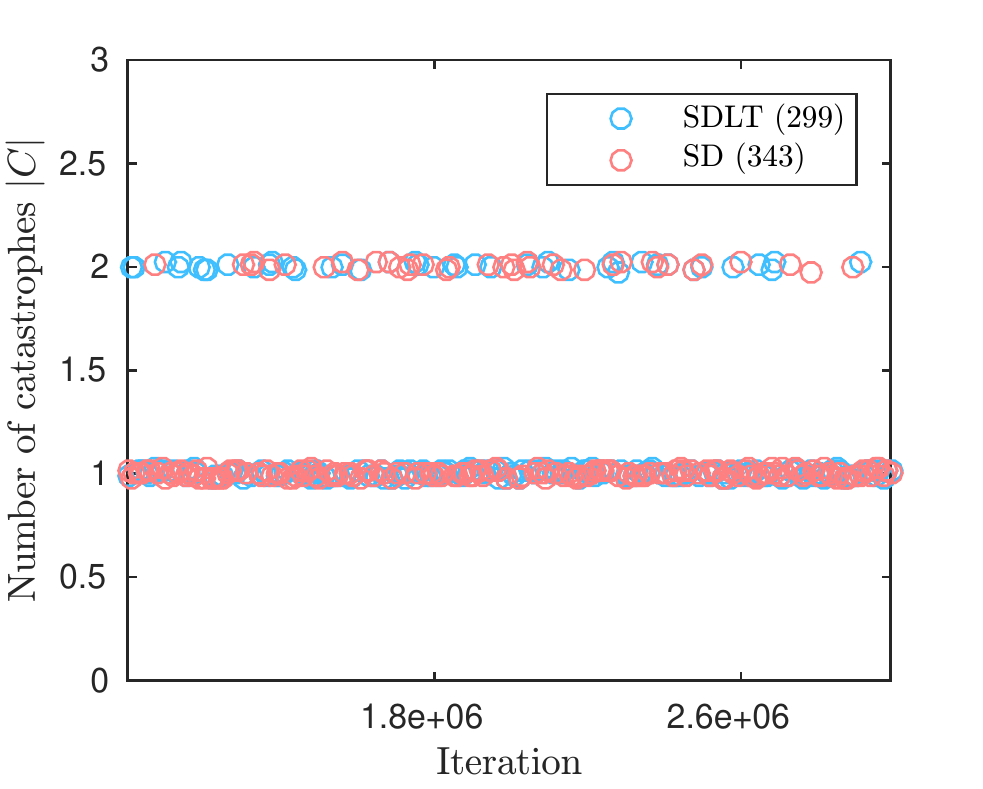} \\
	\includegraphics[width=0.5\textwidth, trim = 0.05cm 0cm 0.6cm 0cm, clip]{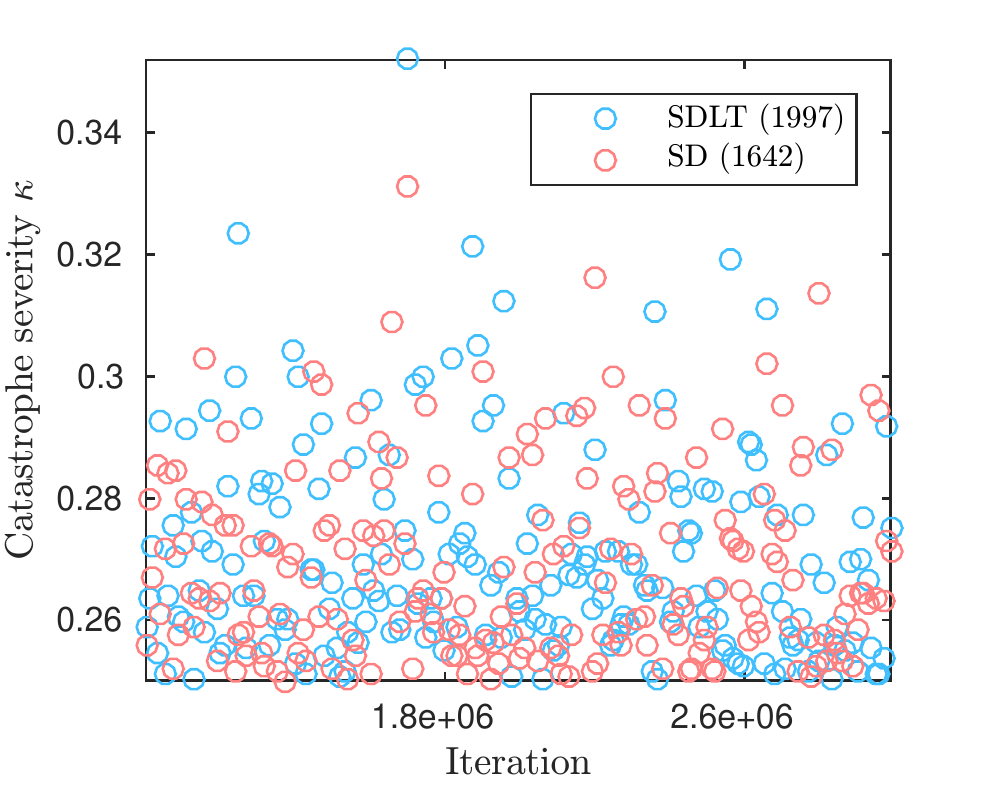} &
	\includegraphics[width=0.5\textwidth, trim = 0.05cm 0cm 0.6cm 0cm, clip]{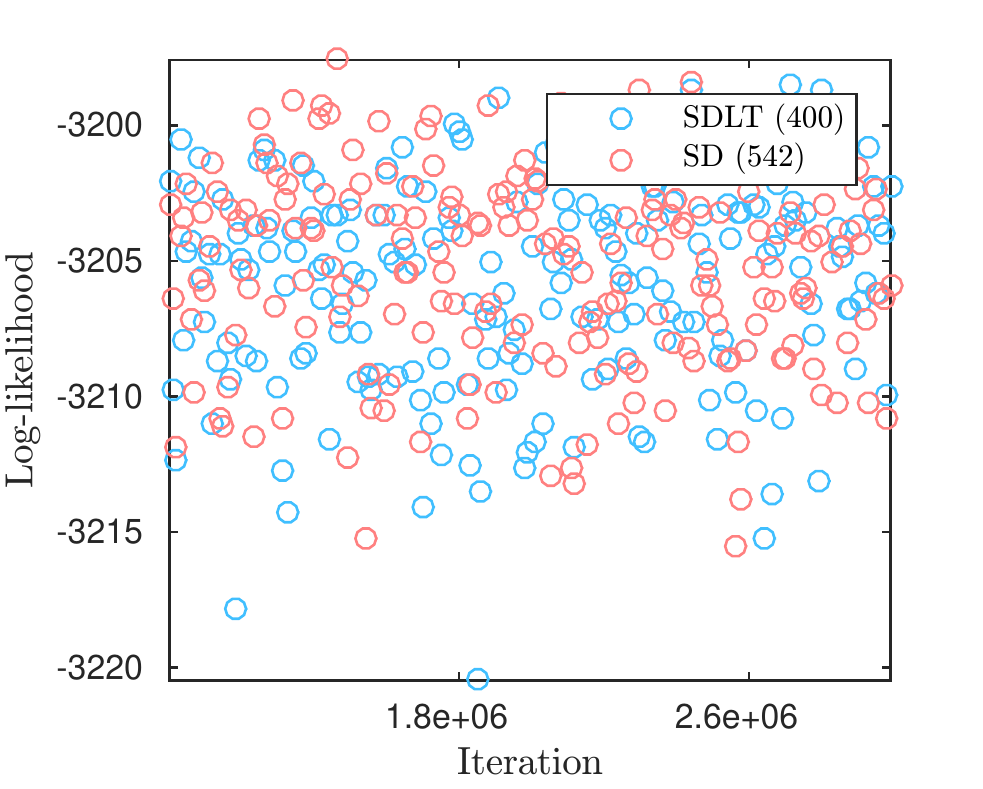}
\end{tabular}
\caption{Trace plots of samples in our analyses of \texttt{SIM-N}.}
\label{trace:simN}
\end{figure}

\begin{figure}[p]
\centering
\begin{tabular}{@{}c@{}@{}c@{}}
	\includegraphics[width=0.5\textwidth, trim = 0.05cm 0cm 0.6cm 0cm, clip]{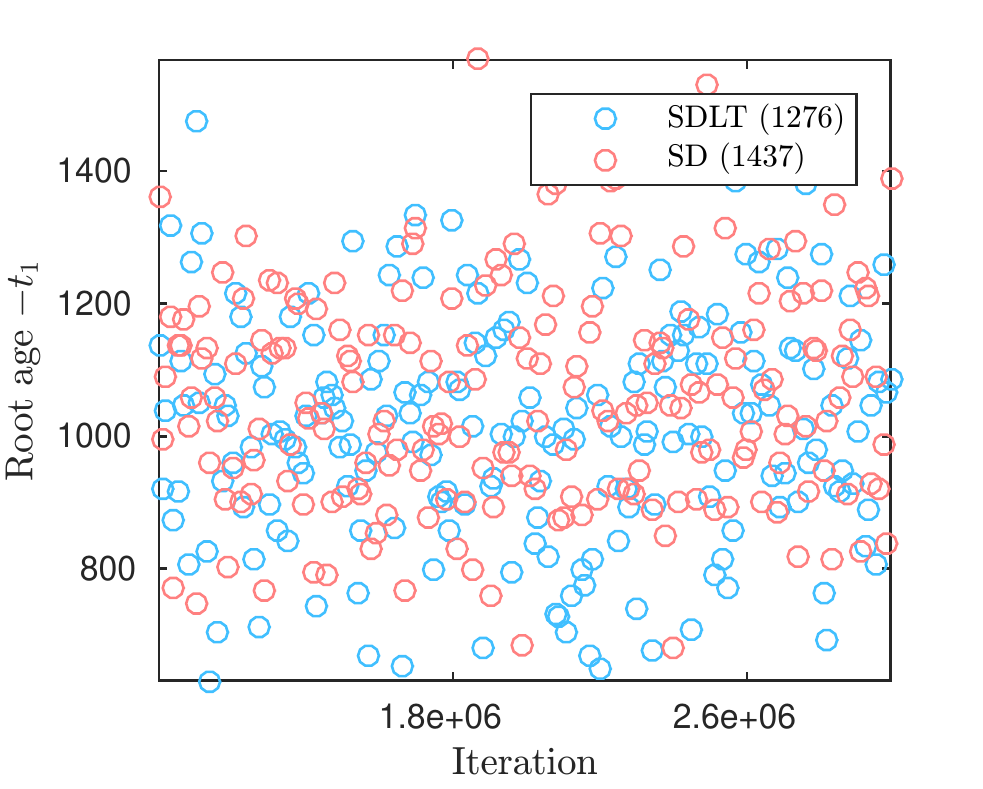} &
	\includegraphics[width=0.5\textwidth, trim = 0.05cm 0cm 0.6cm 0cm, clip]{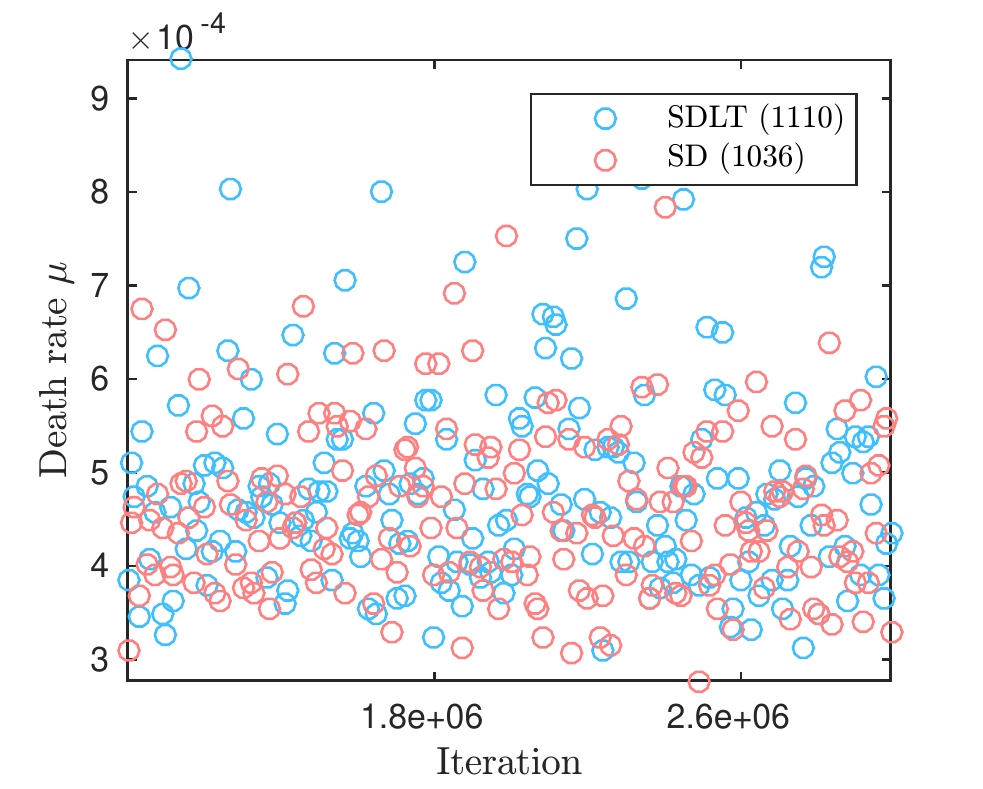} \\
	\includegraphics[width=0.5\textwidth, trim = 0.05cm 0cm 0.6cm 0cm, clip]{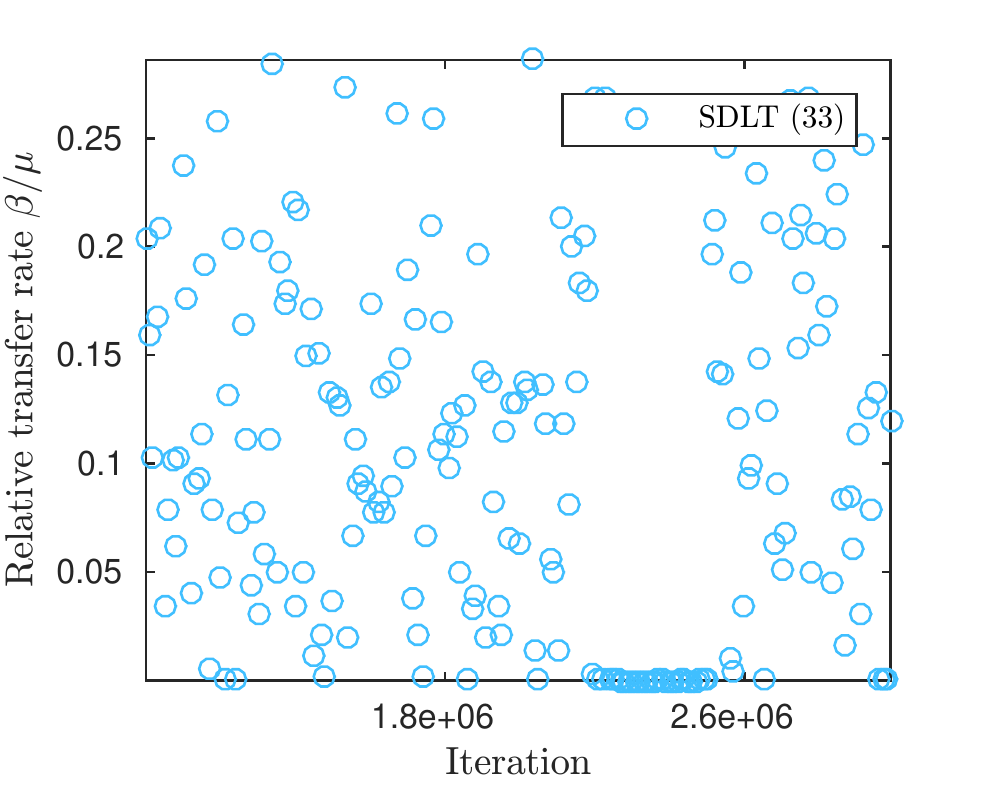} &
	\includegraphics[width=0.5\textwidth, trim = 0.05cm 0cm 0.6cm 0cm, clip]{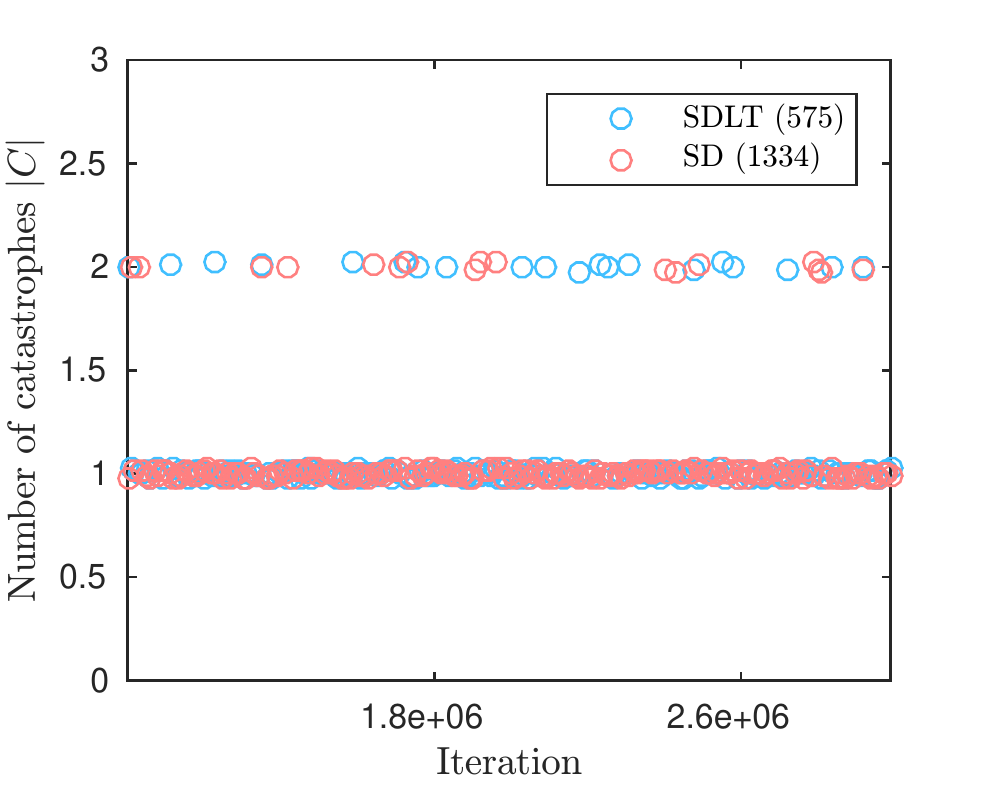} \\
	\includegraphics[width=0.5\textwidth, trim = 0.05cm 0cm 0.6cm 0cm, clip]{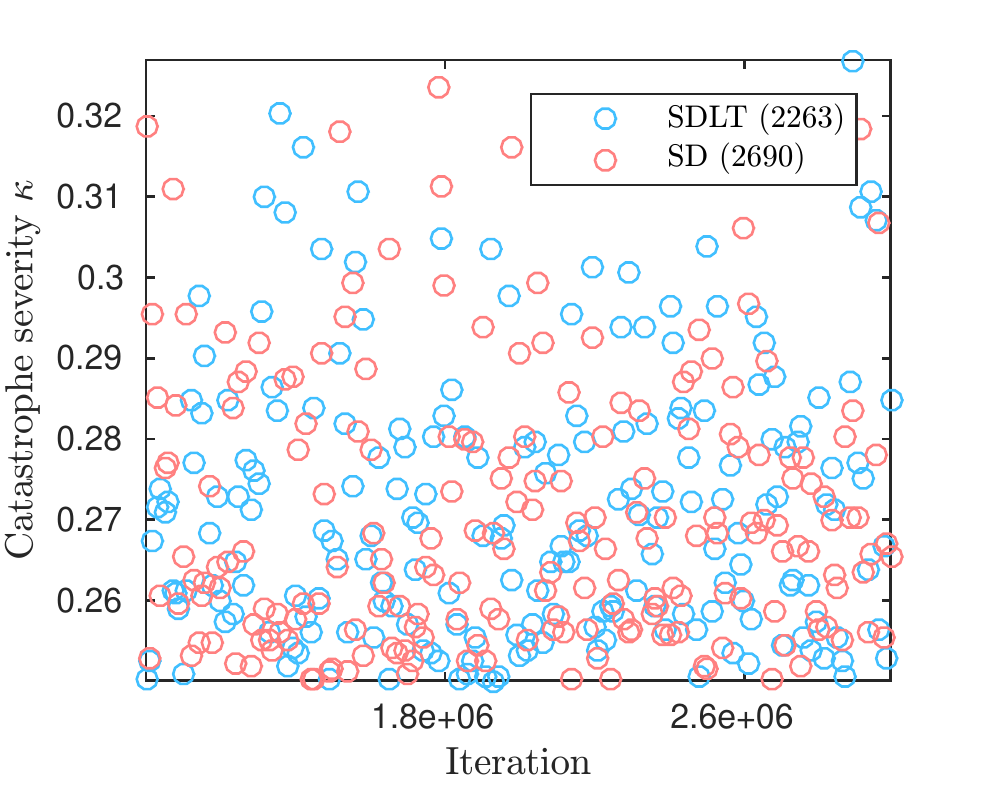} &
	\includegraphics[width=0.5\textwidth, trim = 0.05cm 0cm 0.6cm 0cm, clip]{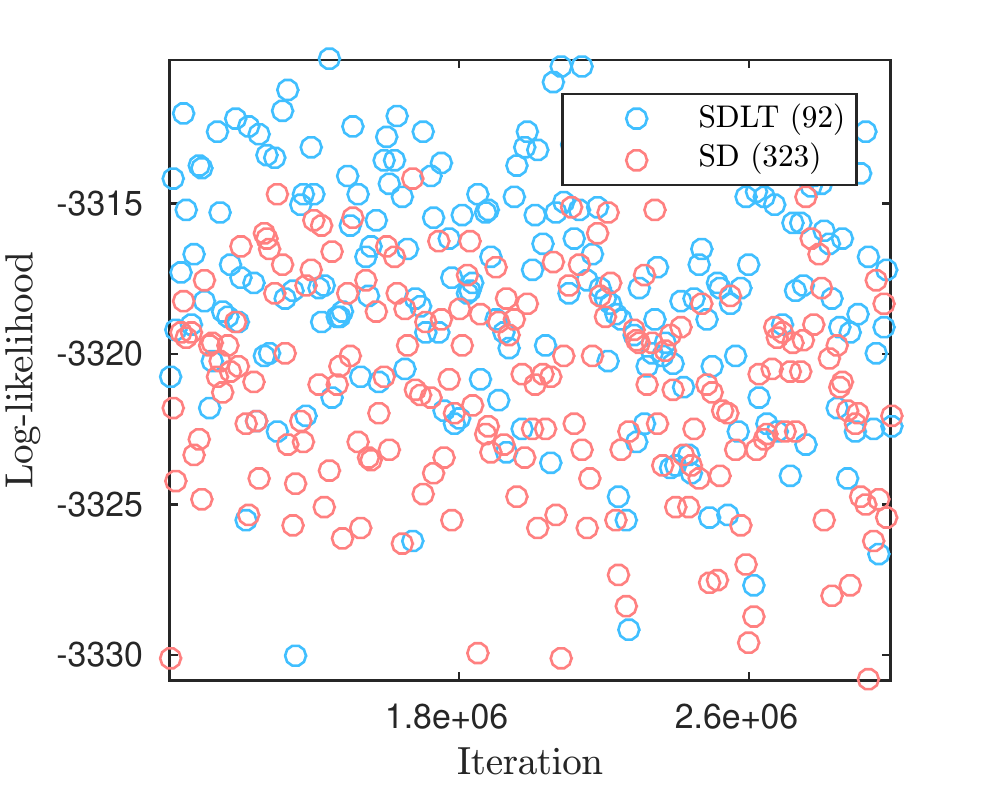}
\end{tabular}
\caption{Trace plots of samples in our analyses of \texttt{SIM-T}.}
\label{trace:simT}
\end{figure}

\begin{figure}[p]
\centering
\begin{tabular}{@{}c@{}@{}c@{}}
	\includegraphics[width=0.5\textwidth, trim = 0.05cm 0cm 0.6cm 0cm, clip]{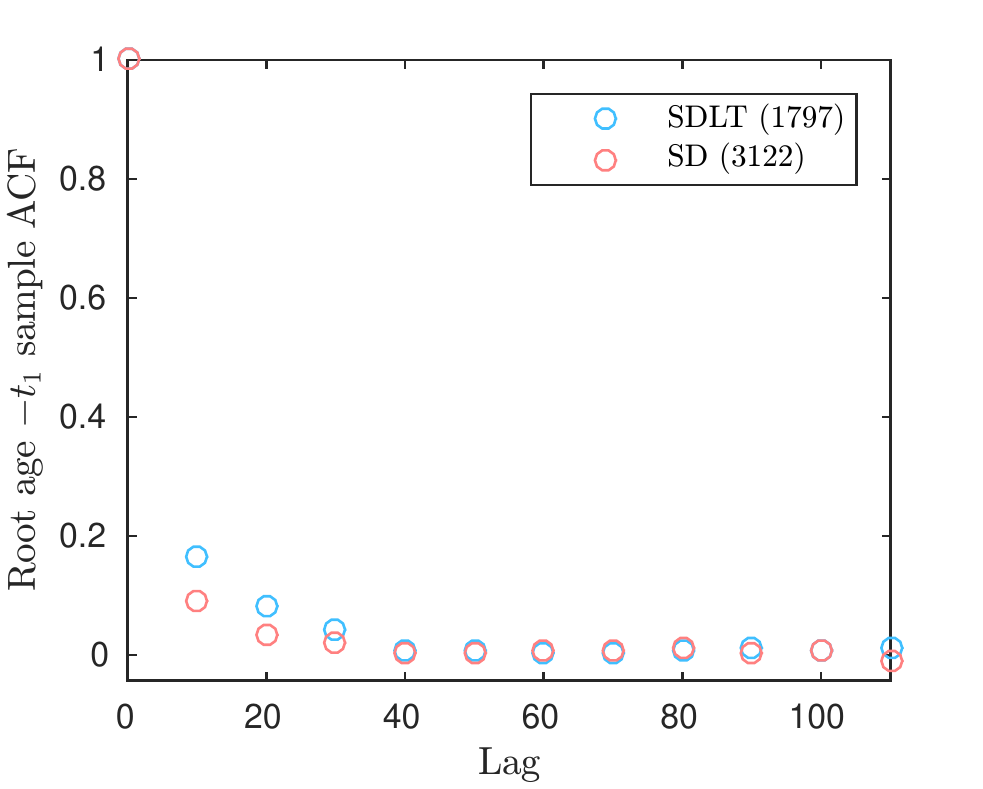} &
	\includegraphics[width=0.5\textwidth, trim = 0.05cm 0cm 0.6cm 0cm, clip]{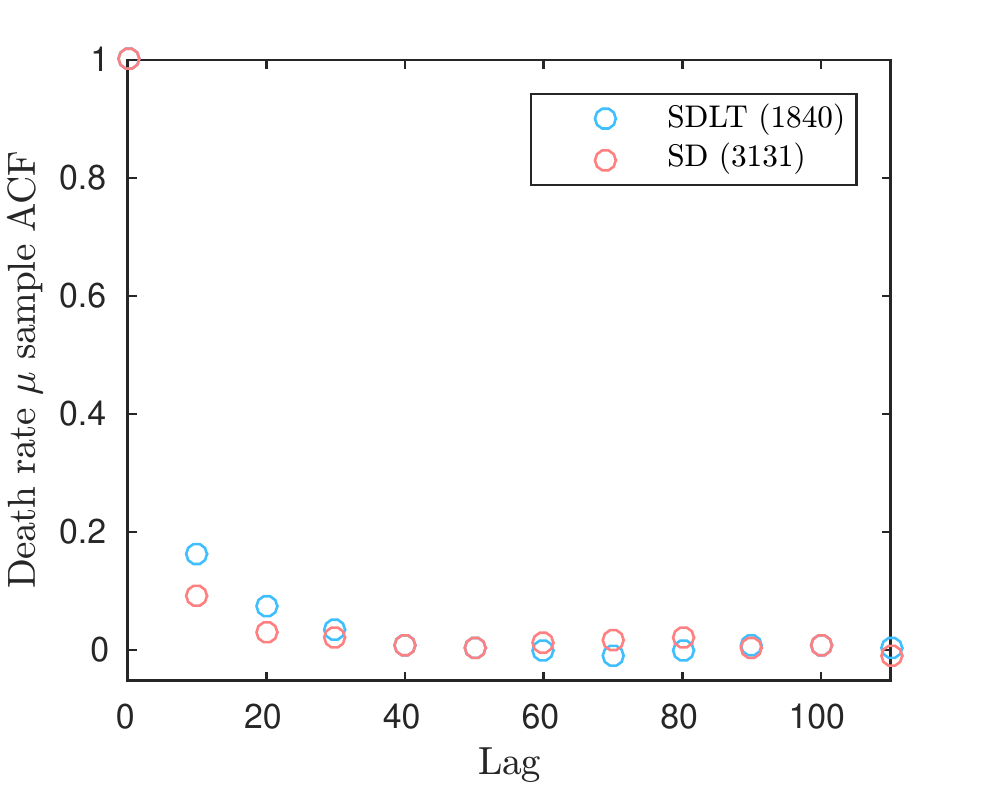} \\
	\includegraphics[width=0.5\textwidth, trim = 0.05cm 0cm 0.6cm 0cm, clip]{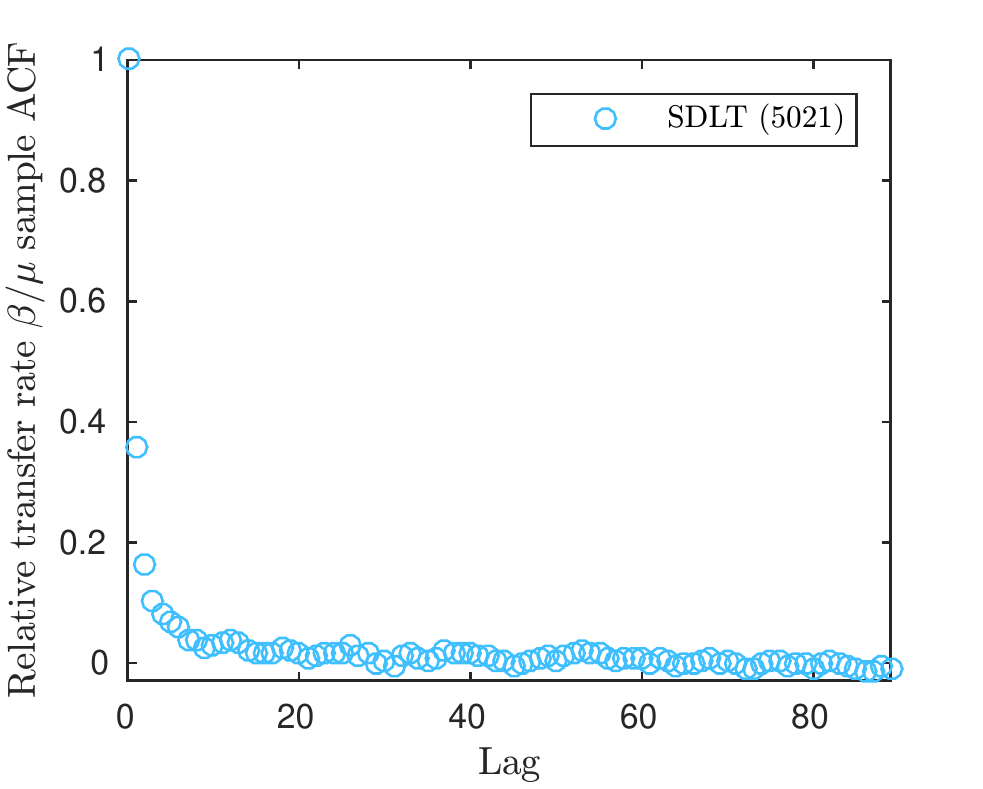} &
	\includegraphics[width=0.5\textwidth, trim = 0.05cm 0cm 0.6cm 0cm, clip]{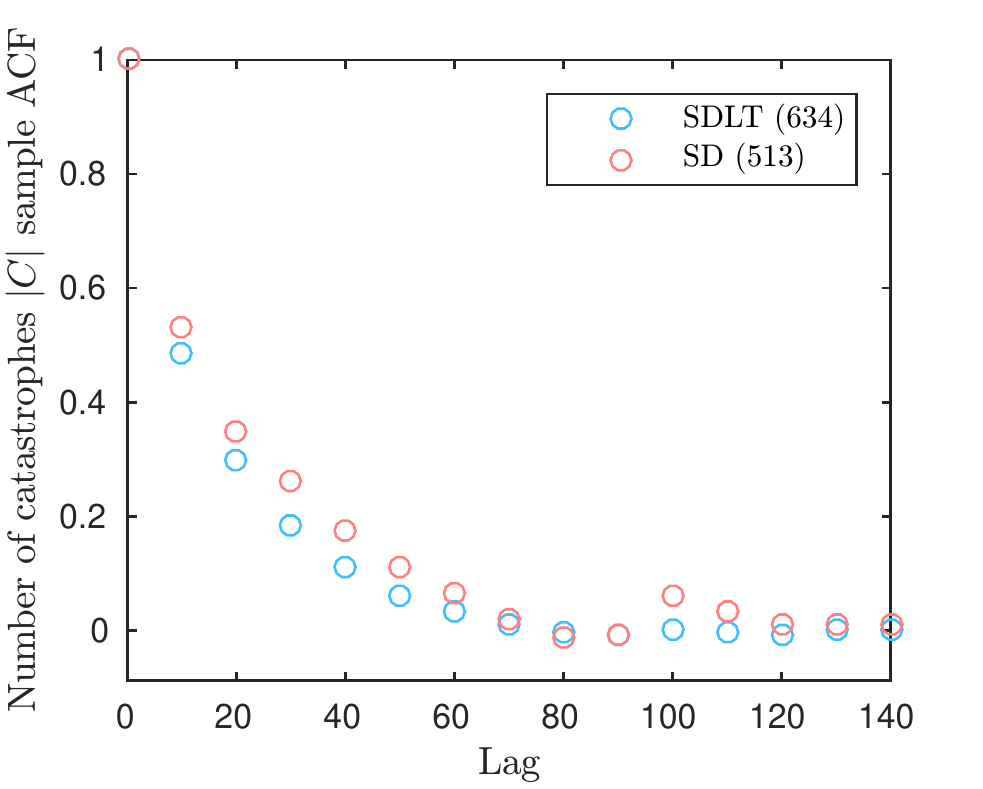} \\
	\includegraphics[width=0.5\textwidth, trim = 0.05cm 0cm 0.6cm 0cm, clip]{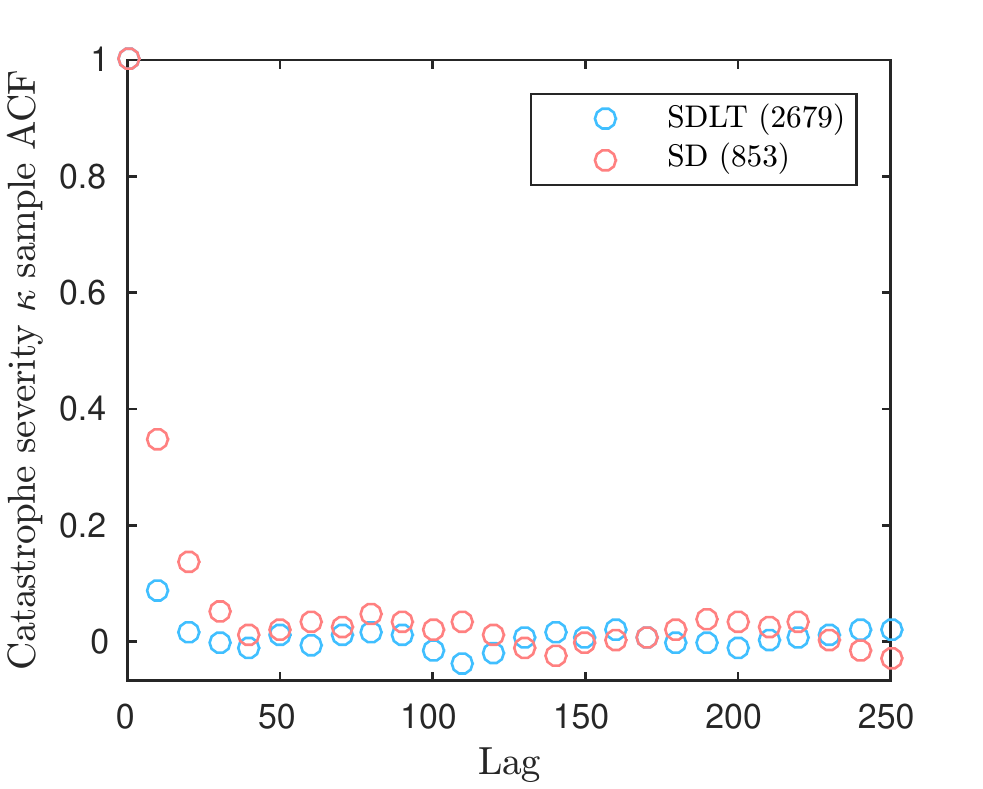} &
	\includegraphics[width=0.5\textwidth, trim = 0.05cm 0cm 0.6cm 0cm, clip]{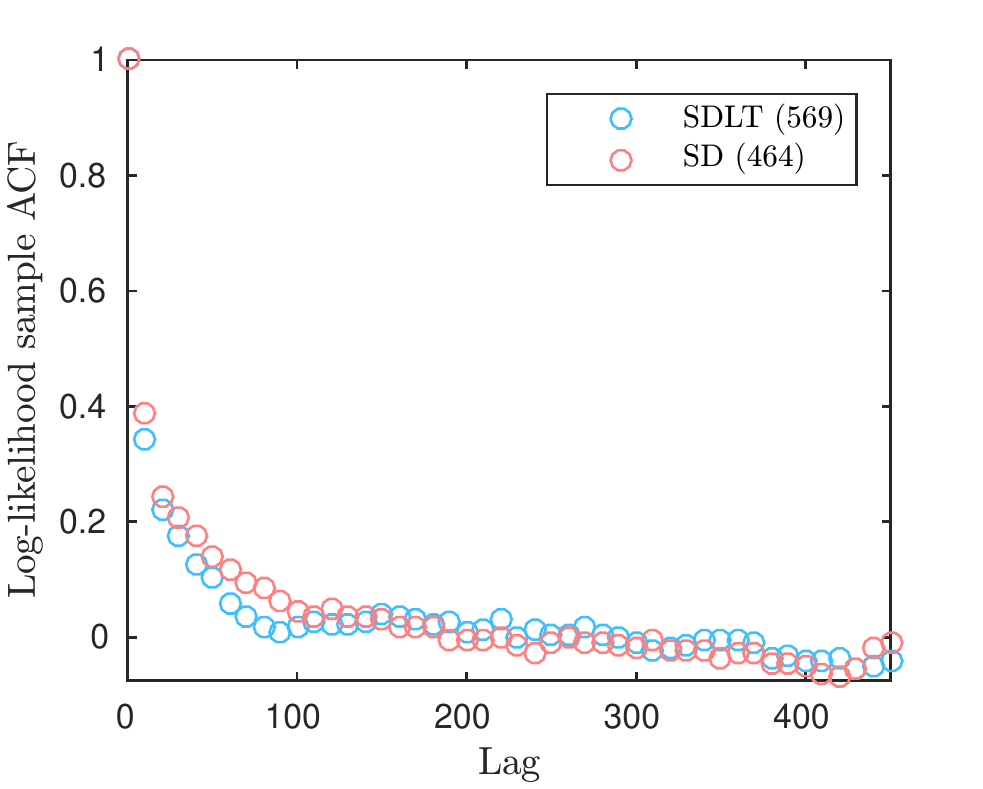}
	\end{tabular}
\caption{Autocorrelation plots of samples in our analyses of \texttt{SIM-B}.}
\label{autocorr:simB}
\end{figure}

\begin{figure}[p]
\centering
\begin{tabular}{@{}c@{}@{}c@{}}
	\includegraphics[width=0.5\textwidth, trim = 0.05cm 0cm 0.6cm 0cm, clip]{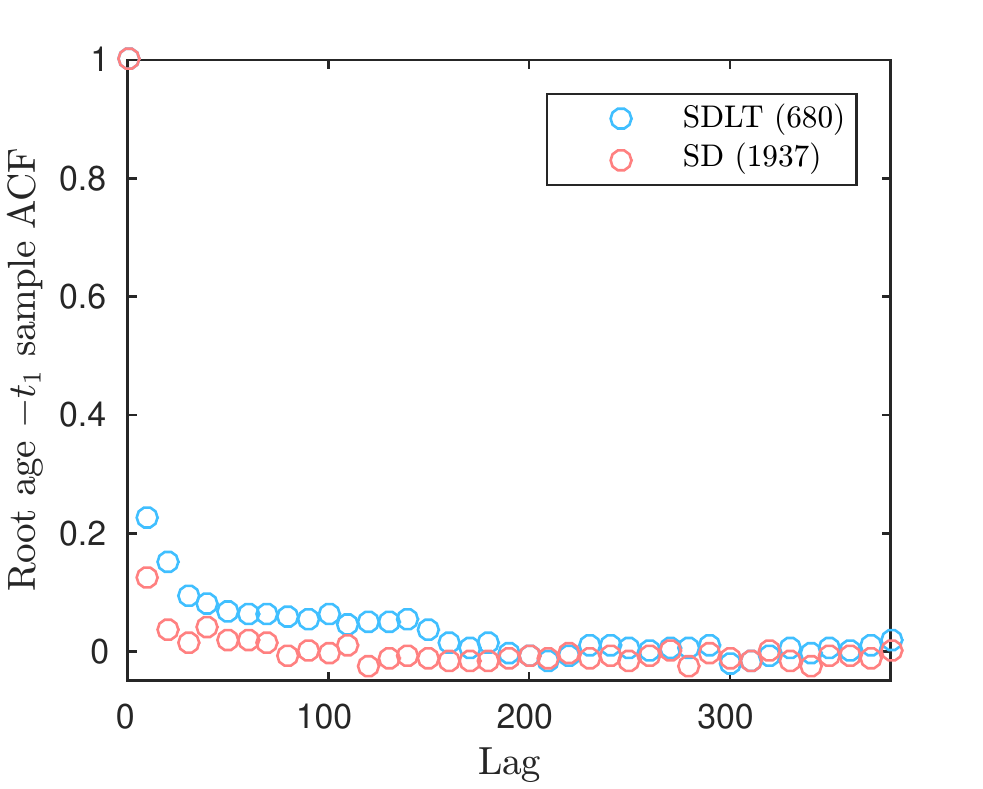} &
	\includegraphics[width=0.5\textwidth, trim = 0.05cm 0cm 0.6cm 0cm, clip]{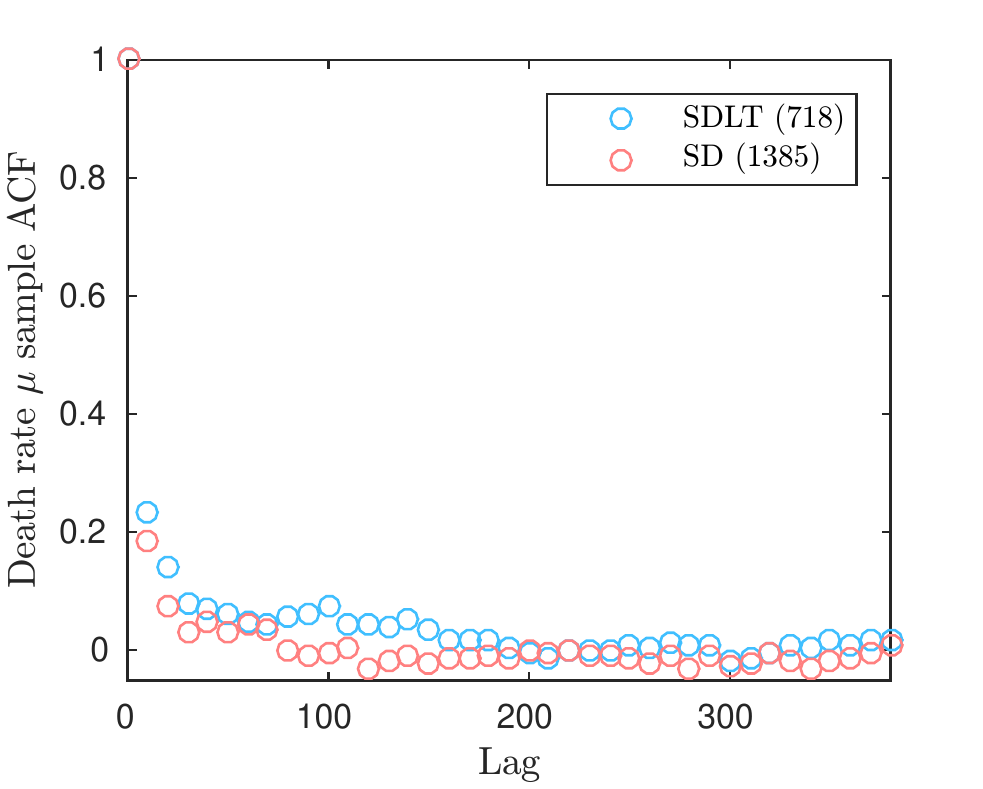} \\
	\includegraphics[width=0.5\textwidth, trim = 0.05cm 0cm 0.6cm 0cm, clip]{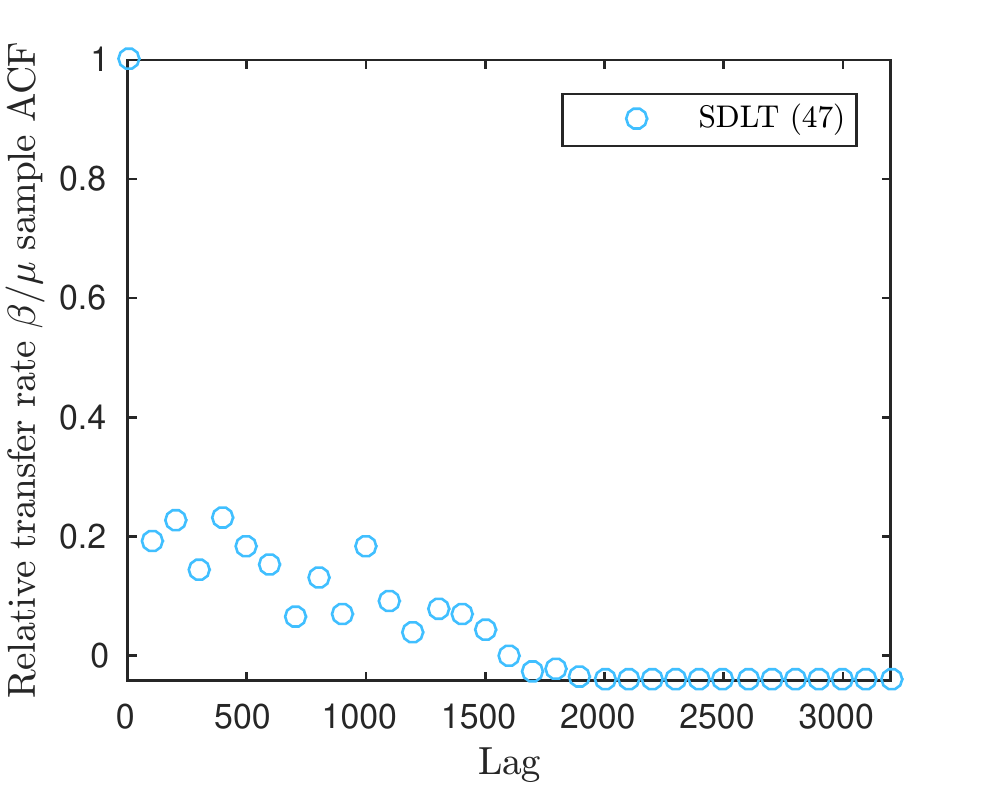} &
	\includegraphics[width=0.5\textwidth, trim = 0.05cm 0cm 0.6cm 0cm, clip]{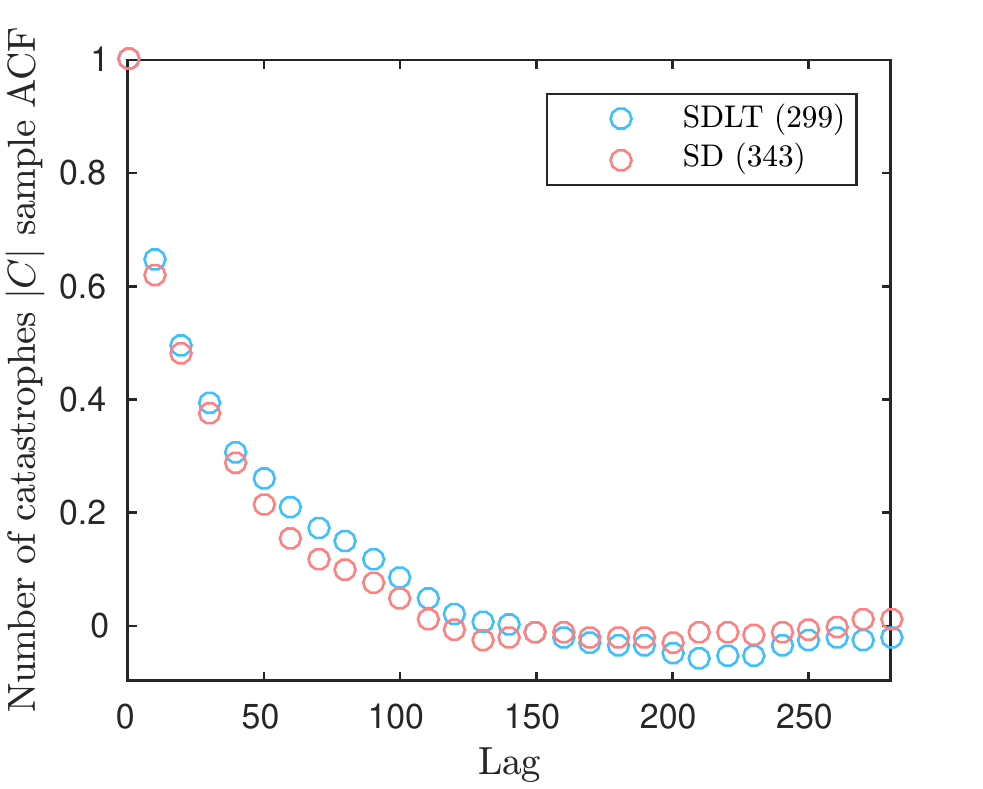} \\
	\includegraphics[width=0.5\textwidth, trim = 0.05cm 0cm 0.6cm 0cm, clip]{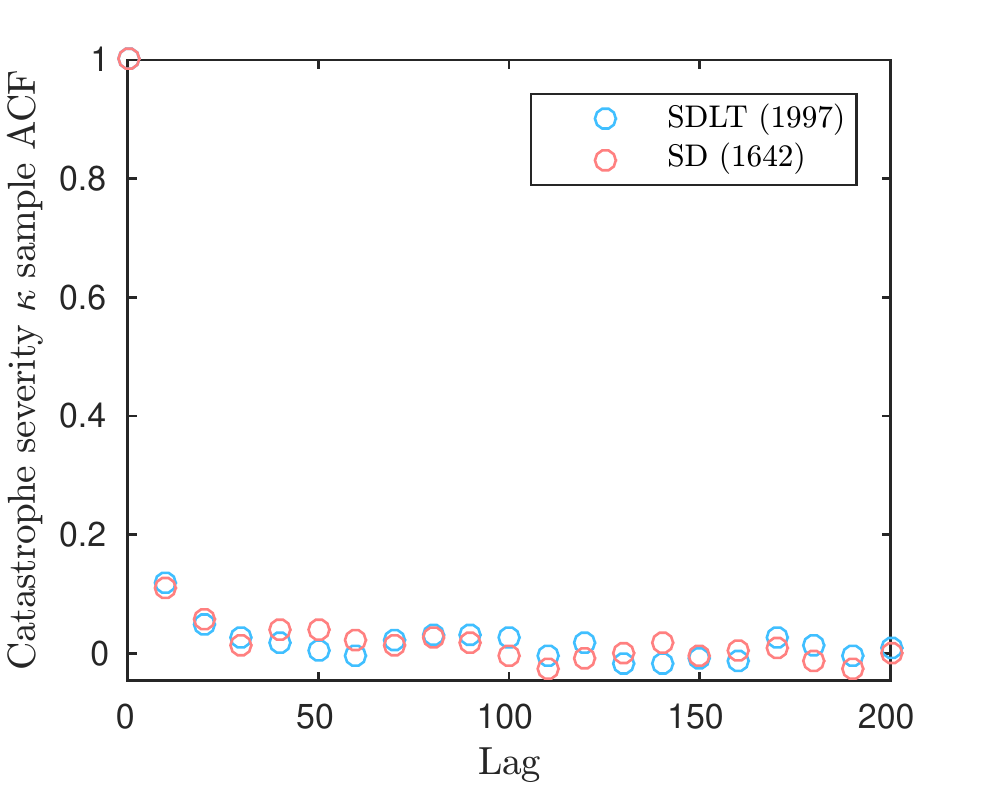} &
	\includegraphics[width=0.5\textwidth, trim = 0.05cm 0cm 0.6cm 0cm, clip]{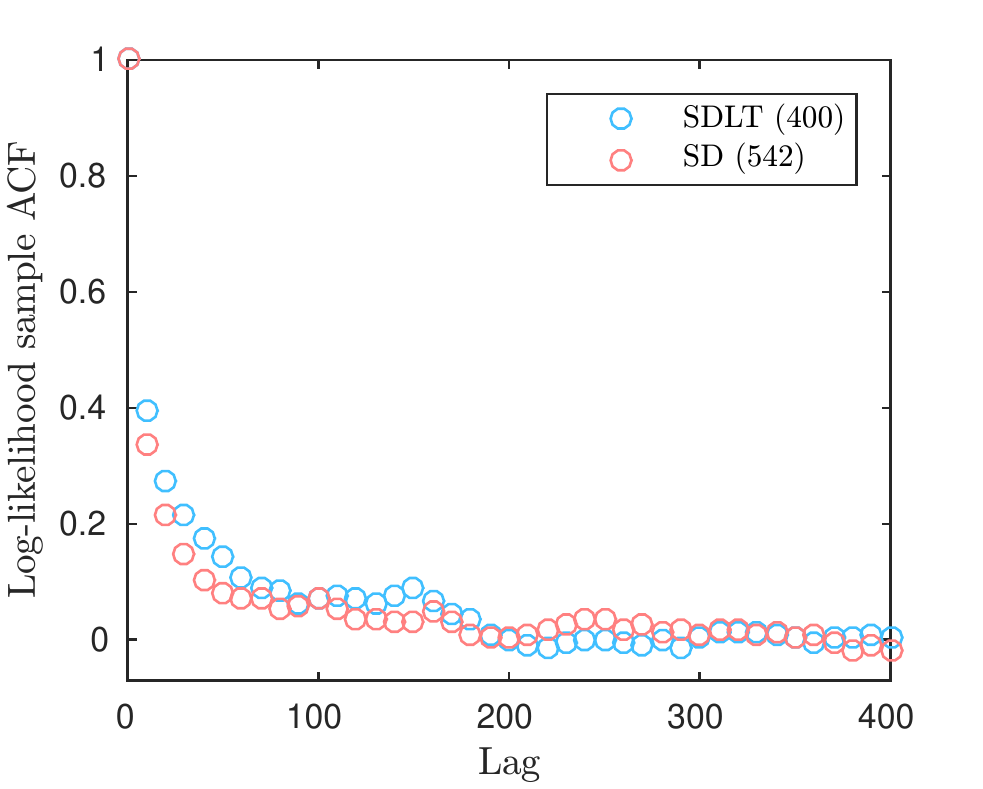}
\end{tabular}
\caption{Autocorrelation plots of samples in our analyses of \texttt{SIM-N}.}
\label{autocorr:simN}
\end{figure}

\begin{figure}[p]
\begin{tabular}{@{}c@{}@{}c@{}}
	\includegraphics[width=0.5\textwidth, trim = 0.05cm 0cm 0.6cm 0cm, clip]{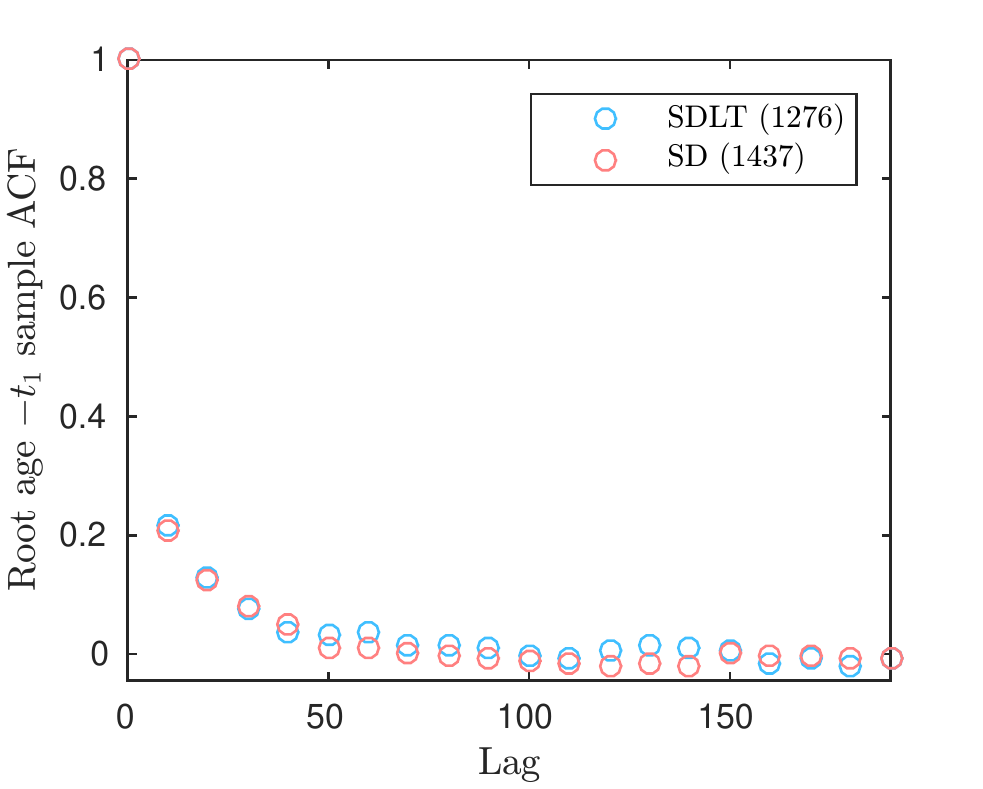} &
	\includegraphics[width=0.5\textwidth, trim = 0.05cm 0cm 0.6cm 0cm, clip]{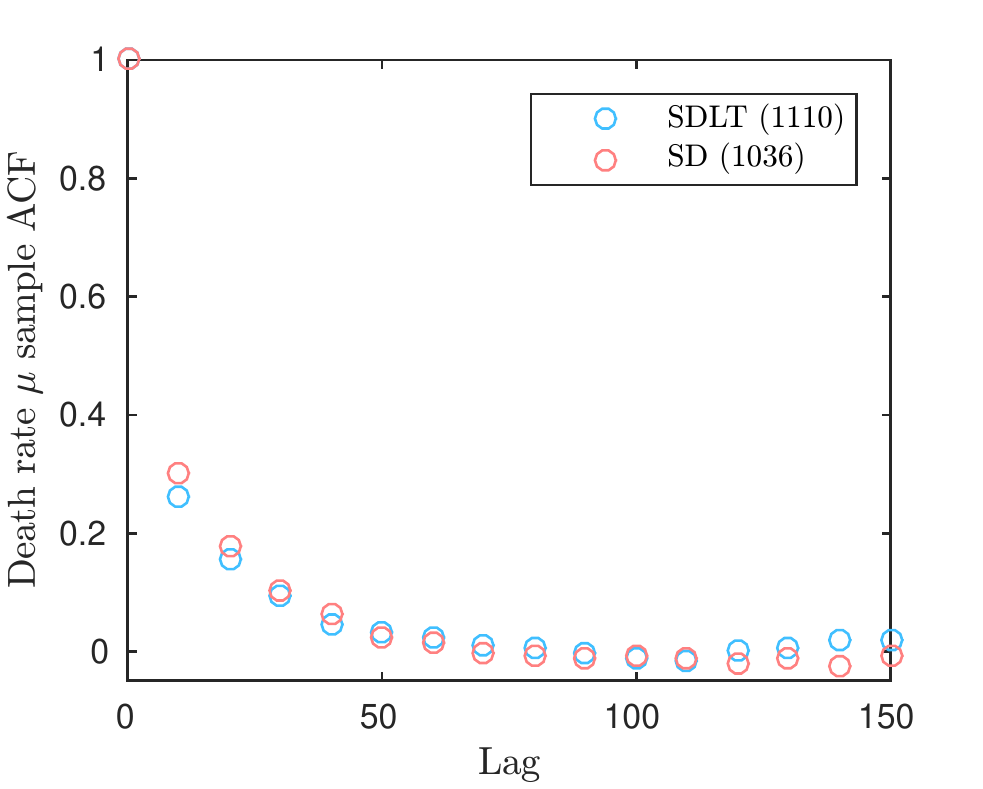} \\
	\includegraphics[width=0.5\textwidth, trim = 0.05cm 0cm 0.6cm 0cm, clip]{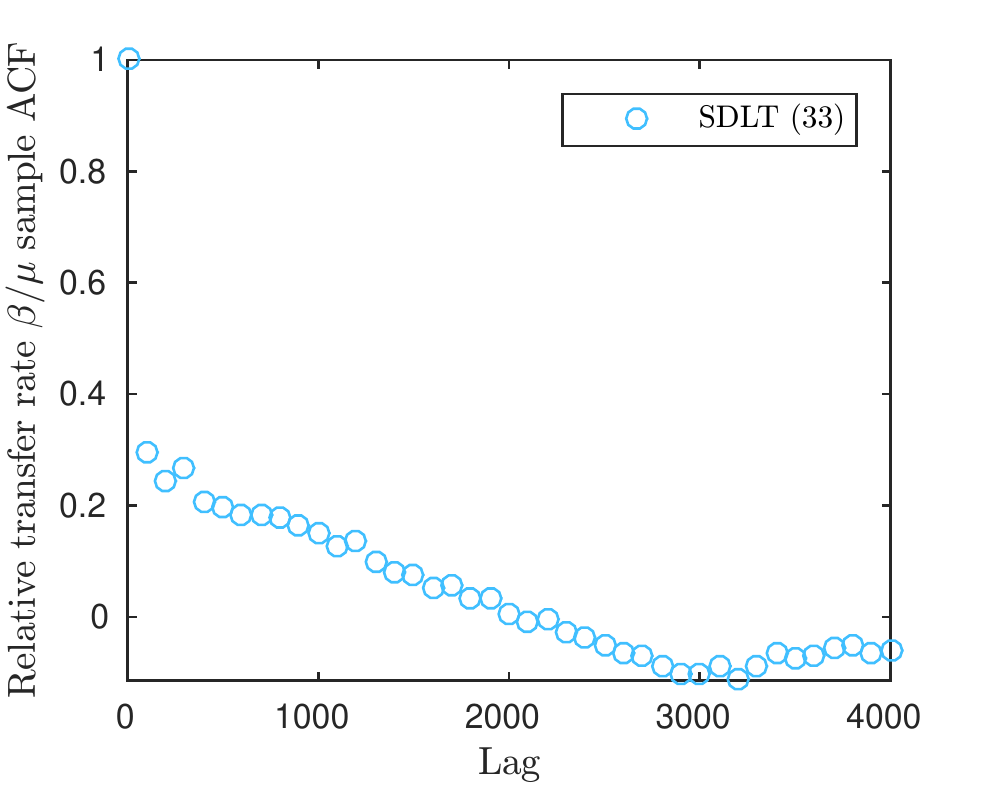} &
	\includegraphics[width=0.5\textwidth, trim = 0.05cm 0cm 0.6cm 0cm, clip]{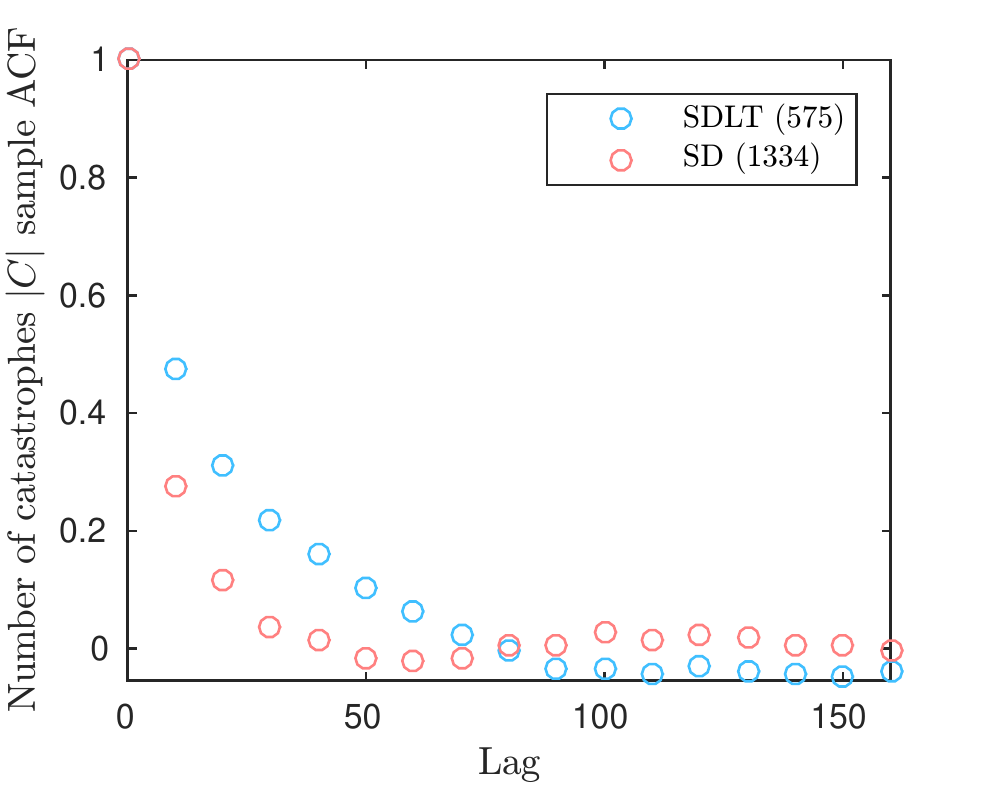} \\
	\includegraphics[width=0.5\textwidth, trim = 0.05cm 0cm 0.6cm 0cm, clip]{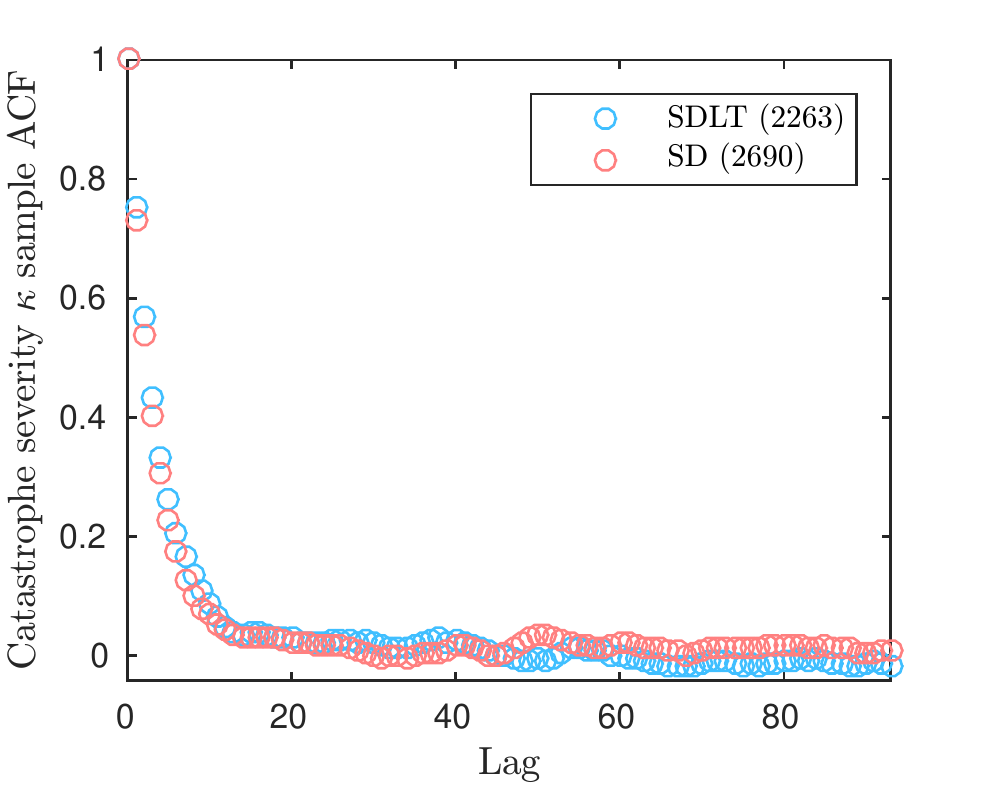} &
	\includegraphics[width=0.5\textwidth, trim = 0.05cm 0cm 0.6cm 0cm, clip]{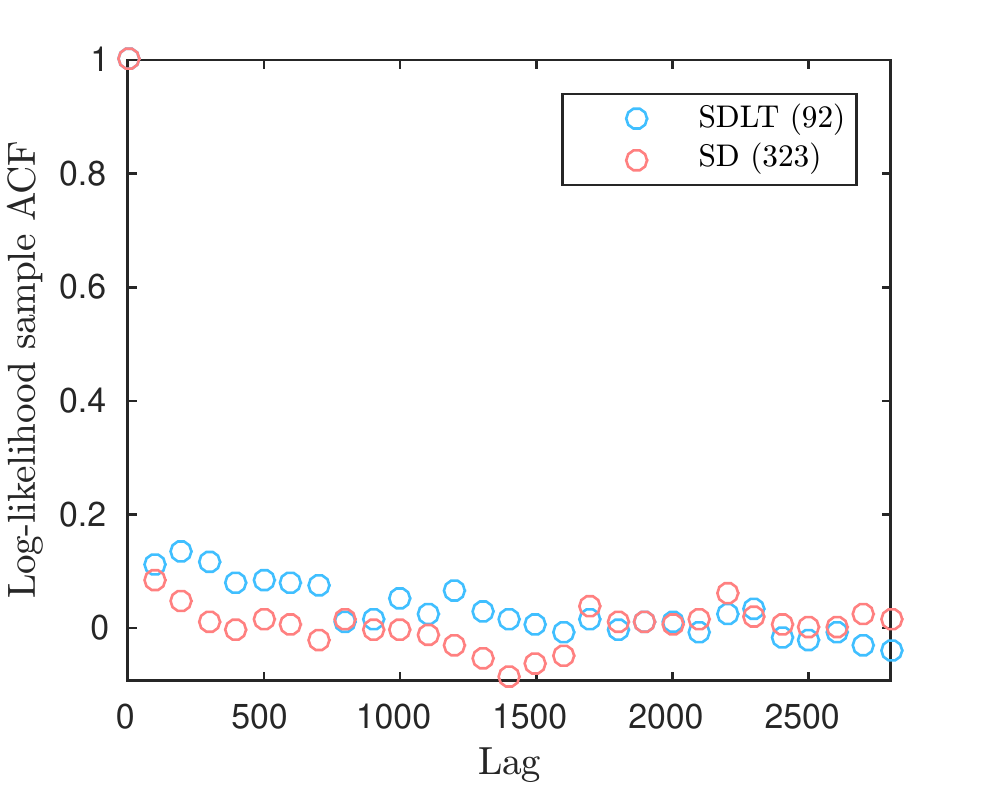}
\end{tabular}
\caption{Autocorrelation plots of samples in our analyses of \texttt{SIM-T}.}
\label{autocorr:simT}
\end{figure}


We repeat the Bayes factor tests in Section~\ref{sec:applications} on the clade constraints in the models applied to the synthetic data sets. Each of the constraints in Figure~\ref{tree:simTrue} restricts the time $ t_i $ of an internal node $ i \in V_A $ to lie on an interval $ [\ubar{t}_i, \bar{t}_i] \subset \R $. We label the clades 1, 2 and 3 and describe their constrained and relaxed versions in Table~\ref{tab:simCons}. We plot histograms of the node ages under each model in Figure~\ref{bf:simLeaves} and report the corresponding Bayes factors~\eqref{eq:SDR} in Figure~\ref{fig:simBF}. Unsurprisingly, the SD model poorly predicts the leaf constraints when fit to \texttt{SIM-B}. There is little to distinguish between the models applied to the other combinations of data sets and constraints as they accurately predict the constraints in the above analyses in each case.

\begin{table}[t]
\centering
\caption[Clade constraints in goodness-of-fit analyses of synthetic data sets.]{Clade constraints in our goodness-of-fit analyses of the synthetic data sets. For constraint 3, we remove the specific time constraint and leave the ancestry constraint.}
\label{tab:simCons}
\begin{tabular}{@{}llll@{}} \toprule
\multirow{2}{*}{\emph{i}}	&	\multirow{2}{*}{Node constraint}
												&	\multicolumn{2}{c}{Time constraint} \\ \cmidrule(l){3-4}
							&					&	\multicolumn{1}{c}{$ \Gamma^{(i)} $}
																		&	\multicolumn{1}{c}{$ \Gamma^{(i')} $} \\ \midrule
1							&	`1'				&	$ [-550, -450] $	&	$ [-800, -200] $	\\
2							&	`8'				&	$ [-150, -50] $		&	$ [-400, +200] $	\\
3							&	$ \pa(\text{`6'}, \dotsc, \text{`10'}) $
												&	$ [-500, -200] $	&	\multicolumn{1}{c}{---}	\\ \bottomrule
\end{tabular}
\end{table}

\begin{figure}[!p]
\centering
\includegraphics[width=\textwidth, trim = 1.25cm 1.5cm 1.25cm 0cm, clip]{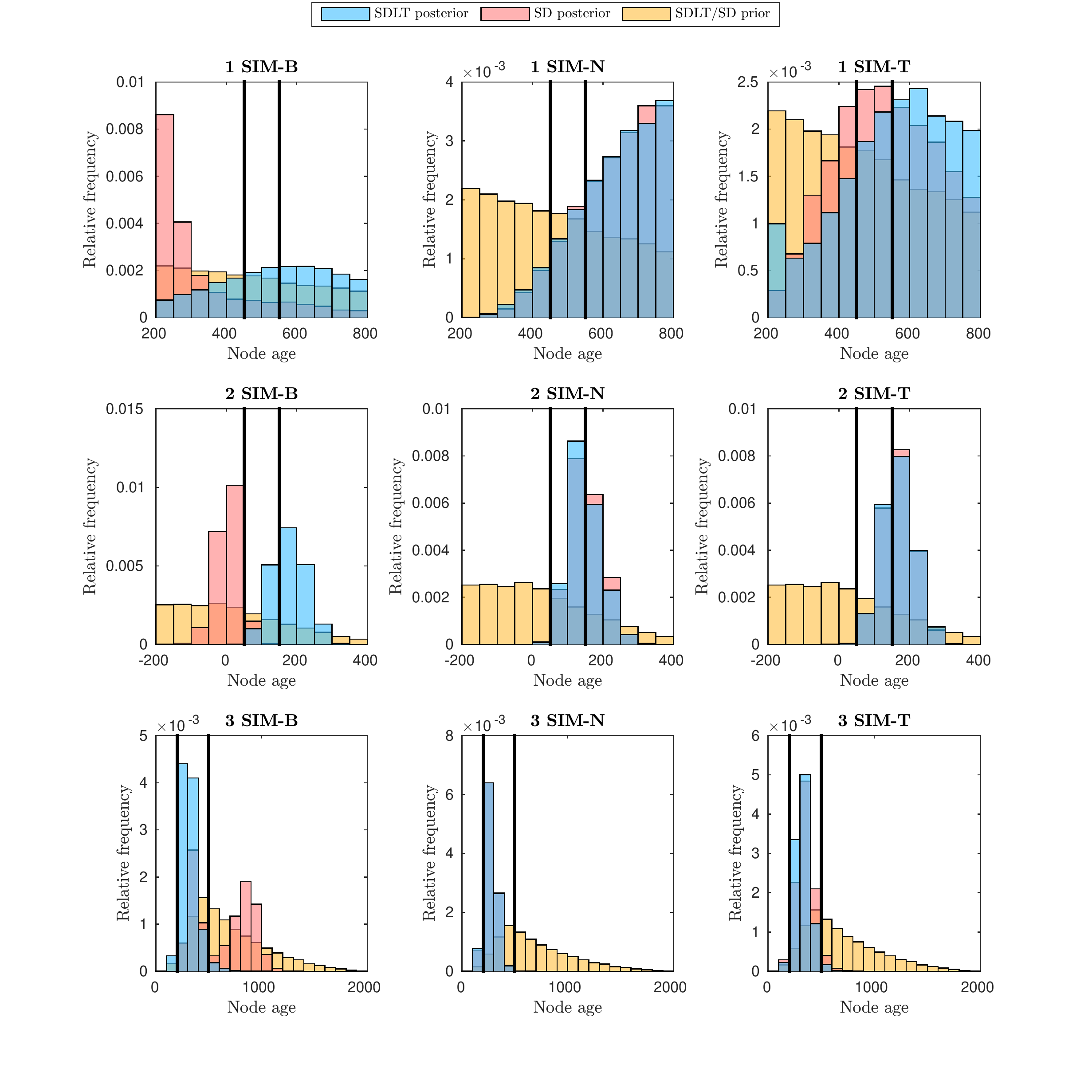}
\caption{We relax each clade constraint in turn and compute the histogram of the corresponding node time under the prior and posterior for each model fit to each of the synthetic data sets. The title of each plot indicates the constraint in Table~\ref{tab:simCons} that we relax for the analysis.}
\label{bf:simLeaves}
\end{figure}

\begin{figure}[t]
\centering
\captionsetup[subfigure]{labelformat=empty}
\subfloat[\texttt{SIM-B}.]{\includegraphics[width=0.32\textwidth, trim = 0.05cm 0cm 0.6cm 0.25cm, clip]{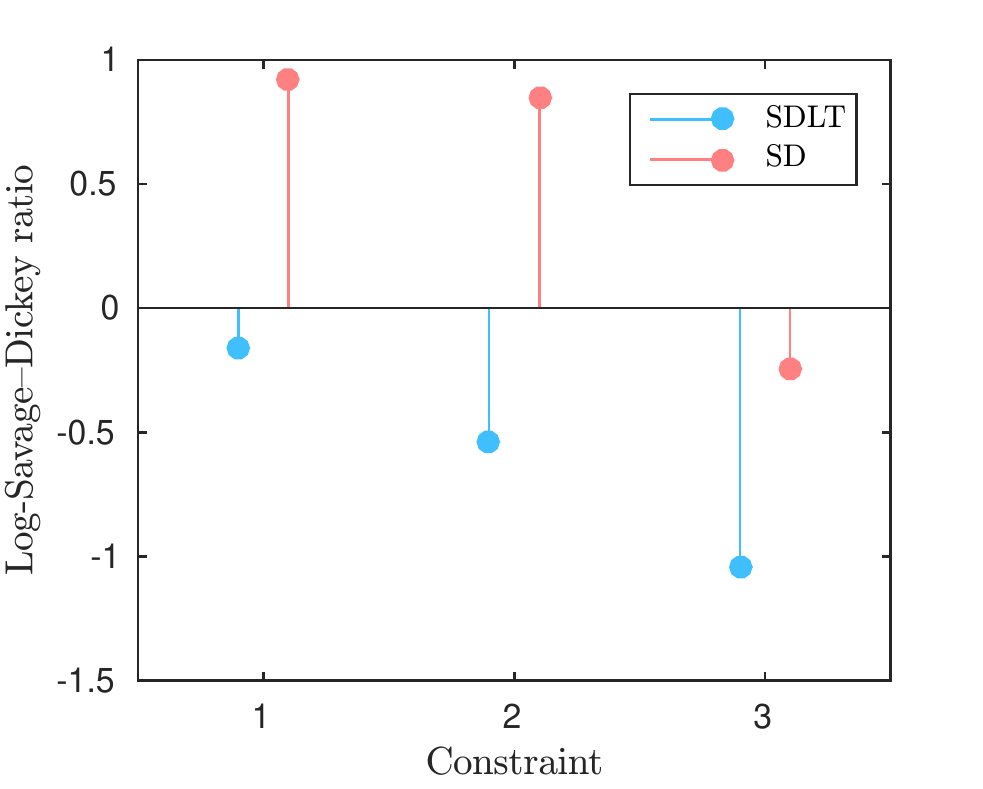}}
\subfloat[\texttt{SIM-N}.]{\includegraphics[width=0.32\textwidth, trim = 0.05cm 0cm 0.6cm 0.25cm, clip]{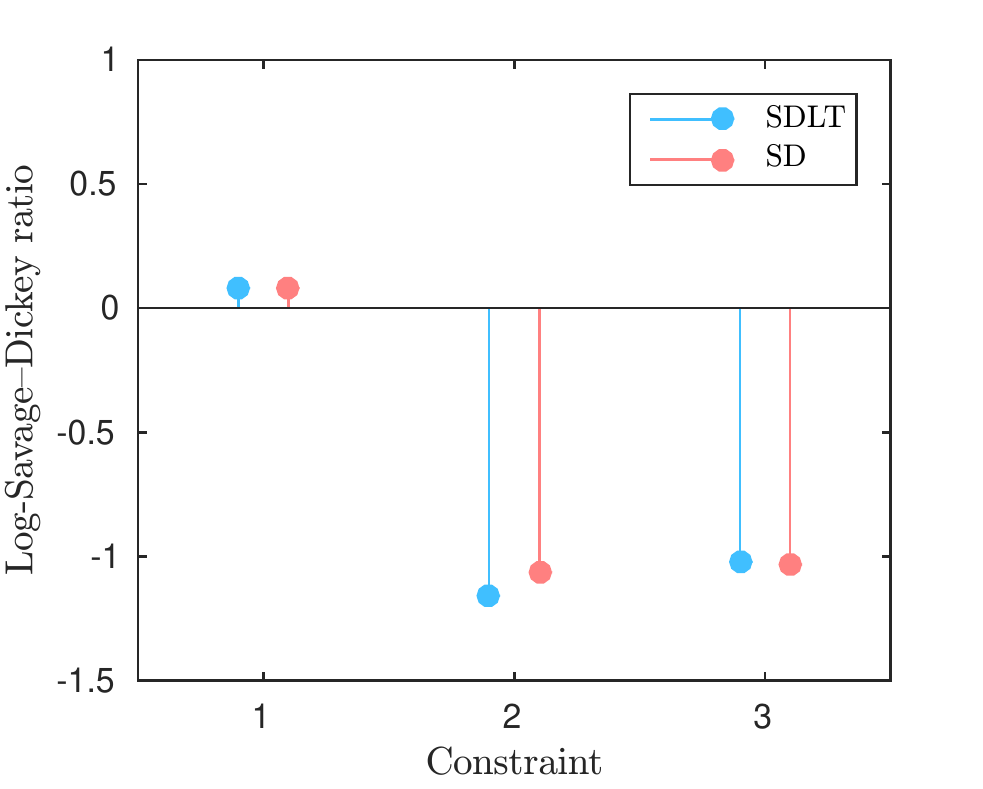}}
\subfloat[\texttt{SIM-N}.]{\includegraphics[width=0.32\textwidth, trim = 0.05cm 0cm 0.6cm 0.25cm, clip]{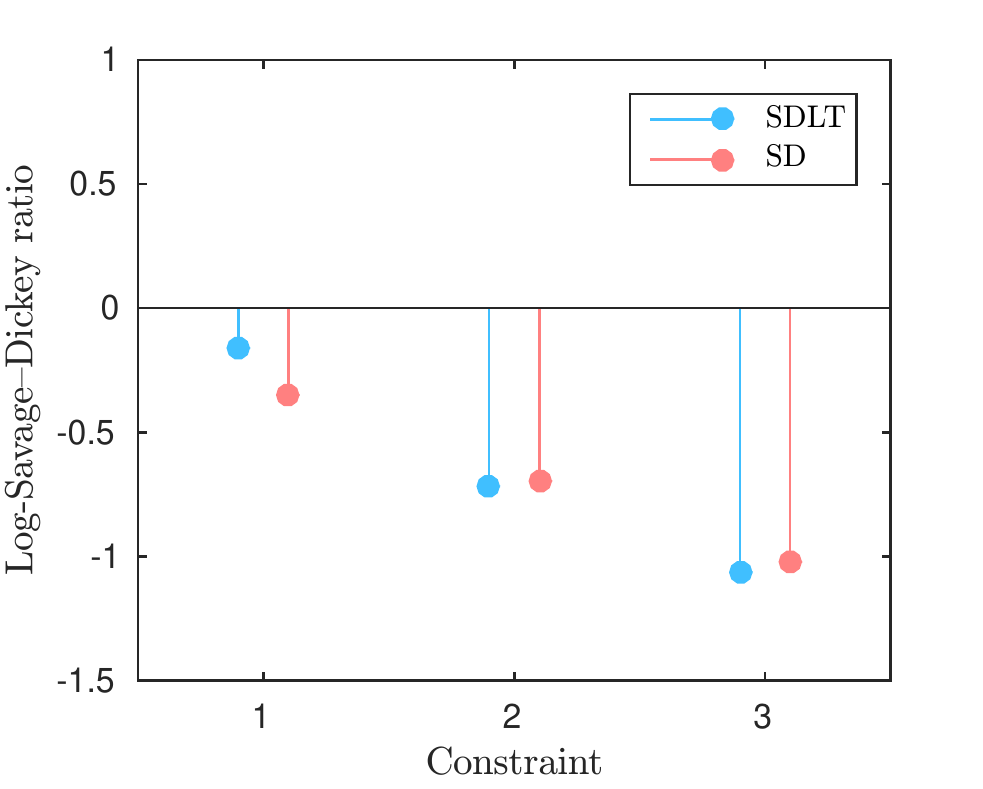}}
\caption[Bayes factors in goodness-of-fit analyses of synthetic data sets]{Bayes factors describing the lack of support for the clade constraints in the SDLT and SD models fit to the synthetic data sets.}
\label{fig:simBF}
\end{figure}

We repeat the predictive performance checks described in Section~\ref{sec:applications} when the registered data $ R(\bD) $ is randomly split into evenly sized training and test portions, $ \bDtr $ and $ \bDte $ respectively. In this case, the training sets are \texttt{SIM-B}, \texttt{SIM-N} and \texttt{SIM-T} from the above analyses, and we draw the test sets from the corresponding distributions. We report the results of this analysis in Table~\ref{tab:pppSim}. The more flexible SDLT model conclusively outperforms the SD model on \texttt{SIM-B}, with little to distinguish the models fit to \texttt{SIM-N} and \texttt{SIM-T}.

\begin{table}[t]
\centering
\caption{Posterior predictive model assessment.}
\label{tab:pppSim}
\begin{tabular}{@{}llll@{}} \toprule
Data set		&	SDLT score	&	SD score 	&	Log-Bayes factor \\ \midrule
\texttt{SIM-B}	&	-3910.0		&	-4057.9		&	147.9	\\
\texttt{SIM-N}	&	-3582.4		&	-3581.9 	&	-0.5	\\
\texttt{SIM-T}	&	-3686.1		&	-3690.7 	&	4.6	\\ \bottomrule
\end{tabular}
\end{table}

On the basis of these analyses, we are satisfied that our inference scheme is correct, and that the SDLT model is identifiable, consistent with the SD model in the absence of lateral transfer, and robust to a common form of model misspecification as in each case parameter samples from the SD model are consistent with their true values in Figure~\ref{tree:simTrue} and Table~\ref{tab:simPars}. We cannot say the same for the SD model fit to \texttt{SIM-B} and our goodness-of-fit analyses confirm this. An alternative approach to model comparison is to use reversible jump MCMC to sample from a posterior on models and parameters. The SD model is nested within the SDLT model so it is not difficult to implement such an algorithm. However, with our current set of proposal distributions, we are unable to bridge the gap between the SDLT and SD models fit to \texttt{SIM-B} to obtain an accurate estimate of the corresponding Bayes factor.

\section{Applications}
\label{app:applications}

This section contains figures to support our analyses of the Polynesian data set \texttt{POLY-0} in Section~\ref{sec:applications}. We summarise the sampled trees with majority-rule consensus trees in Figure~\ref{cons:poly0}. We plot histograms of parameter samples in Figure~\ref{hist:poly0}, trace plots in Figure~\ref{trace:poly0}, autocorrelation plots in Figure~\ref{autocorr:poly0} and marginal leaf times when constraints are relaxed in Figure~\ref{bf:polyLeaves}. Figures in parentheses denote effective sample sizes.

\begin{figure}[p!ht]
\captionsetup[subfigure]{labelformat=empty}
\centering
\subfloat[SDLT.]{\includegraphics[width=0.475\textwidth, trim = 0cm 0cm 0cm 0cm, clip]{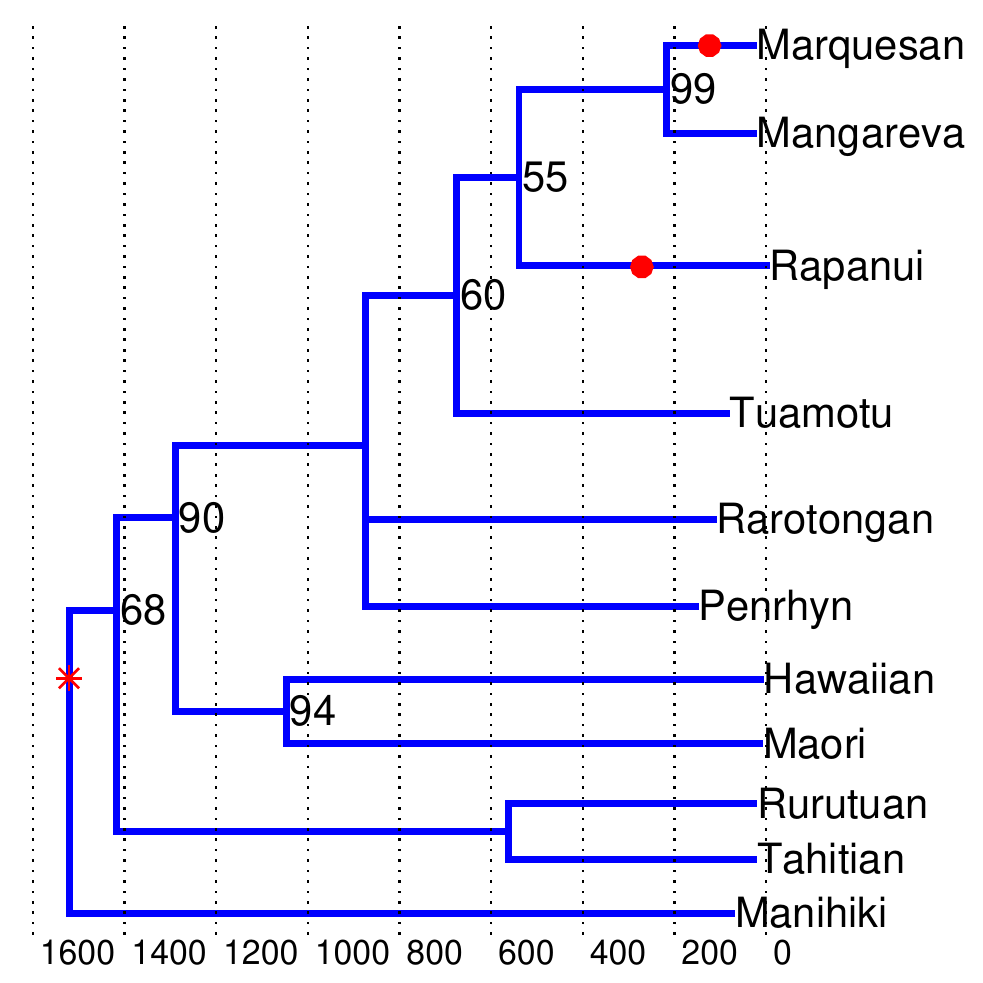}} ~
\subfloat[SD.]{\includegraphics[width=0.475\textwidth, trim = 0cm 0cm 0cm 0cm, clip]{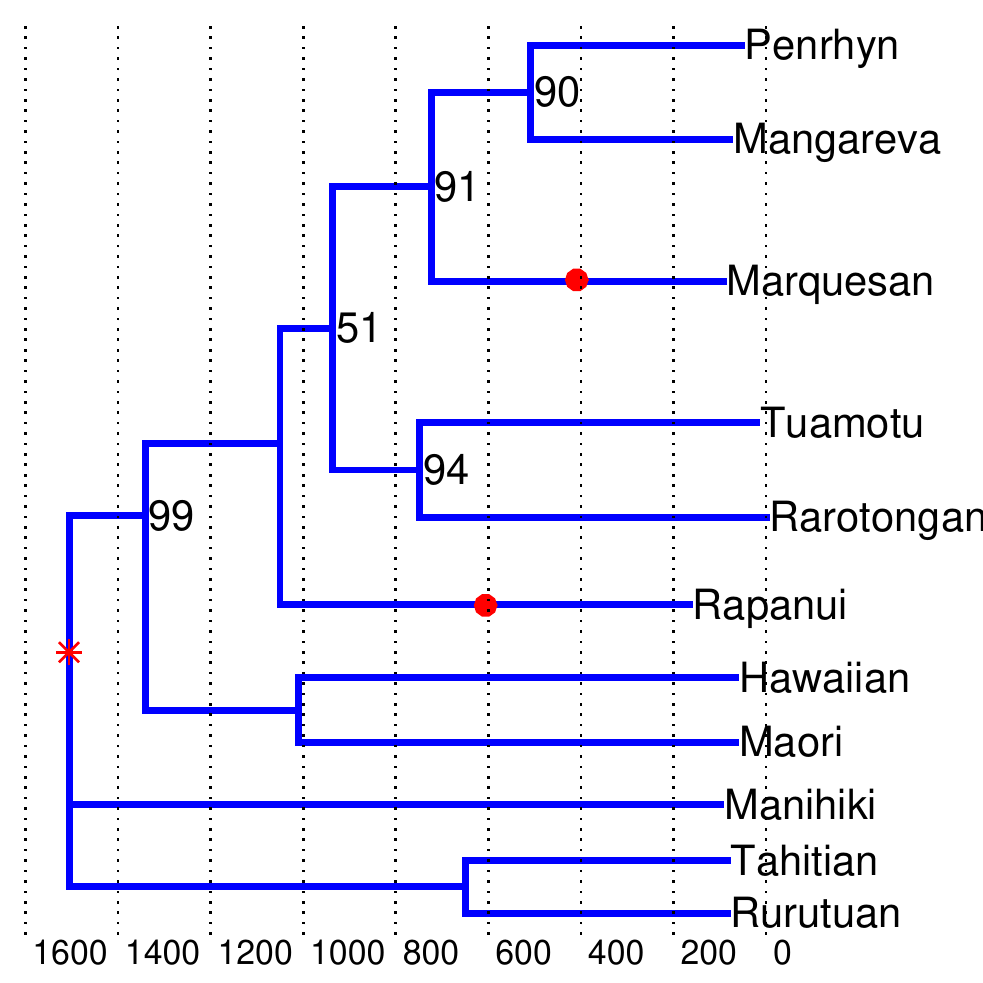}}
\caption{Majority-rule consensus trees for our analyses of \texttt{POLY-0}. A majority rule consensus tree depicts the relationships which appear in the majority of the sampled trees. The tree is multifurcating when a given edge does not appear in the majority of the samples. Red circles represent the number of catastrophes on a branch to the nearest integer.}
\label{cons:poly0}
\end{figure}

\begin{figure}[p!ht]
\centering
\begin{tabular}{@{}c@{}@{}c@{}}
	\includegraphics[width=0.5\textwidth, trim = 0.05cm 0cm 0.6cm 0cm, clip]{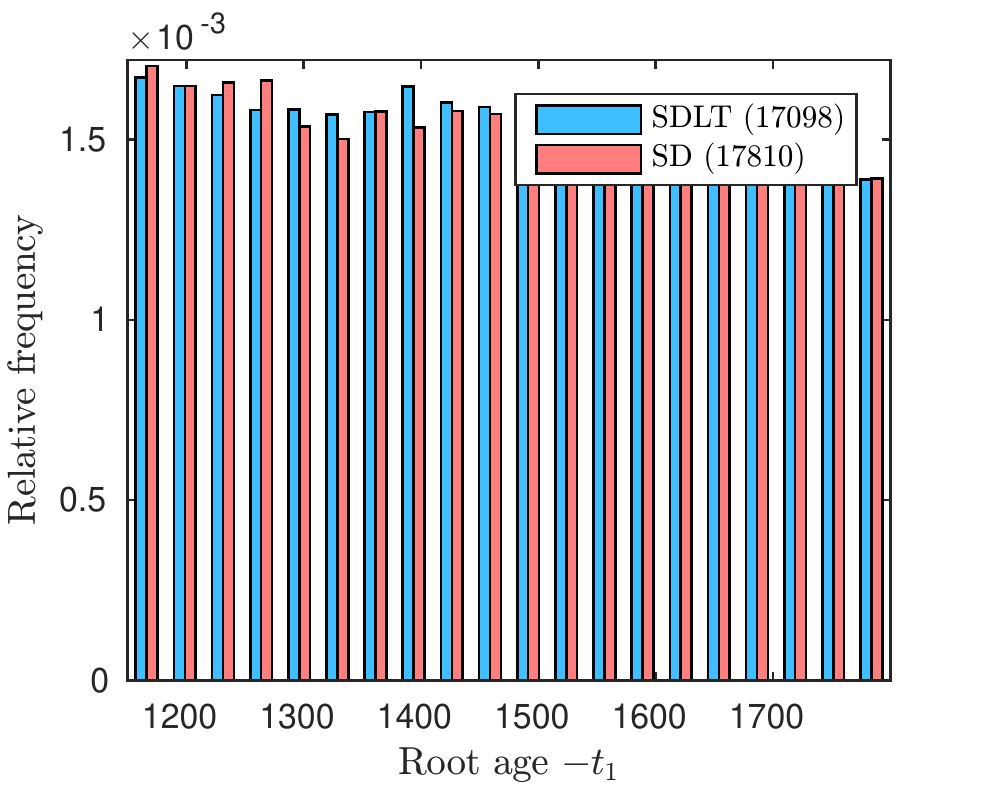} &
	\includegraphics[width=0.5\textwidth, trim = 0.05cm 0cm 0.6cm 0cm, clip]{POLY/POLY0/Exact/CR/histMu.pdf} \\
	\includegraphics[width=0.5\textwidth, trim = 0.05cm 0cm 0.6cm 0cm, clip]{POLY/POLY0/Exact/CR/histBM.pdf} &
	\includegraphics[width=0.5\textwidth, trim = 0.05cm 0cm 0.6cm 0cm, clip]{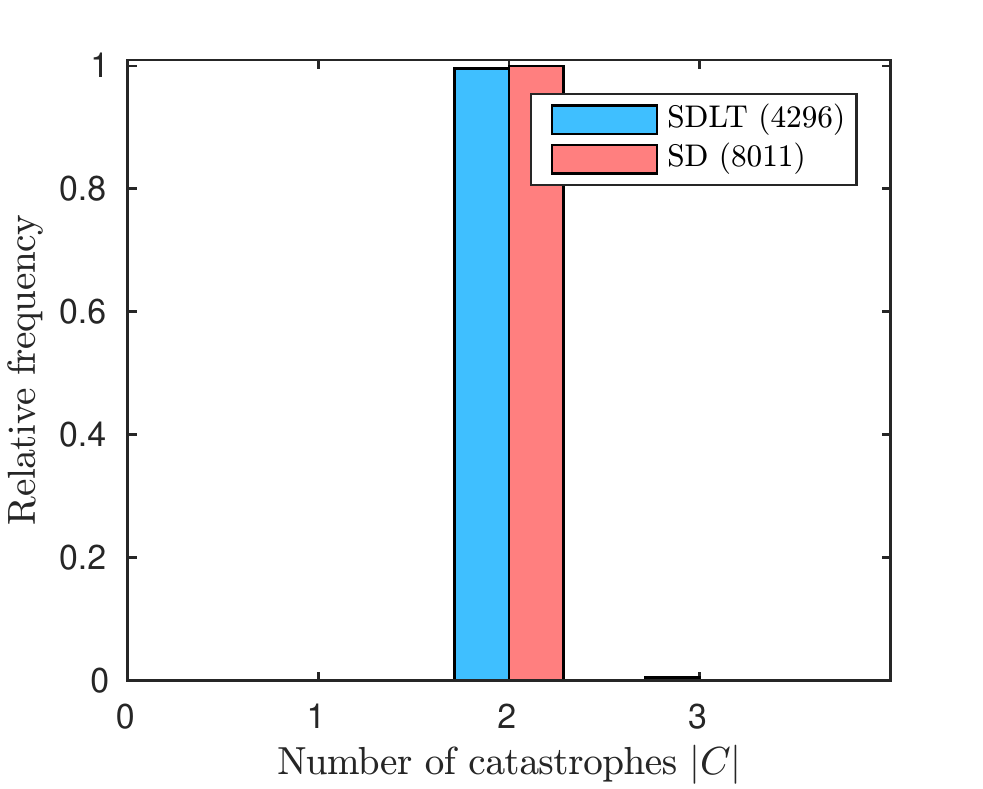} \\
	\includegraphics[width=0.5\textwidth, trim = 0.05cm 0cm 0.6cm 0cm, clip]{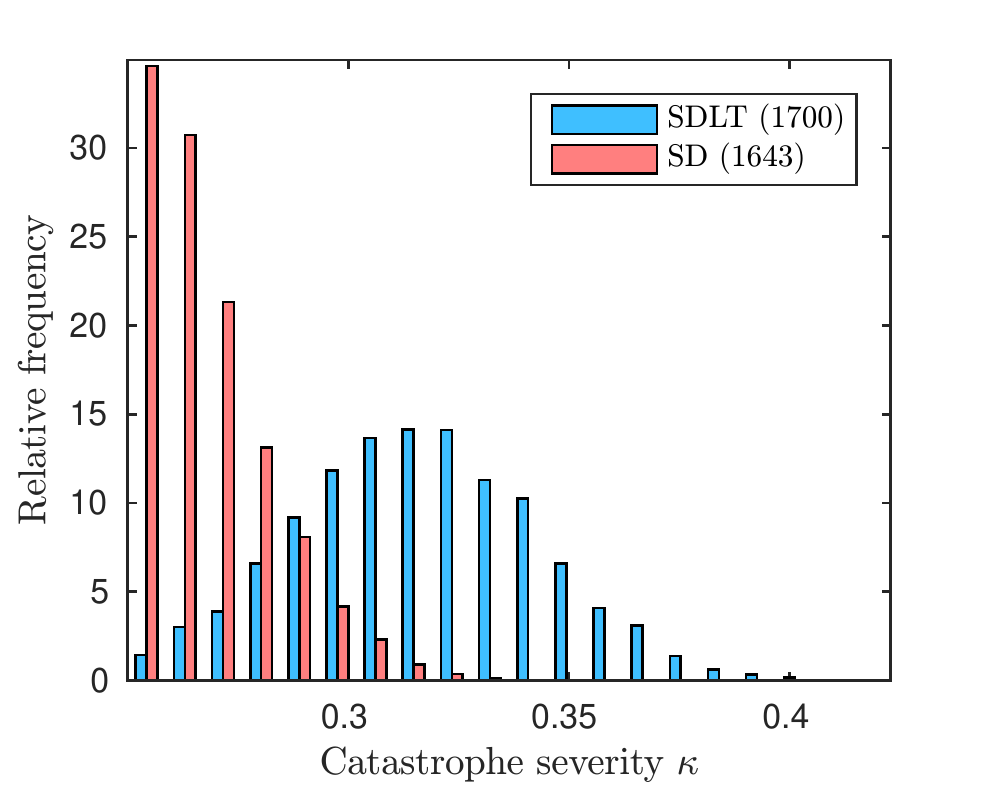} &
	\includegraphics[width=0.5\textwidth, trim = 0.05cm 0cm 0.6cm 0cm, clip]{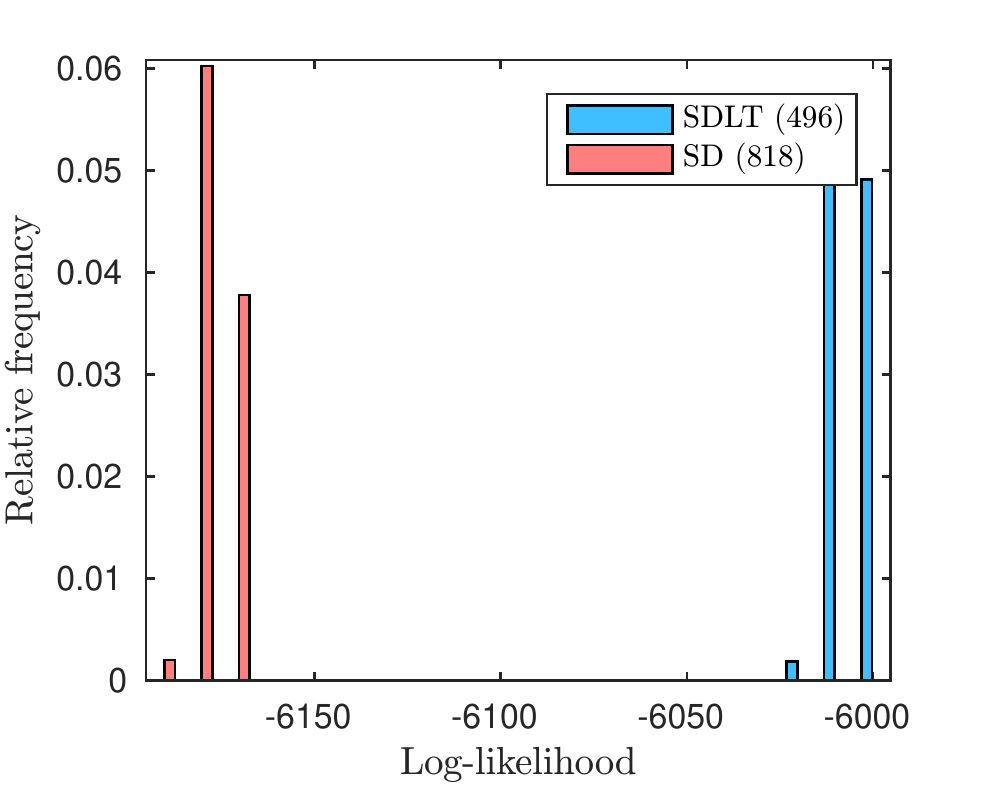}
\end{tabular}
\caption{Histograms of samples in our analyses of \texttt{POLY-0}.}
\label{hist:poly0}
\end{figure}

\begin{figure}[p!ht]
\centering
\begin{tabular}{@{}c@{}@{}c@{}}
	\includegraphics[width=0.5\textwidth, trim = 0.05cm 0cm 0.6cm 0cm, clip]{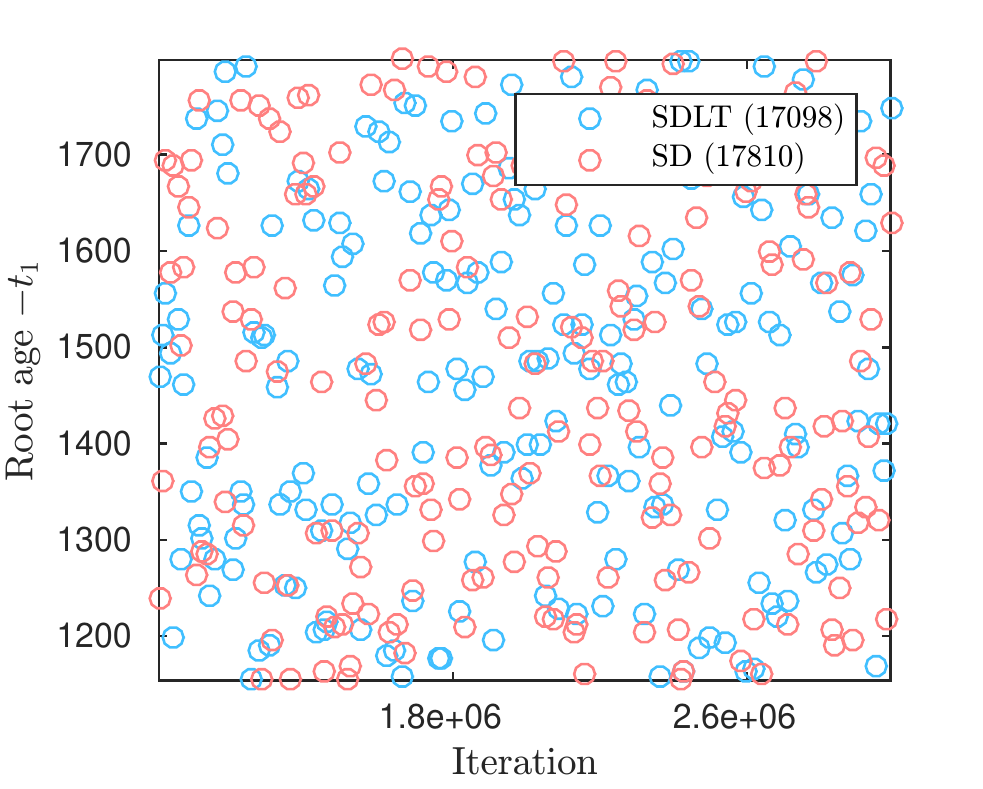} &
	\includegraphics[width=0.5\textwidth, trim = 0.05cm 0cm 0.6cm 0cm, clip]{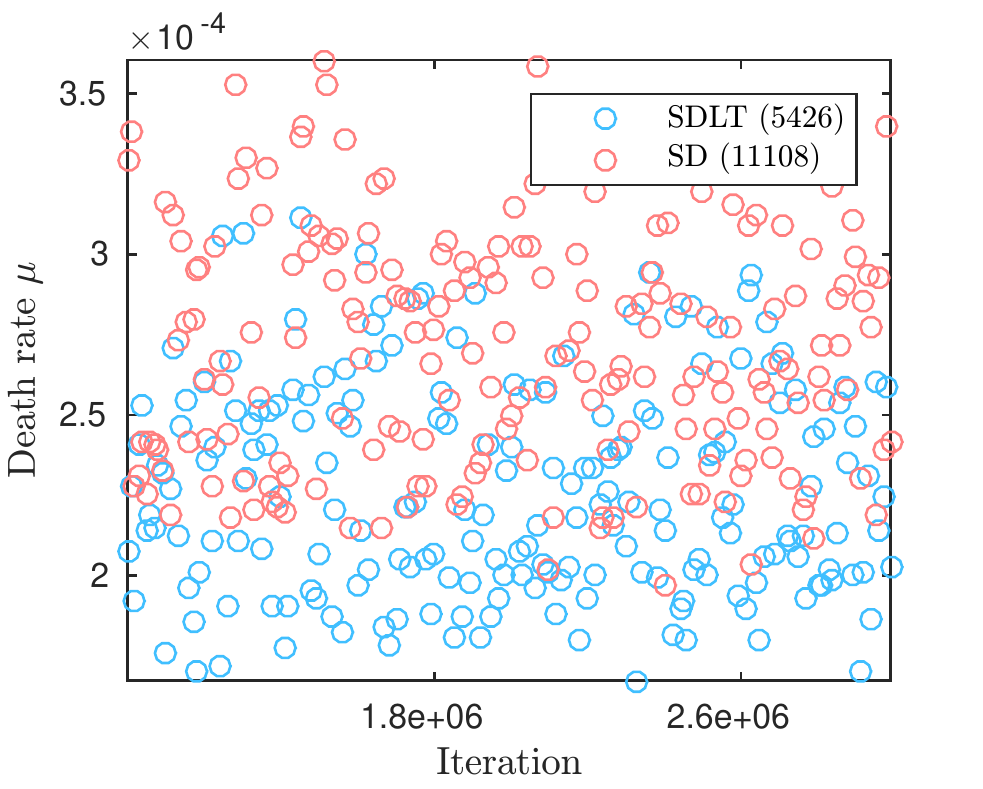} \\
	\includegraphics[width=0.5\textwidth, trim = 0.05cm 0cm 0.6cm 0cm, clip]{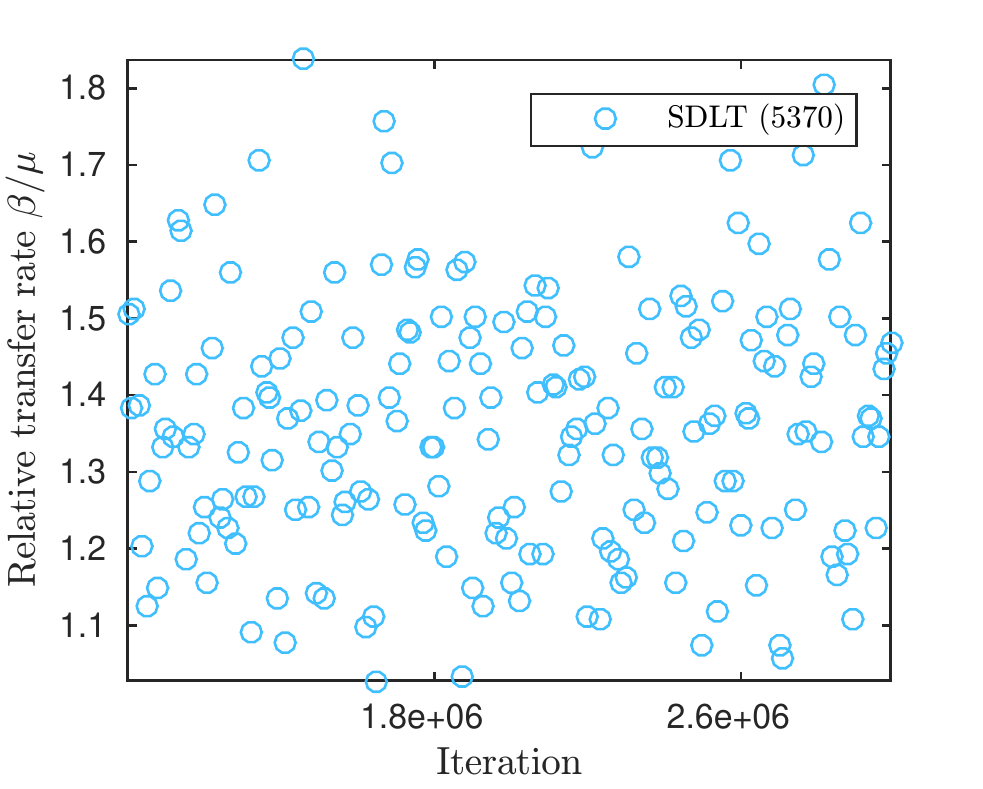} &
	\includegraphics[width=0.5\textwidth, trim = 0.05cm 0cm 0.6cm 0cm, clip]{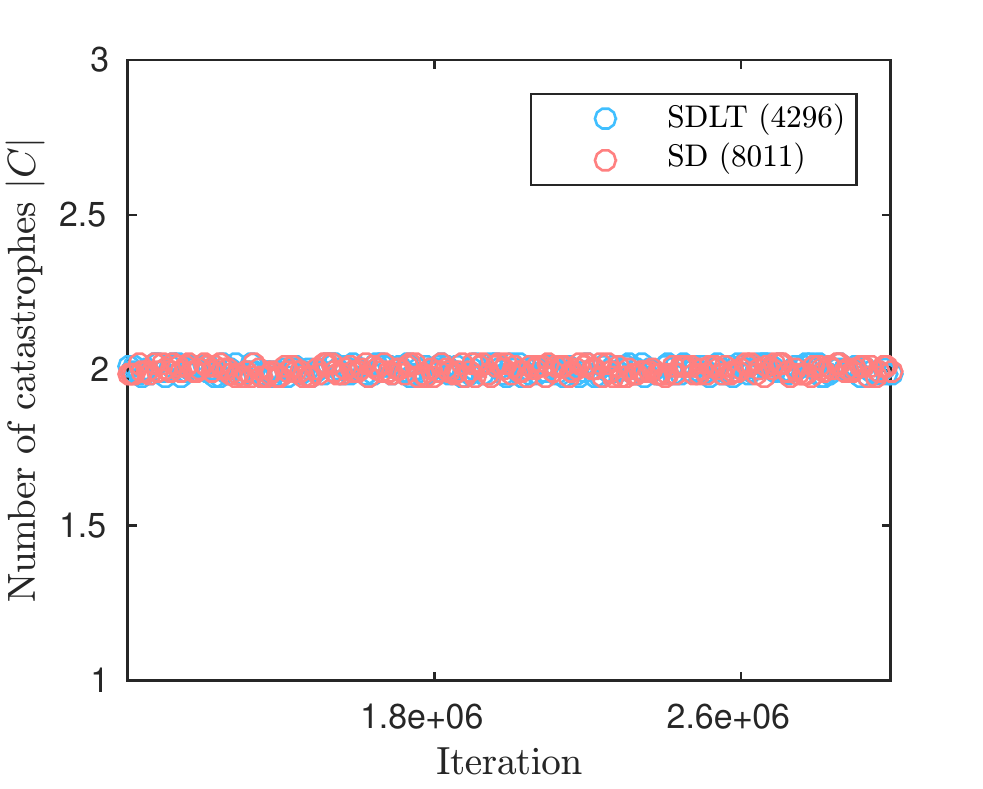} \\
	\includegraphics[width=0.5\textwidth, trim = 0.05cm 0cm 0.6cm 0cm, clip]{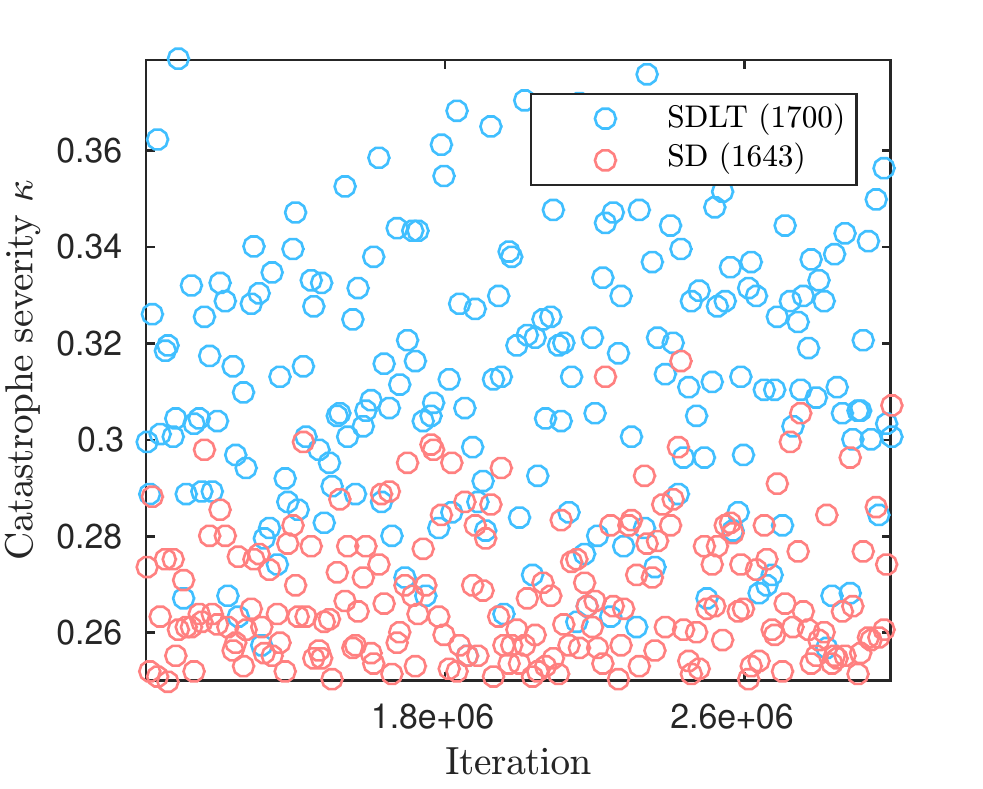} &
	\includegraphics[width=0.5\textwidth, trim = 0.05cm 0cm 0.6cm 0cm, clip]{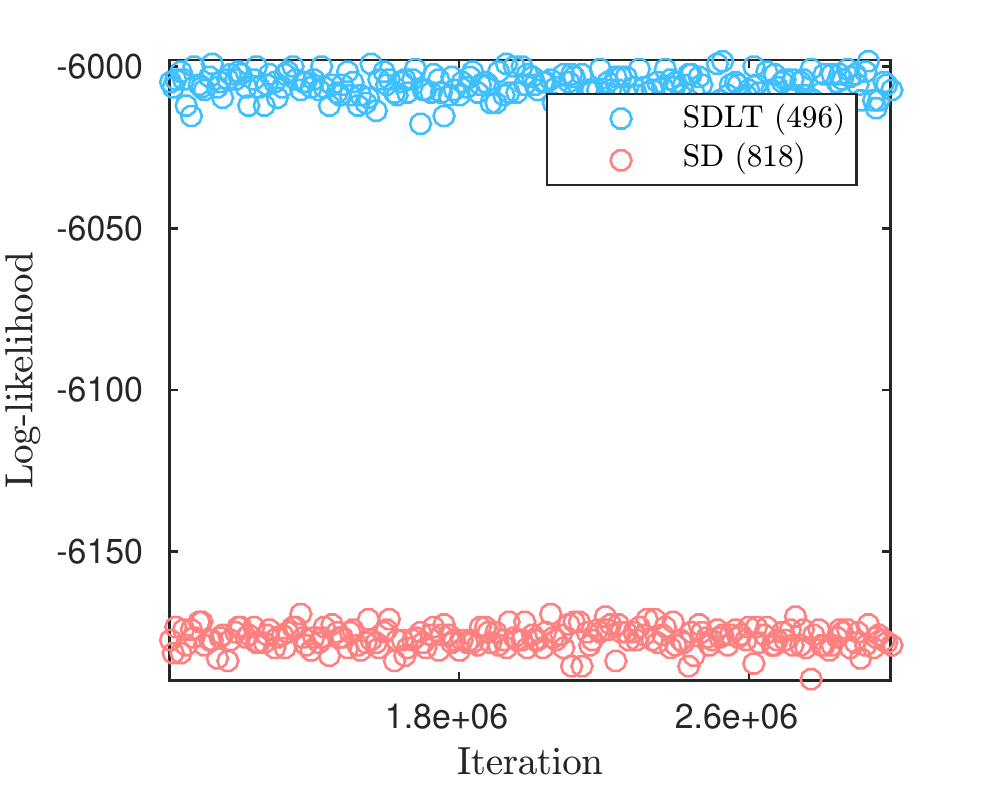}
\end{tabular}
\caption{Trace plots of samples in our analyses of \texttt{POLY-0}. For clarity, the trace plots only depict a subset of the samples and this explains the discrepancy between the effective sample size for the number of catastrophes and the corresponding plot.}
\label{trace:poly0}
\end{figure}

\begin{figure}[p!ht]
\centering
\begin{tabular}{@{}c@{}@{}c@{}}
	\includegraphics[width=0.5\textwidth, trim = 0.05cm 0cm 0.6cm 0cm, clip]{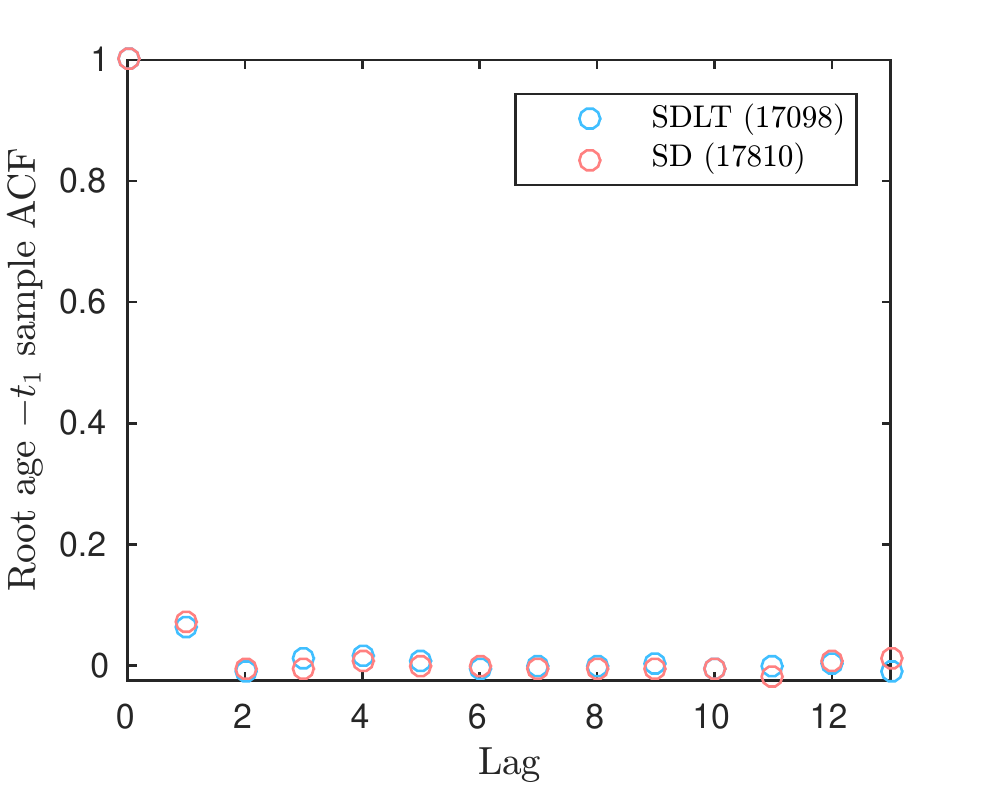} &
	\includegraphics[width=0.5\textwidth, trim = 0.05cm 0cm 0.6cm 0cm, clip]{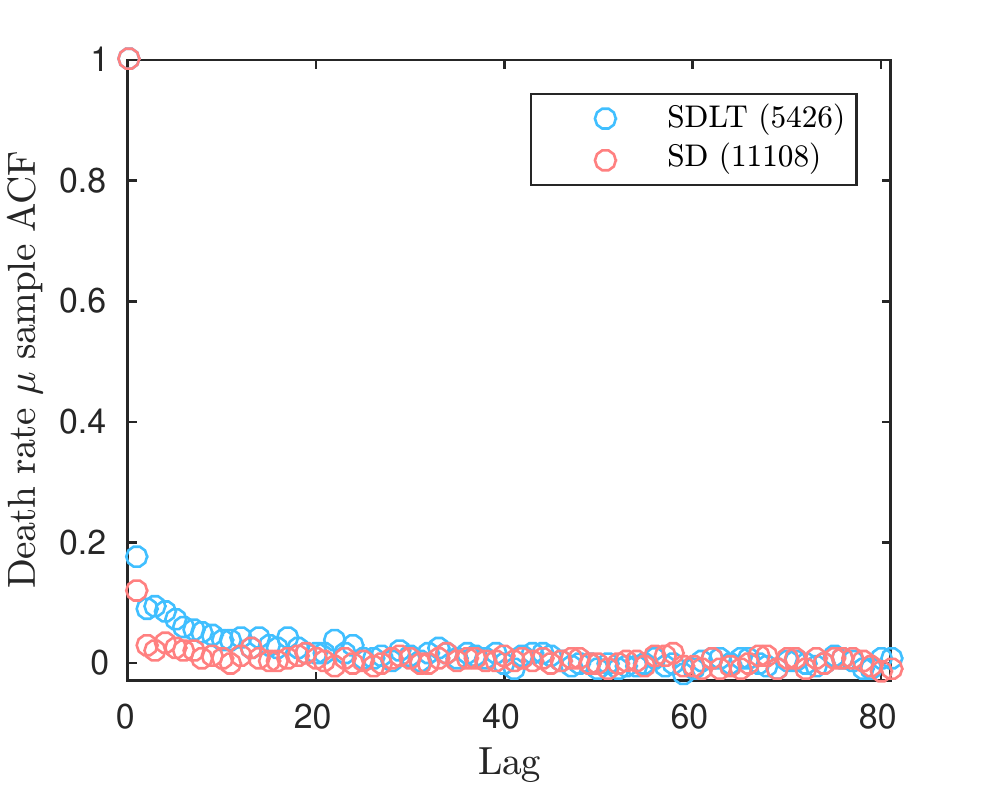} \\
	\includegraphics[width=0.5\textwidth, trim = 0.05cm 0cm 0.6cm 0cm, clip]{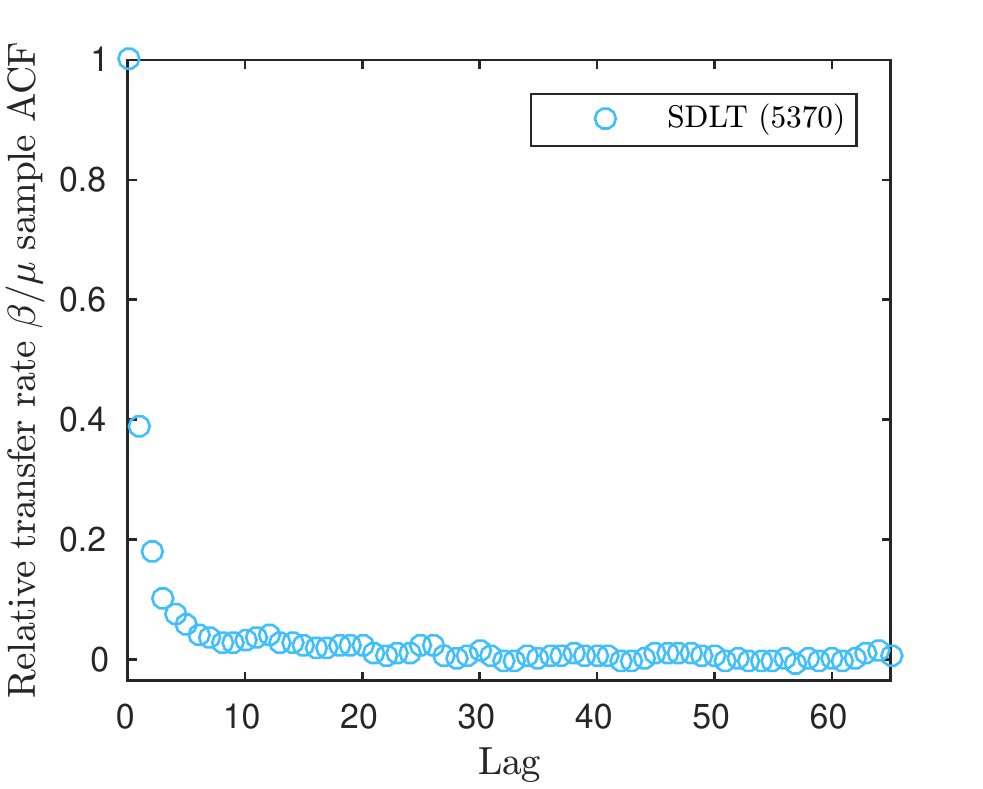} &
	\includegraphics[width=0.5\textwidth, trim = 0.05cm 0cm 0.6cm 0cm, clip]{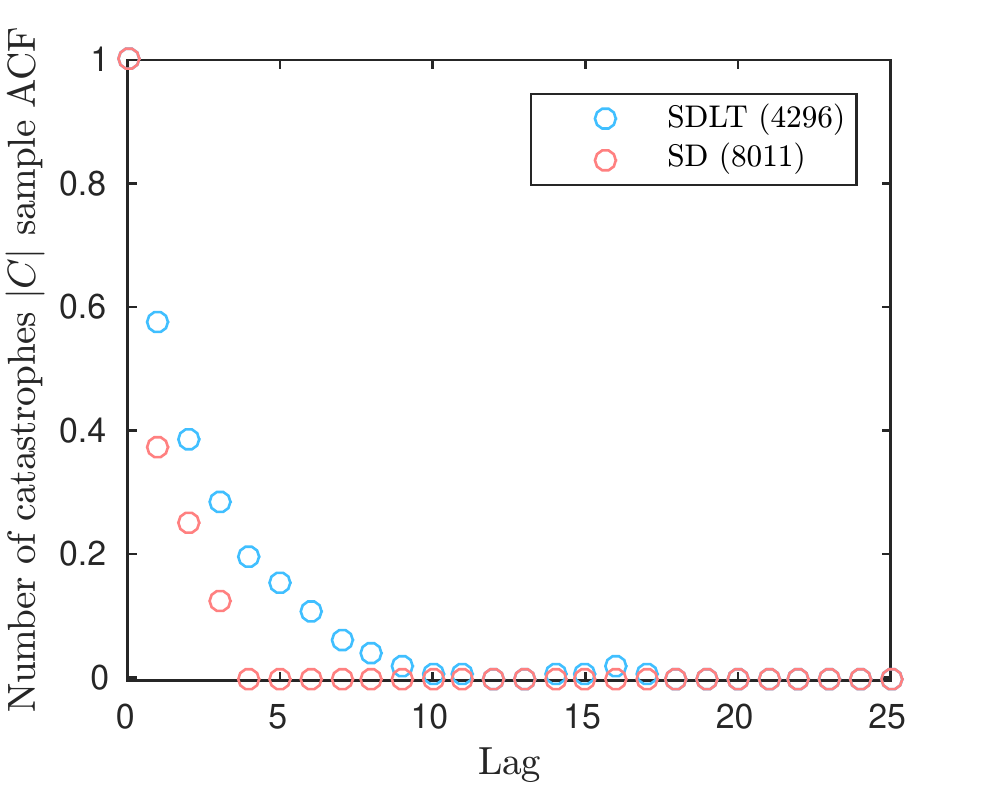} \\
	\includegraphics[width=0.5\textwidth, trim = 0.05cm 0cm 0.6cm 0cm, clip]{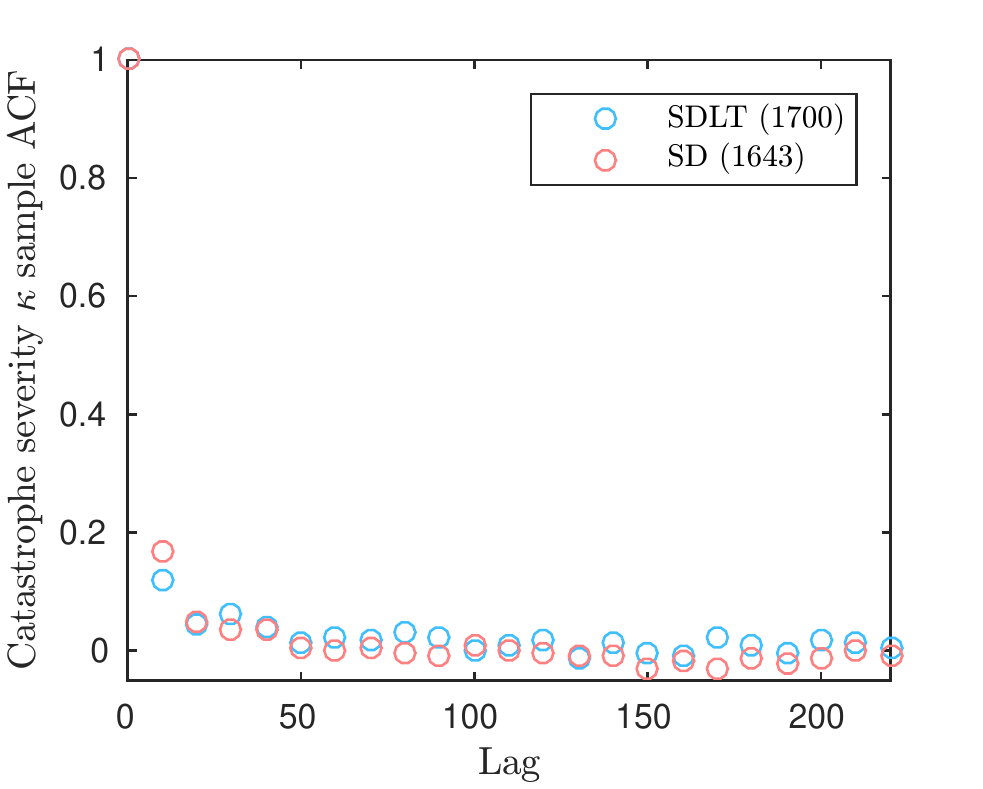} &
	\includegraphics[width=0.5\textwidth, trim = 0.05cm 0cm 0.6cm 0cm, clip]{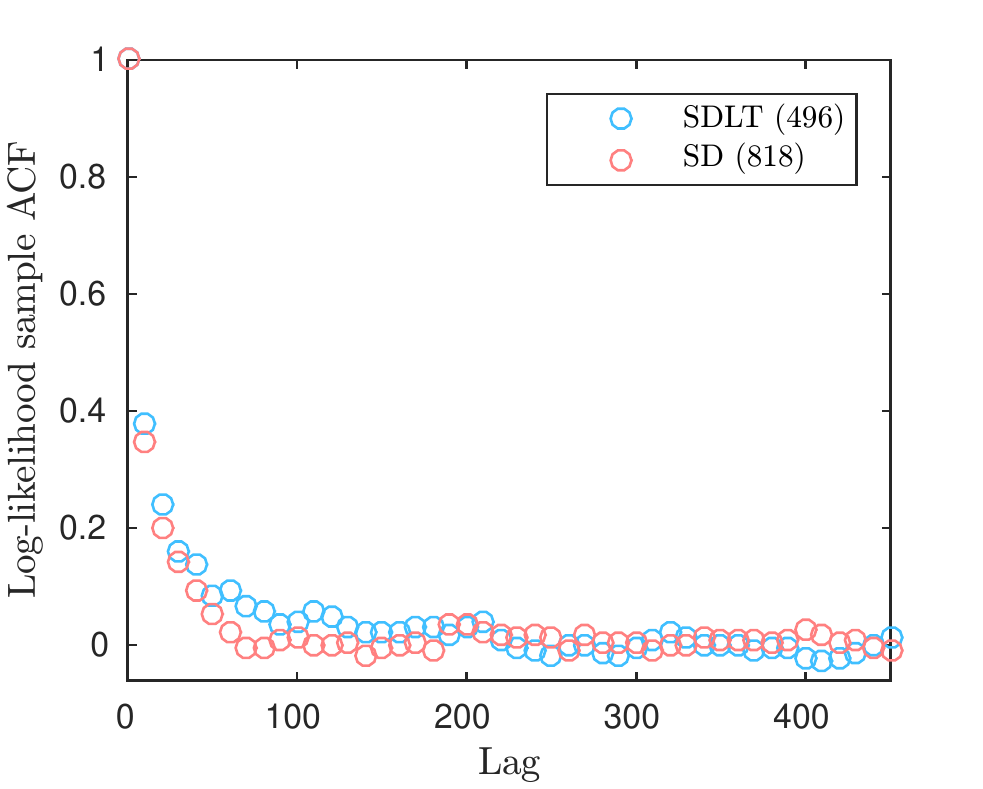}
\end{tabular}
\caption{Sample autocorrelation plots in our analyses of \texttt{POLY-0}.}
\label{autocorr:poly0}
\end{figure}

\begin{figure}[p!ht]
\centering
\includegraphics[width=\textwidth, trim = 1.25cm 1.5cm 1.25cm 1cm, clip]{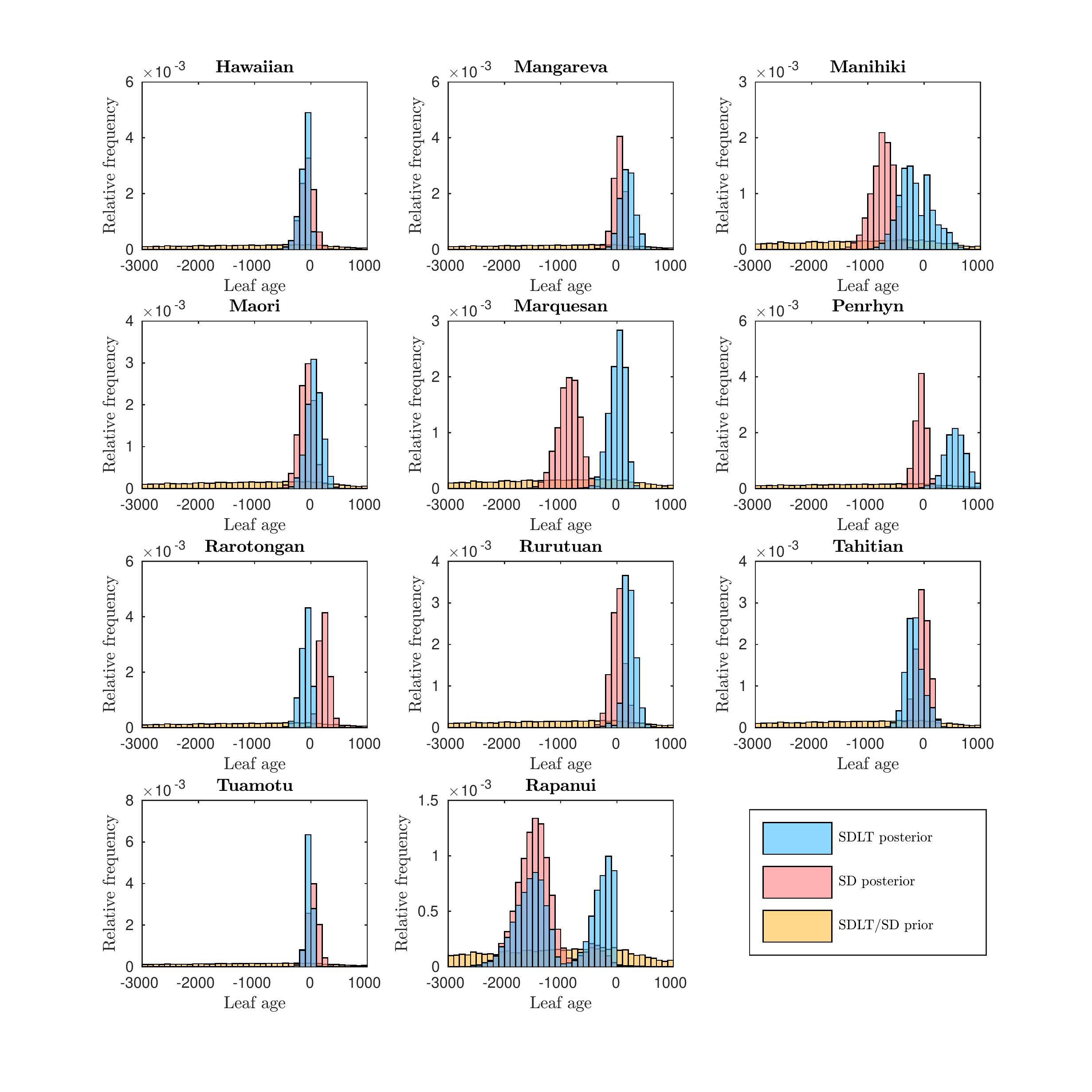}
\caption{We relax the constraint on each leaf in turn and compute the histogram of its time under the prior and posterior for each model fit to \texttt{POLY-0}. Time in years is on the horizontal axis and relative frequency on the vertical axis.}
\label{bf:polyLeaves}
\end{figure}

\end{document}